# A comparison of methods for designing hybrid type 2 cluster-randomized trials with continuous effectiveness and implementation endpoints

**Melody A. Owen**[1], **Fan Li**[1], **Ruyi Liu**[1], **Donna Spiegelman**[1]
[1] melody.owen@yale.edu Center for Methods in Implementation and Prevention Science, Yale University, New Haven, CT, USA

May 14, 2026

## Abstract
Hybrid type 2 studies are gaining popularity for their ability to assess both implementation and health outcomes as co-primary endpoints. Often conducted as cluster-randomized trials (CRTs), five design methods can validly power these studies: p-value adjustment methods, combined outcomes approach, single weighted 1-DF test, disjunctive 2-DF test, and conjunctive test. We compared these methods theoretically and numerically. Theoretical comparisons of power equations allowed us to identify when one method had more or less power than another globally. We showed that p-value adjustment methods are always less powerful than both the combined outcomes approach and the single 1-DF test, and identified conditions where the disjunctive 2-DF test is less powerful than the single 1-DF test. To further investigate when power advantages shift, we conducted a large-scale numerical study using our novel `crt2power` R package, which calculates power or sample size for CRTs with two continuous co-primary endpoints using these methods. Across 45,000 input scenarios, we found specific patterns: when treatment effects are unequal, the disjunctive 2-DF test tends to be most powerful; when treatment effects are equal, the single 1-DF test tends to dominate. Together, these comparisons offer practical guidance for powering hybrid type 2 studies.

## 1. Introduction

Cluster randomized trials (CRTs) are studies in which the unit of randomization is a cluster rather than an individual. Clusters can be villages, towns, hospitals, wards of a hospital, etc. CRTs can offer logistical and administrative convenience, reduce treatment group contamination, and are advantageous when the intervention in question is best administered at a cluster level.[1] In implementation science, our fundamental research goal is to understand how best to deliver an intervention effectively; in these endeavors, CRTs are often utilized. This is because many implementation outcomes are often measured and assessed at the cluster level. Effectiveness-implementation hybrid designs offer simultaneous assessment of a health (or effectiveness) outcome and an implementation outcome. On one end of the spectrum, hybrid type 1 studies consider the health outcome as the primary outcome, while the implementation outcome is the secondary outcome. Hybrid type 3 studies consider the implementation outcome as the primary outcome, and the health outcome as the secondary outcome. The focus of this manuscript is the hybrid type 2 study, which considers both outcomes as co-primary outcomes.[2]

Hybrid type 2 studies are very advantageous because they allow for simultaneous analysis of both an effectiveness outcome and an implementation outcome in one study. With this added efficiency are various statistical complications. The first is the issue of multiple testing, where one must control the overall type I error rate. The management of this rate will have important implications on the overall study design parameters that result from the power calculations. There is also the complication of clustering – this introduces various correlations that come from the hierarchical structure of the data. In a recent study by Owen et al., five study design methods (specifically for sample size and power calculation) that can be applied to hybrid type 2 studies were identified through a literature review.[3] These methods were used on data motivated by a real-world hybrid study in order to showcase how the calculations for these methods could be conducted and how the considerations differ. However, to date, no formal analytic or simulation-based comparisons were made across the five methods in the context of hybrid type 2 designs. To fill in this gap, we aim to thoroughly examine the performance of these methods via theoretical comparisons and a numerical evaluation, where different scenarios have different assumptions and input parameters.

In order to conduct the numerical evaluation, and to make these methods more widely available for practice with hybrid type 2 designs, we also created an R package called `crt2power`.[4] This package allows users to calculate the statistical power or sample size requirements for CRTs with two continuous co-primary endpoints given a set of input parameters. This package includes code for each of the five methods identified in Owen et al., namely the p-value adjustment methods, the combined outcomes approach, the single weighted 1-degree of freedom (DF) combined test, the disjunctive 2-DF test, and the conjunctive intersection-union test.[3] Each method has three functions: 1) a function to calculate the statistical power (the probability that a given test correctly rejects the null hypothesis) given the number of clusters, cluster size (number of individuals per cluster), and other input parameters, 2) a function to calculate the required number of clusters given the statistical power and cluster size, and 3) a function to calculate the cluster size given the statistical power and number of clusters. This R package and the usage information is publicly available at https://github.com/melodyaowen/crt2power. It is also available on The Comprehensive R Archive Network (CRAN) at https://cran.r-

project.org/web/packages/crt2power/index.html.[4] Accompanying this R package is a ShinyApp, `crt2powerApplication`, which provides a user-friendly interface where input values can be entered, and the resulting power or sample size will be calculated and displayed for all five methods. This ShinyApp is available at https://mowen17.shinyapps.io/crt2powerApplication/.

In this manuscript, we begin by introducing notation for CRTs and hybrid type 2 studies, as well as describing the study design methods that are examined in the numerical evaluation and introduced in Owen et al.[3] in Section 2. Then, a thorough description of the software package and ShinyApp are provided along with usage examples in Section 3. Next, a theoretical comparison of the power equations is provided in Section 4. Lastly, a numerical evaluation comparing the performance of these methods is conducted in Section 5.

## 2. Methods

### 2.1 Notation

We begin by introducing notation for CRTs in a hybrid type 2 setting. In this setting, we have two continuous primary outcomes, denoted as $Q = 2$ with outcome index $q = 1, 2$. We consider two treatment groups; in this manuscript, we assume equal treatment allocation, but note that our software accommodates unequal treatment allocation. Then, we denote $K_1$ as the number of clusters in the treatment group, and $K_2$ as the number of clusters in the control group (under equal allocation, $K_1 = K_2$). We can also simply refer to it as $K$ under equal treatment allocations, and the clusters are indexed as $k = 1, \dots, 2K$. The total number of clusters in the study is $K_1 + K_2$, or $2K$, and the number of individuals in each cluster is $m$, indexed as $j = 1, \dots, m$. The two continuous outcome vectors are $Y_{1,kj}$ and $Y_{2,kj}$, which we will generically refer to as $Y_1$ and $Y_2$, whenever there is no ambiguity by omitting the cluster and individual subscripts.

When using the p-value adjustment methods, which we will discuss in the next section, the power, cluster size, and number of clusters are calculated separately for each outcome. Thus, we have $K^{(q)}$ as the number of clusters in each treatment group based on the $q^{\text{th}}$ outcome, and $m^{(q)}$ as the number of individuals per cluster based on the $q^{\text{th}}$ outcome. We assume the continuous outcomes follow a bivariate linear mixed model, given by:

$$\begin{pmatrix} Y_{1,kj} \\ Y_{2,kj} \end{pmatrix} = \begin{pmatrix} \gamma_1 \\ \gamma_2 \end{pmatrix} + \begin{pmatrix} \beta_1^* \\ \beta_2^* \end{pmatrix} X_k + \begin{pmatrix} b_{1,k} \\ b_{2,k} \end{pmatrix} + \begin{pmatrix} e_{1,kj} \\ e_{2,kj} \end{pmatrix},$$

where $X_k$ denotes the treatment group for cluster $k$ ($X_k = 1$ for treatment, $X_k = 0$ for control), $\beta_1^*$ and $\beta_2^*$ are the treatment effects on the first and second outcome, respectively, and $\gamma_1$ and $\gamma_2$ are the intercepts. Then, $\boldsymbol{b}_k = \left(b_{1,k}, b_{2,k}\right)^T$ is the vector of random intercepts for cluster $k$ for each outcome $q$, and is assumed to follow $\mathcal{N}(\mathbf{0}_{2\times 1}, \boldsymbol{\Sigma}_{\boldsymbol{b}})$, and $\boldsymbol{e}_{kj} = \left(e_{1,kj}, e_{2,kj}\right)^T$ is the random error for each subject $j$ for each outcome $q$, and follows $\mathcal{N}(\mathbf{0}_{2\times 1}, \boldsymbol{\Sigma}_{\boldsymbol{e}})$. Note that $\boldsymbol{\Sigma}_{\boldsymbol{b}}$ and $\boldsymbol{\Sigma}_{\boldsymbol{e}}$ must be positive definite. Then, the diagonal element of $\boldsymbol{\Sigma}_{\boldsymbol{b}}$ and $\boldsymbol{\Sigma}_{\boldsymbol{e}}$ are $\sigma_{q,B}^2$ and $\sigma_{q,W}^2$ respectively, and the off-diagonal elements are $\sigma_{12,B}$ and $\sigma_{12,W}$. The total variances of $Y_1$ and $Y_2$ are thus $Var(Y_1) = \sigma_1^2 = \sigma_{1,B}^2 + \sigma_{1,W}^2$ and $Var(Y_2) = \sigma_2^2 = \sigma_{2,B}^2 + \sigma_{2,W}^2$.

Throughout the manuscript, we consider four key correlation coefficients. These are the endpoint specific intraclass correlation coefficients (ICCs) for $Y_1$ and $Y_2$, the inter-subject between-endpoint ICC, and the intra-subject between-endpoint ICC, defined as $\rho_0^{(1)} = ICC(Y_1) = Corr(Y_{1,kj}, Y_{1,kj'}) = \frac{\sigma_{1,B}^2}{\sigma_{1,B}^2+\sigma_{1,W}^2}$ (i.e. the correlation for $Y_1$ for two different individuals in the same cluster); $\rho_0^{(2)} = ICC(Y_2) = Corr(Y_{2,kj}, Y_{2,kj'}) = \frac{\sigma_{2,B}^2}{\sigma_{2,B}^2+\sigma_{2,W}^2}$ (i.e. the correlation for $Y_2$ for two different individuals in the same cluster); $\rho_1^{(1,2)} = Corr(Y_{1,kj}, Y_{2,kj'}) = \frac{\sigma_{12,B}}{(\sigma_{1B}^2+\sigma_{1W}^2)^{1/2}(\sigma_{2B}^2+\sigma_{2W}^2)^{1/2}}$ (i.e. the correlation between $Y_1$ and $Y_2$ for different individuals in the same cluster); $\rho_2^{(1,2)} = Corr(Y_{1,kj}, Y_{2,kj}) = \frac{\sigma_{12,B}+\sigma_{12,W}}{(\sigma_{1B}^2+\sigma_{1W}^2)^{1/2}(\sigma_{2B}^2+\sigma_{2W}^2)^{1/2}}$ (i.e. the correlation between $Y_1$ and $Y_2$ for the same individual), respectively.[5] These are used to calculate the variance inflation factors (VIFs) used in many of the equations for power that quantify how clustering inflates the variance of each outcome and their correlations. These are defined as $VIF_1 = 1 + (m-1)\rho_0^{(1)}$, $VIF_2 = 1 + (m-1)\rho_0^{(2)}$, and $VIF_{12} = \rho_2^{(1,2)} + (m-1)\rho_1^{(1,2)}$. Information regarding each parameter used in these equations and the software package are summarized in Table 1.

The non-centrality parameter used for power calculations is $\lambda$, and the statistical power is denoted as $\pi$. Then, under this model, the standard design formulas for a CRT for the $q$-th outcome and equal treatment allocation are

$$m^{(q)} = \frac{2\left(Z_{1-\alpha/2} + Z_\beta\right)^2 \sigma_q^2 \left(1 - \rho_0^{(q)}\right)}{\left(\beta_q^*\right)^2 K^{(q)} - 2\left(Z_{1-\alpha/2} + Z_\beta\right)^2 \sigma_q^2 \rho_0^{(q)}}, K^{(q)} = \frac{2\left(Z_{1-\alpha/2} + Z_\beta\right)^2 \sigma_q^2 \left[1 + \left(m^{(q)} - 1\right)\rho_0^{(q)}\right]}{m^{(q)}\left(\beta_q^*\right)^2},$$

$$\lambda^{(q)} = \frac{\left(\beta_q^*\right)^2}{2\,\frac{\sigma_q^2}{K^{(q)} m^{(q)}}\left[1 + \left(m^{(q)} - 1\right)\rho_0^{(q)}\right]},$$

as shown in Donner and Klar, where $\lambda$ is the non-centrality parameter used for power calculations, and $\lambda^{(q)}$ is the non-centrality parameter based on the $q^{\text{th}}$ outcome. They also extended these to accommodate unequal treatment allocation.[6] Note here that for $\left(Z_{1-\alpha/2} + Z_\beta\right)^2$, $Z_{1-\alpha/2}$ is the $\left(1 - \frac{\alpha}{2}\right) \times 100^{th}$ lower percentile of the standard normal distribution with error rate $\alpha$, and $Z_\beta$ is the critical value corresponding to $\beta$. So, ignoring multiple testing, for example, at 80% power and a significance level of 5%, we have $\left(Z_{1-\alpha/2} + Z_\beta\right)^2 = (1.96 + 0.84)^2 = 7.84$. Next, we describe each of the study design methods that are available in the `crt2power` R package[4] and ShinyApp, all of which are included in the numerical evaluation in Section 5.

Table 1. Description of required input parameters

| Parameter | Statistical Notation | Variable Name in R Package | Description |
|---|---|---|---|
| Statistical Power | $\pi$ | `power` | Probability of detecting a true effect under the alternative hypothesis |
| Number of clusters | $K$ | `K` | Number of clinics in each treatment arm |
| Cluster Size | $m$ | `m` | Number of patients in each clinic |
| Overall (family-wise) False Positive Rate | $\alpha$ | `alpha` | Probability of one or more Type I error(s) |
| Effect for $Y_1$ | $\beta_1^*$ | `beta1` | Estimated intervention effect on $Y_1$, in percentage point increase |
| Effect for $Y_2$ | $\beta_2^*$ | `beta2` | Estimated Intervention effect on $Y_2$, in percentage point increase |
| Total Variance of $Y_1$ | $\text{Var}(Y_1) = \sigma_1^2$ | `varY1` | Total variance of the first outcome |
| Total Variance of $Y_2$ | $\text{Var}(Y_2) = \sigma_2^2$ | `varY2` | Total variance of the second outcome |
| Endpoint-specific ICC for $Y_1$ | $ICC(Y_1) = \rho_0^{(1)}$ | `rho01` | Correlation for $Y_1$ for two different individuals in the same cluster |
| Endpoint-specific ICC for $Y_2$ | $ICC(Y_2) = \rho_0^{(2)}$ | `rho02` | Correlation for $Y_2$ for two different individuals in the same cluster |
| Inter-subject between-endpoint ICC | $Corr(Y_{1,kj}, Y_{2,kj'}) = \rho_1^{(1,2)}$ | `rho1` | Correlation between $Y_1$ and $Y_2$ for two *different* individuals in the same cluster |
| Intra-subject between-endpoint ICC | $Corr(Y_{1,kj}, Y_{2,kj}) = \rho_2^{(1,2)}$ | `rho2` | Correlation between $Y_1$ and $Y_2$ for the *same* individual |
| Treatment allocation ratio | $r$ | `r` | Treatment allocation ratio: $K_2 = rK_1$ where $K_1$ is the number of clusters in the experimental group |
| Statistical distribution | -- | `dist` | Specification of which distribution to base calculation on, either the $\chi^2$-distribution or $F$-distribution[1] |

[1]When selecting the $\chi^2$-distribution, all methods will use this distribution with the exception of the conjunctive IU test, which will use the multivariate normal (MVN) distribution; when selecting the $F$-distribution, all methods will use this distribution with the exception of the conjunctive IU test, which will use the $t$-distribution.

### 2.2 Study Design Methods

Here, we briefly describe the study design methods that are available in the `crt2power` package[4] and accompanying ShinyApp. Table 2 summarizes each of the methods, including the hypothesis test, non-centrality parameter, equation for statistical power, and test statistic in which the power calculations are based. We denote statistical power as $\pi$, and the statistical power based on the $q^{\text{th}}$ outcome as $\pi^{(q)}$. For more details, we refer the reader to Owen et al.[3]

Throughout the following sections, we denote which tests utilize the disjunctive hypothesis, and which utilize a conjunctive hypothesis. For a disjunctive test, the null hypothesis is $H_0: \beta_1^* = 0$ and $\beta_2^* = 0$ and the alternative hypothesis is $H_A: \beta_1^* \neq 0$ or $\beta_2^* \neq 0$. To reject the null hypothesis, the intervention must have an effect on at least one outcome but not necessarily both. For a conjunctive test, the null hypothesis is $H_0: \beta_1^* = 0$ or $\beta_2^* = 0$, and the alternative hypothesis is $H_A: \beta_1^* \neq 0$ and $\beta_2^* \neq 0$. That is, to reject the null hypothesis, the intervention must have an effect on both outcomes. This distinction is important when understanding the study design methods that follow, and especially when deciding which method best suits one's study goals.

**2.2.1 P-value Adjustment Methods** – The most popular method for addressing multiple testing is adjusting the p-value. There are three key ways to adjust for the p-value, namely the Bonferroni correction,[7] the Sidak method,[8] and the D/AP approach.[9] These p-value adjustment methods are used for a disjunctive hypothesis framework, where we have $H_0: \beta_1^* = 0$ and $\beta_2^* = 0$ vs. $H_A: \beta_1^* \neq 0$ or $\beta_2^* \neq 0$.[10] Although the design calculations are carried out for each outcome separately, using adjusted significance levels ensures that the overall decision rule – rejecting the null if at least one outcome is significant – validly corresponds to a disjunctive hypothesis while maintaining control of the family-wise error rate. The study design calculations are marginal, where either $K$, $m$, or power for both of the $Q = 2$ outcomes are calculated independent of the other, and the final result is $K = \max(K^{(1)}, K^{(2)})$, $m = \max(m^{(1)}, m^{(2)})$, and $\pi = \min(\pi^{(1)}, \pi^{(2)})$. The resulting design parameters $(K, m, \pi)$ reflect the conservative marginal power framework typically used in multi-endpoint designs, ensuring that each outcome achieves the desired power under the adjusted significance level.[10] This marginal approach provides a guaranteed lower bound that aligns with standard practice in multiple testing scenarios.

**2.2.2 Combined Outcomes** – The combined, or composite, outcomes approach combines the two outcome vectors into a single outcome.[11] A popular way of combining the outcomes is to sum them to produce $\beta_c^*$. To eliminate the possibility that $\beta_1^* = -\beta_2^*$, it is assumed that both elements of $\beta^*$ are in the same direction, and data can be transformed to ensure that this is the case. In this approach, the hypothesis setup is $H_0: \beta_c^* = 0$ vs. $H_A: \beta_c^* \neq 0$. Note that rejecting the null hypothesis in this case is concluding that the treatment is efficacious on the combined outcome, which means that the treatment could be effective on just one or both of the treatments. Thus, this test considers a disjunctive hypothesis. Power calculations for this approach require specification of the combined treatment effect, $\beta_c^*$, the endpoint specific ICC for the combined outcome, $\rho_0^{(c)}$, and the total variance of the combined outcome, $\sigma_c^2$. Expressions for these quantities as a function of other, more intuitive or readily available ones, are derived in Owen et al.,[3] and are

$$\rho_0^{(c)} = \frac{\rho_0^{(1)}\sigma_1^2 + \rho_0^{(2)}\sigma_2^2 + 2\rho_1^{(1,2)}\sigma_1\sigma_2}{\sigma_1^2 + \sigma_2^2 + 2\rho_2^{(1,2)}\sigma_1\sigma_2}; \;\; \sigma_c^2 = \sigma_1^2 + \sigma_2^2 + 2\rho_2^{(1,2)}\sigma_1\sigma_2 ,$$

where $\rho_0^{(1)}$ and $\rho_0^{(2)}$ are the endpoint specific ICCs for $Y_1$ and $Y_2$ respectively, $\sigma_1^2$ and $\sigma_2^2$ are the total variances of $Y_1$ and $Y_2$ respectively, $\rho_1^{(1,2)}$ is the inter-subject between-endpoint ICC, and $\rho_2^{(1,2)}$ is the intra-subject between-endpoint ICC.

**2.2.3 Single Weighted 1-DF Combined Test** – In this approach, two separate test statistics are weighted to create a single test statistic. Originally proposed by Pocock et al. (1987) and O'Brien et al. (1984) for the individual randomized controlled trial,[12, 13] this method was extended in Owen et al. to accommodate clustering.[3] Here, we are testing $H_0: \beta_1^* + \beta_2^* = 0$ vs. $H_A: \beta_1^* + \beta_2^* \neq 0$, similar to the combined outcomes approach. Thus, this is also a disjunctive test. For conciseness, this test is also referred to as the "single weighted 1-DF test" or "single 1-DF test".

**2.2.4 Disjunctive 2-DF Test** – In this 2-DF test, we simultaneously test both outcomes for any departure from the null hypothesis. This test utilizes a linear hypothesis, and is written as $H_0: \boldsymbol{L\beta}^* = \boldsymbol{0}$ vs. $H_A: \boldsymbol{L\beta}^* \neq \boldsymbol{0}$. For the hybrid type 2 scenario, $\boldsymbol{L}$ is a $2 \times 2$ contrast matrix whose rows represent linearly independent hypotheses concerning the treatment effect parameter, $\boldsymbol{\beta}^*$.[5] When $\boldsymbol{L} = \begin{bmatrix} 1 & 0 \\ 0 & 1 \end{bmatrix}$, as would usually be the case, $H_0: \boldsymbol{L\beta}^* = \boldsymbol{0} \implies H_0: \begin{bmatrix} \beta_1^* = 0 \\ \beta_2^* = 0 \end{bmatrix}$. This test is also disjunctive; to reject the null hypothesis, the treatment needs to be effective on at least one outcome. The F-distribution was proposed for the distribution of the test statistic, but one may also utilize the $\chi^2$-distribution in larger sample size settings.

**2.2.5 Conjunctive Intersection-Union Test** – The conjunctive test, or intersection-union (IU) test, requires that the treatment be effective on both outcomes in order to reject the null hypothesis. Thus, the hypothesis setup is written as $H_0: \beta_1^* = 0$ or $\beta_2^* = 0$ vs. $H_A: \beta_1^* \neq 0$ and $\beta_2^* \neq 0$. It was proposed that the t-distribution be used as the distribution of the test statistic, but one may also utilize the multivariate normal distribution in larger sample size settings.[5]

When referring to the various power, number of clusters, and cluster size parameters for each method, we use acronyms in the superscripts of these variables. Method 1: p-value adjustment methods uses the acronym "PADJ", Method 2: combined outcomes approach uses the acronym "COMB", Method 3: single weighted 1-DF test uses the acronym "W1DF", Method 4: disjunctive 2-DF test uses the acronym "DIS2DF", and Method 5: conjunctive IU test uses the acronym "CONJ". So, for example, $\pi^{\mathrm{DIS2DF}}$ refers to the statistical power of a study using the disjunctive 2-DF test.

Table 2. Summary of study design formulas for hybrid type 2 CRTs

| Method | Hypothesis Setup | Non-Centrality Parameter and Power | Number of clusters in the treatment group (K) |
|---|---|---|---|
| P-value Adjustments | $H_0: \beta_1^* = 0$ and<br>$\beta_2^* = 0$<br>$H_A: \beta_1^* \neq 0$ or<br>$\beta_2^* \neq 0$ | $\lambda^{(q)} = \frac{\left(\beta_q^*\right)^2}{2\,\frac{\sigma_q^2}{Km}\left[1 + (m-1)\rho_0^{(q)}\right]}$<br>$\pi^{(q)} = 1 - \chi^2\left[\lambda^{(q)}, 1\right]; \pi = \min\left(\pi^{(1)}, \pi^{(2)}\right)$ | $K^{(q)} = \frac{2\left(Z_{1-\alpha/2} + Z_\beta\right)^2 \sigma_q^2\,[1 + \left(m^{(q)} - 1\right)\rho_0^{(q)}]}{m^{(q)}\left(\beta_q^*\right)^2}$ |
| Combined Outcomes | $H_0: \beta_c^* = 0$<br>$H_A: \beta_c^* \neq 0$ | $\lambda = \frac{(\beta_c^*)^2}{2\,\frac{\sigma_c^2}{Km}\left[1 + (m-1)\rho_0^{(c)}\right]}$<br>$\pi = 1 - \chi^2[\lambda, 1]$ | $K = \frac{2\left(Z_{1-\alpha/2} + Z_\beta\right)^2 \sigma_c^2\,\left[1 + (m-1)\rho_0^{(c)}\right]}{m\,(\beta_c^*)^2}$<br>$\sigma_c^2 = \sigma_1^2 + \sigma_2^2 + 2\rho_2^{(1,2)}\sigma_1\sigma_2; \quad \rho_0^{(c)} = \frac{\rho_0^{(1)}\sigma_1^2 + \rho_0^{(2)}\sigma_2^2 + 2\rho_1^{(1,2)}\sigma_1\sigma_2}{\sigma_1^2 + \sigma_2^2 + 2\rho_2^{(1,2)}\sigma_1\sigma_2}$ |
| Single Weighted 1-DF Combined Test | $H_0: \beta_1^* + \beta_2^* = 0$<br>$H_A: \beta_1^* + \beta_2^* \neq 0$ | $\lambda = \left[\frac{\sqrt{\frac{(\beta_1^*)^2}{\frac{2\sigma_1^2}{Km}\left[1 + (m-1)\rho_0^{(1)}\right]}} + \sqrt{\frac{(\beta_2^*)^2}{\frac{2\sigma_2^2}{Km}\left[1 + (m-1)\rho_0^{(2)}\right]}}}{\sqrt{2\left(1 + \frac{\left(\rho_2^{(1,2)} + (m-1)\rho_1^{(1,2)}\right)}{\sqrt{\left(1 + (m-1)\rho_0^{(1)}\right)\left(1 + (m-1)\rho_0^{(2)}\right)}}\right)}}\right]^2$<br>$\pi = 1 - \chi^2[\lambda, 1]$ | $K = \frac{2\left(Z_{1-\alpha/2} + Z_\beta\right)^2\left(1 + \frac{\left(\rho_2^{(1,2)} + (m-1)\rho_1^{(1,2)}\right)}{\sqrt{\left(1 + (m-1)\rho_0^{(1)}\right)\left(1 + (m-1)\rho_0^{(2)}\right)}}\right)}{\left[\sqrt{\frac{(\beta_1^*)^2}{\frac{2\sigma_1^2}{m}\left[1 + (m-1)\rho_0^{(1)}\right]}} + \sqrt{\frac{(\beta_2^*)^2}{\frac{2\sigma_2^2}{m}\left[1 + (m-1)\rho_0^{(2)}\right]}}\right]^2}$ |
| Disjunctive 2-DF Test | $H_0: \boldsymbol{L\beta}^* = \mathbf{0}$<br>$H_A: \boldsymbol{L\beta}^* \neq \mathbf{0}$ | $\lambda = \left[\frac{Km[(\beta_1^*)^2\,\sigma_2^2\,VIF_2 - 2\,\beta_1^*\,\beta_2^*\,\sigma_1\,\sigma_2\,VIF_{12} + (\beta_2^*)^2\,\sigma_1^2\,VIF_1]}{2\,\sigma_1^2\,\sigma_2^2\,[VIF_1\,VIF_2 - VIF_{12}^2]}\right]$<br>F-distribution:<br>$\pi = \int_{F_{1-\alpha}(S, 2K-S-Q)}^{\infty} f(x; \lambda, S, 2K - S - Q)dx$<br>$\chi^2$ distribution: $\pi = 1 - \int_0^{5.99} \chi^2(x;\, 2, \lambda)dx$ | $K = \left[\frac{2\left(Z_{1-\frac{\alpha}{2}} + Z_\beta\right)^2 \sigma_1^2\,\sigma_2^2\,[VIF_1 VIF_2 - VIF_{12}^2]}{m[(\beta_1^*)^2\,\sigma_2^2\,VIF_2 - 2\,\beta_1^*\,\beta_2^*\,\sigma_1\,\sigma_2\,VIF_{12} + (\beta_2^*)^2\,\sigma_1^2\,VIF_1]}\right]$<br>(for $\chi^2$-distribution only) |
| Conjunctive IU Test | $H_0: \beta_1^* = 0$ or<br>$\beta_2^* = 0$<br>$H_A: \beta_1^* \neq 0$ and<br>$\beta_2^* \neq 0$ | $[\zeta_1, \zeta_2]^T = \left[\frac{\beta_1^*\sqrt{2K}}{\sqrt{\frac{4\sigma_1^2 VIF_1}{m}}}, \quad \frac{\beta_2^*\sqrt{2K}}{\sqrt{\frac{4\sigma_2^2 VIF_2}{m}}}\right]^T$<br>$\pi = \int_{c_1}^{\infty}\int_{c_2}^{\infty} f_{\boldsymbol{W}}(w_1, w_2)\,dw_1 dw_2$ | -- |

### 2.2.6 Motivation and Interpretation of Methods

**2.2.6 Motivation and Interpretation of Methods** – Different study goals and questions of interest can naturally lead to different hypothesis structures, and the choice of study design method depends on how evidence across outcomes will be interpreted in decision-making. When improvement in either outcome would support moving an intervention forward, a disjunctive formulation is appropriate. P-value adjustment methods are a popular approach in practice and are often straightforward to conduct and communicate; the combined outcomes approach and single 1-DF approach summarize both outcomes into one signal, which can be useful when outcomes are measured on similar scales or when simplicity (i.e., a single test) is needed for planning. In contrast, when success with both outcomes is necessary to consider an intervention successful, a conjunctive hypothesis is appropriate, as targeted by the intersection-union test. No single method is globally preferable across different contexts. Instead, the appropriate approach depends on the scientific aims, resource constraints, measurement considerations, and the role each outcome plays in subsequent decisions. The results that follow therefore focus on illustrating how different design parameters affect the achievable power for each approach, clarifying the relative performance of different tests, and enabling researchers to select the method that best aligns with their goals in light of the power properties. These design methods, to are knowledge, are the methods one can possibly consider for this setting, and were identified through an extensive literature review in Owen et al.[3]

## 3. Software Description

### 3.1 Description of the R Package and ShinyApp

The `crt2power` package is an R package that allows users to calculate the statistical power or sample size requirements for a cluster-randomized trial with two continuous co-primary outcomes given a set of input parameters.[4] More precisely, there are three classes of functions: 1) functions that calculate the statistical power given cluster size, number of clusters, and other necessary input parameters; 2) functions that calculate the required number of clusters given the desired statistical power, cluster size, and other necessary input parameters; and 3) functions that calculate the required cluster size given the desired statistical power, number of clusters, and other necessary input parameters. For each of these three classes of functions, there are five functions, which correspond to the five study design methods that are currently available for a clustered hybrid type 2 study design. In each function name, it is specified which parameter the function calculates followed by which study design method is being used. In addition, there is also a function that allows the user to calculate the specified design parameter (either "power", "K", or "m") for all of the study design methods at once, outputting a table. This software is intended to support researchers during the design phase that meets study goals, which include, but are not limited to, statistical power. As always, the choice of method should be finalized prior to data collection, rather than selected post-hoc based on observed power or results. Table 3 summarizes the list of functions available in the `crt2power` package.

### 3.2 Specification of Design Parameters

In order to calculate either $K$, the number of clusters in the experimental group, $m$, the number of individuals in each cluster, or power ($\pi$), the probability of detecting a true effect under the alternative hypothesis, the user must specify various study design parameters so that their desired study is adequately powered. Table 1 describes each of the input parameters that are required for these calculations, along with each parameter's statistical notation and variable name in the package. Depending on the study design method that is used, not all of these input parameters are utilized. For example, the p-value adjustment methods do not require the intra- and inter-subject between-endpoint ICCs, with the exception of the D/AP p-value adjustment method utilizing the intra-subject between-endpoint ICC.

### 3.3 ShinyApp and Usage Examples

Figure 1 shows R code using the `crt2power` for four different function calls one might conduct using this package, and Figure *2* shows the homepage and a usage example for the `crt2powerApplication` ShinyApp. The ShinyApp has four main tabs – an overview tab, a tab for calculating statistical power, a tab for calculating the number of clusters ($K$), and a tab for calculating the number of individuals per cluster ($m$). For each of the calculation tabs, the user may choose to display the results in either a plot or a table. The ShinyApp can be accessed via https://mowen17.shinyapps.io/crt2powerApplication/. Note that validation checks are conducted in both the ShinyApp and R package, and include but are not limited to ensuring all inputs are numeric, ensuring the resulting covariance matrix is positive definite, ensuring variance and sample size values are non-negative, as well as many others. The authors plan to monitor and address any future bugs promptly to ensure that the tools remain stable and informative for the user.

Table 3. List of `crt2power` functions[4]

| Design Method | List of Functions |
|---|---|
| 1. P-Value Adjustment Methods | a. `calc_pwr_pval_adj(…)`<br>b. `calc_K_pval_adj(…)`<br>c. `calc_m_pval_adj(…)` |
| 2. Combined Outcomes Approach | a. `calc_pwr_comb_outcome(…)`<br>b. `calc_K_comb_outcome(…)`<br>c. `calc_m_comb_outcome(…)` |
| 3. Single Weighted 1-DF Combined Test | a. `calc_pwr_single_1dftest(…)`<br>b. `calc_K_single_1dftest(…)`<br>c. `calc_m_single_1dftest(…)` |
| 4. Disjunctive 2-DF Test | a. `calc_pwr_disj_2dftest(…)`<br>b. `calc_K_disj_2dftest(…)`<br>c. `calc_m_disj_2dftest(…)` |
| 5. Conjunctive Intersection-Union Test | a. `calc_pwr_conj_test(…)`<br>b. `calc_K_conj_test(…)`<br>c. `calc_m_conj_test(…)` |
| All five methods | a. `run_crt2_design(output = "power", …)`<br>b. `run_crt2_design(output = "K", …)`<br>c. `run_crt2_design(output = "m", …)` |

Figure 1. Usage examples for the `crt2power` R package

```
# Example of using the combined outcomes approach for calculating power
calc_pwr_comb_outcome(     dist = "Chi2", K = 8, m = 50, alpha = 0.05,
                           beta1 = 0.2, beta2 = 0.4, varY1 = 0.5, varY2 = 1,
                           rho01 = 0.05, rho02 = 0.1, rho1 = 0.01, rho2 = 0.1,
                           r = 1)
```

```
[1] 0.8308
```

```
# Example of using the single weighted 1-DF test for calculating K
calc_K_single_1dftest(     dist = "F", power = 0.9, m = 70, alpha = 0.05,
                           beta1 = 0.4, beta2 = 0.3, varY1 = 1.5, varY2 = 0.5,
                           rho01 = 0.1, rho02 = 0.07, rho1 = 0.05, rho2  = 0.3,
                           r = 2)
```

```
# A tibble: 1 × 2
       Treatment (K1)      Control (K2)
       <int>               <dbl>
       9                   18
```

```
# Example of using conjunctive IU test for calculating m
calc_m_conj_test(   dist = "MVN", power = 0.8, K = 10, alpha = 0.05,
                    beta1 = 0.4, beta2 = 0.4, varY1 = 0.5, varY2 = 1,
                    rho01 = 0.05, rho02 = 0.1, rho1 = 0.07, rho2  = 0.9,
                    r = 1, two_sided = TRUE)
```

```
[1] 465
```

```
# Example of calculating power based on all five methods
run_crt2_design(    output = "power", K = 6, m = 70, alpha = 0.05,
                    beta1 = 0.4, beta2 = 0.4, varY1 = 0.5, varY2 = 0.5,
                    rho01 = 0.1, rho02 = 0.1, rho1 = 0.07, rho2 = 0.9, r = 1)
```

```
# A tibble: 10 × 3
Design Method                  Power (Chi2-distribution)  Power (F-distribution)
<chr>                          <dbl>                      <dbl>
1. P-Value Adjustments
       a. Bonferroni           0.750                      0.585
       b. Sidak                0.752                      0.587
       c. D/AP                 0.823                      0.711
2. Combined Outcomes           0.881                      0.785
3. Single Weighted 1-DF Test   0.881                      0.785
4. Disjunctive 2-DF Test       0.810                      0.634
5. Conjunctive IU Test
       a. 1-Sided Hypothesis   0.847                      0.781
       b. 2-Sided Hypothesis   0.756                      0.638
```

Figure 2. Overview page of the `crt2powerApplication` ShinyApp and usage examples

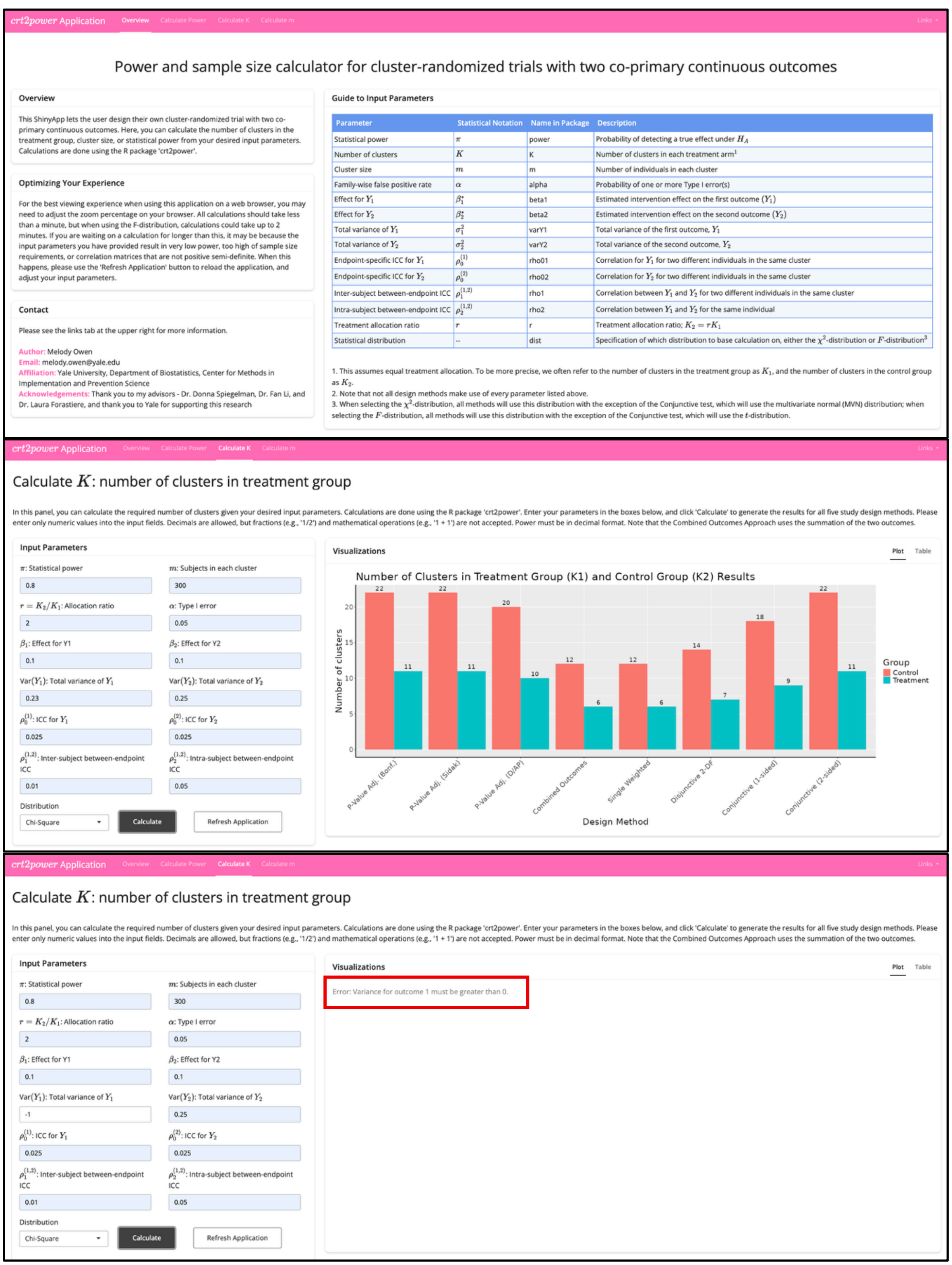

## 4. Theoretical Relationships Between the Methods

It is of interest to determine under which conditions some of these methods might be identical to others, if at all. This would simplify the choice of which method to use in the design phase of a hybrid type 2 CRT. In addition, it is of interest to determine if any method is globally more powerful than any other method, or if not, under which circumstances this would be the case. Thus, we conducted theoretical comparisons of the statistical power whenever possible, and when the mathematics was not tractable, we conducted a numerical analysis. The theoretical analyses, which we conducted first, helped guide the design of the numerical analysis, investigating the behavior of the design characteristics as input parameters.

In order to examine power in the theoretical analyses, we begin by comparing the equations for the non-centrality parameter for the design methods – namely, $\lambda$, and then assessing the underlying distributions to discern the exact relationship between the methods. We examine the non-centrality parameter because the statistical power is a function of $\lambda$; understanding the relationship between the non-centrality parameters of two methods is necessary for understanding the relationship between their statistical power.

### 4.1 Relationship Between Methods 2, 3, and 4

Among the five study design methods, the methods that are the most similar in form are the ones that utilize a single test statistic that combines the two outcomes in some way, namely the combined outcomes approach, single weighted 1-DF test, and disjunctive 2-DF test. The p-value adjustment methods and the conjunctive test consider two test statistics – one for each outcome, and are thus different in form. For this reason, we began by first comparing the aforementioned methods to understand if any among them are equivalent under certain scenarios, or if one is globally more powerful. To compare the methods, we use the $\chi^2$-distribution.

**4.1.1 Method 2: Combined Outcomes vs. Method 3: Single Weighted 1-DF Combined Test and Method 4: Disjunctive 2-DF Test** – In Owen et al., it was found that if $\rho_0^{(1)} = \rho_0^{(2)}$ and $\sigma_1^2 = \sigma_2^2$, then $\lambda^{\text{COMB}} = \lambda^{\text{W1DF}}$, resulting in $\pi^{\text{COMB}} = \pi^{\text{W1DF}}$.[3] We further examined the relationship between $\lambda^{\text{COMB}}$ and $\lambda^{\text{W1DF}}$, and could not identify any other theoretical relationship or constraint on the input parameters that resulted in $\lambda^{\text{COMB}} > \lambda^{\text{W1DF}}$, $\lambda^{\text{COMB}} < \lambda^{\text{W1DF}}$, or $\lambda^{\text{COMB}} = \lambda^{\text{W1DF}}$. Similarly, we also compared the combined outcomes test to the disjunctive 2-DF test. Collapsing the contrast matrix into a single row (e.g. $\boldsymbol{L} = [1 \quad 1]$) effectively tests whether a weighted combination of the two treatment effects differs from zero, yielding a 1-DF test that is algebraically equivalent to the combined outcomes approach ($\lambda^{\text{COMB}} = \lambda^{\text{DIS2DF,DF=1}}$). Because the motivation of the 2-DF test is to jointly assess departures across both outcomes, this 1-DF reduction is primarily of theoretical interest and detailed further in Appendix A.5 . When the disjunctive 2-DF test is not changed into a 1-DF test, no theoretical relationship or constraint on the input parameters were found that resulted in $\lambda^{\text{COMB}} > \lambda^{\text{DIS2DF}}$, $\lambda^{\text{COMB}} < \lambda^{\text{DIS2DF}}$, or $\lambda^{\text{COMB}} = \lambda^{\text{DIS2DF}}$. These findings motivated the need to examine these methods in a numerical evaluation in order to further investigate whether there are study design scenarios that result in one of the methods being more powerful than the other methods.

**4.1.2 Method 3: Single Weighted 1-DF Combined Test vs. Method 4: Disjunctive 2-DF Test** – For the single weighted 1-DF combined test and the disjunctive 2-DF test, we compared the equations for the non-centrality parameter, $\lambda^{\mathrm{W1DF}}$ and $\lambda^{\mathrm{DIS2DF}}$, respectively. We set $\lambda^{\mathrm{W1DF}} = \lambda^{\mathrm{DIS2DF}}$ to identify a necessary condition for this equation, and in Appendix A.1 we show that $\lambda^{\mathrm{W1DF}} = \lambda^{\mathrm{DIS2DF}}$ when $\sigma_2 \beta_1^* \sqrt{VIF_2} = \sigma_1 \beta_2^* \sqrt{VIF_1}$, where $VIF_1 = 1 + (m-1)\rho_0^{(1)}$ and $VIF_2 = 1 + (m-1)\rho_0^{(2)}$. We rule out the cases when $\sigma_1$, $\sigma_2$, $\beta_1^*$, or $\beta_2^*$ equal 0 as they would not be plausible assumptions.

Even when $\lambda^{\mathrm{W1DF}} = \lambda^{\mathrm{DIS2DF}}$, it follows that $\pi^{\mathrm{W1DF}} \neq \pi^{\mathrm{DIS2DF}}$ under the $\chi^2$-distribution. This is because the single weighted 1-DF test uses 1-DF and disjunctive 2-DF test uses 2-DF, which determines the bounds, i.e. the critical values ($c^{\mathrm{W1DF}}$ and $c^{\mathrm{DIS2DF}}$) of the integrals used for calculating power. For example, for an overall false-positive rate of $\alpha = 0.05$, the critical value for the single weighted 1-DF test is calculated using the central $\chi^2$-distribution with 1-DF: $\chi^2_{1-\alpha}(1) = c^{\mathrm{W1DF}} = 3.84$. For the disjunctive 2-DF test, the critical value uses the central $\chi^2$-distribution with 2-DF: $\chi^2_{1-\alpha}(2) = c^{\mathrm{DIS2DF}} = 5.99$. In fact , for all $\alpha \in (0,1)$, $\chi^2_{1-\alpha}(1) < \chi^2_{1-\alpha}(2)$, and so $c^{\mathrm{W1DF}} < c^{\mathrm{DIS2DF}}$. Knowing this relationship between the critical values between the single weighted 1-DF test and the disjunctive 2-DF test will help us further understand the relationship of statistical power when $\lambda^{\mathrm{W1DF}} = \lambda^{\mathrm{DIS2DF}}$.

Note that the equations for power can also be written in terms of their cumulative distribution functions (CDF) under the alternative. The CDF of the $\chi^2$-distribution makes use of the "Marcum Q-function", denoted $Q_{d/2}(\sqrt{\lambda}, \sqrt{x})$; for random variables $X \sim \chi^2(k, \lambda)$ (i.e. non-central $\chi^2$-distribution with $d$ degrees-of-freedom and non-centrality parameter $\lambda$), the CDF is $F(x; d, \lambda) = 1 - Q_{d/2}(\sqrt{\lambda}, \sqrt{x})$ by definition. Using this fact, the statistical power for the single weighted 1-DF test and the disjunctive 2-DF test can be written as

$$\pi^{\mathrm{W1DF}} = \int_{c^{\mathrm{W1DF}}}^{\infty} \chi^2(x;\ 1, \lambda^{\mathrm{W1DF}})dx = Q_{1/2}\left(\sqrt{\lambda^{\mathrm{W1DF}}}, \sqrt{c^{\mathrm{W1DF}}}\right)$$

$$\pi^{\mathrm{DIS2DF}} = \int_{c^{\mathrm{DIS2DF}}}^{\infty} \chi^2(x;\ 2, \lambda^{\mathrm{DIS2DF}})dx = Q_1\left(\sqrt{\lambda^{\mathrm{DIS2DF}}}, \sqrt{c^{\mathrm{DIS2DF}}}\right).$$

The Marcum Q-Function, $Q_{d/2}(\sqrt{\lambda}, \sqrt{c})$, is strictly increasing in $d/2$ and $\sqrt{\lambda}$ for all $\sqrt{\lambda} \geq 0$ and $\sqrt{c}$, $d/2 > 0$. It is strictly decreasing in $\sqrt{c}$ for all $\sqrt{\lambda}$, $\sqrt{c} \geq 0$ and $d/2 > 0$. In other words, $Q_{d/2}(\sqrt{\lambda}, \sqrt{c})$ increases as $d/2$ increases, and $Q_{d/2}(\sqrt{\lambda}, \sqrt{c})$ decreases as $\sqrt{c}$ increases. Due to these competing effects, it is not clear how $Q_{1/2}\left(\sqrt{\lambda^{\mathrm{W1DF}}}, \sqrt{c^{\mathrm{W1DF}}}\right)$ compares to $Q_1\left(\sqrt{\lambda^{\mathrm{DIS2DF}}}, \sqrt{c^{\mathrm{DIS2DF}}}\right)$ when the non-centrality parameters are the same. Furthermore, although the function $Q_{1/2}$ can be reduced nicely using the complimentary error functions (see Appendix A.1 ), the integrals in in $Q_1$ cannot be reduced in the same way. This is because $Q_1$ includes $I_0$, the modified Bessel function of the first kind, which complicates the integral and prevents it from being written as the complementary error function, and so we cannot directly compare the equations.

To better understand the case when $\lambda^{\mathrm{W1DF}} = \lambda^{\mathrm{DIS2DF}}$, R was used to visualize the difference between $\pi^{\mathrm{W1DF}} = Q_{1/2}\left(\sqrt{\lambda^{\mathrm{W1DF}}}, \sqrt{c^{\mathrm{W1DF}}}\right)$ and $\pi^{\mathrm{DIS2DF}} = Q_1\left(\sqrt{\lambda^{\mathrm{DIS2DF}}}, \sqrt{c^{\mathrm{DIS2DF}}}\right)$. It was found that when the non-centrality parameters are the same, the single weighted 1-DF test approach always yields more power than the disjunctive 2-DF test approach, regardless of the choice of $\alpha$. This is shown in Appendix A.1 . Then, taking a closer look at this scenario theoretically when $\lambda^{\text{W1DF}} = \lambda^{\text{DIS2DF}}$, i.e. when $\sigma_2\beta_1^*\sqrt{VIF_2} = \sigma_1\beta_2^*\sqrt{VIF_1}$, we can rewrite the expression as follows:

$$\sigma_2\beta_1^*\sqrt{VIF_2} = \sigma_1\beta_2^*\sqrt{VIF_1} \Longrightarrow \frac{\beta_1^*}{\sigma_1\sqrt{VIF_1}} = \frac{\beta_2^*}{\sigma_2\sqrt{VIF_2}}.$$

In other words, when the cluster-corrected standardized effect sizes, $\frac{\beta_1^*}{\sigma_1\sqrt{VIF_1}}$ and $\frac{\beta_2^*}{\sigma_2\sqrt{VIF_2}}$, are the same, then the non-centrality parameters between these methods will be the same. So, we can conclude that when the standardized effect sizes that also account for clustering through the variance inflation factor are the same, then the single weighted 1-DF test will be more powerful than the disjunctive 2-DF test. In a real study, especially in a hybrid 2 study where the outcomes are different and on different scales, the occurrence of this equality is unlikely to happen.

We have established that when $\sigma_2\beta_1^*\sqrt{VIF_2} = \sigma_1\beta_2^*\sqrt{VIF_1}$, and thus $\lambda^{\text{W1DF}} = \lambda^{\text{DIS2DF}}$, it follows that $\pi^{\text{W1DF}} > \pi^{\text{DIS2DF}}$ across all $\alpha \in (0.01, 0.025, 0.05, 0.1)$ and $\lambda^{\text{W1DF}} = \lambda^{\text{DIS2DF}} \in [0, 30]$. However, when $\sigma_2\beta_1^*\sqrt{VIF_2} \neq \sigma_1\beta_2^*\sqrt{VIF_1}$, then $\left(\sigma_2\beta_1^*\sqrt{VIF_2} - \sigma_1\beta_2^*\sqrt{VIF_1}\right)^2$ will always be greater than 0. In this case, $0 < \left(\sigma_2\beta_1^*\sqrt{VIF_2} - \sigma_1\beta_2^*\sqrt{VIF_1}\right)^2$ implies $\lambda^{\text{W1DF}} < \lambda^{\text{DIS2DF}}$. We are again faced with the issue as before where the relationship between $Q_{1/2}\left(\sqrt{\lambda^{\text{W1DF}}}, \sqrt{c^{\text{W1DF}}}\right)$ and $Q_1\left(\sqrt{\lambda^{\text{DIS2DF}}}, \sqrt{c^{\text{DIS2DF}}}\right)$ is unclear due to the competing effects that the inputs have on the function. The threshold of when $Q_{1/2}\left(\sqrt{\lambda^{\text{W1DF}}}, \sqrt{c^{\text{W1DF}}}\right) > Q_1\left(\sqrt{\lambda^{\text{DIS2DF}}}, \sqrt{c^{\text{DIS2DF}}}\right)$ or vice versa now depends on many variables, since we are under the constraint of $\lambda^{\text{W1DF}} \neq \lambda^{\text{DIS2DF}}$. Since the relationship of the non-centrality parameters depend on $\sigma_1$, $\sigma_2$, $\beta_1^*$, $\beta_2^*$, $VIF_1$, and $VIF_2$, we must examine this case in the numerical evaluation.

### 4.2 Examination of Method 1: P-value Adjustments

Based on the illustrative example shown in Owen et al., it was hypothesized that the p-value adjustment methods were less powerful than the combined outcomes test, the single weighted 1-DF test, and the disjunctive 2-DF test.[3] We compared their equations for the non-centrality parameter and statistical power to formally prove if this is the case.

For all of the p-value adjustment methods, the statistical power is $\pi^{\text{PADJ}} = \min\left(\pi^{(1)}, \pi^{(2)}\right)$. Since a smaller value of $\lambda$ corresponds to a smaller statistical power, this is the same as writing $\pi^{\text{PADJ}} = 1 - \chi^2\left[\lambda^{\text{PADJ}} = \min\left(\lambda^{(1)}, \lambda^{(2)}\right), 1\right]$. Recall that here, $\pi^{(q)}$ and $\lambda^{(q)}$ refer to the statistical power and non-centrality parameter calculated based on the $q^{th}$ outcome for the marginal tests. Thus, when comparing the p-value adjustment methods to methods 2, 3, and 4, we have two cases: 1) $\lambda^{(1)} > \lambda^{(2)}$; and 2) $\lambda^{(1)} < \lambda^{(2)}$. Under the first case, $\lambda^{\text{PADJ}} = \lambda^{(2)}$, and under the second case, $\lambda^{\text{PADJ}} = \lambda^{(1)}$. If they are equal, then power can be calculated using parameters from either outcome. The proofs for each case are logically equivalent, so we show the proof for one case. We compare the methods by comparing their equations for the non-centrality parameters. We know that $\rho_2^{(1,2)} < 1$, and assume that $\rho_1^{(1,2)} < \rho_0^{(1)}$, $\rho_1^{(1,2)} < \rho_0^{(2)}$, implying that $VIF_{12} < VIF_1$ and $VIF_{12} < VIF_2$. Lastly, we also assume that the treatment effects are non-negative. If this is not the case, they can be transformed in order to meet this assumption.

#### 4.2.1 Method 1: P-value Adjustments vs. Method 2: Combined Outcomes

**4.2.1 Method 1: P-value Adjustments vs. Method 2: Combined Outcomes** – We hypothesize that the non-centrality parameter for the p-value adjustment is smaller than that of the combined outcomes approach. That is,

$$\min\left(\frac{(\beta_1^*)^2}{\frac{2}{Km}\sigma_1^2 VIF_1}, \frac{(\beta_2^*)^2}{\frac{2}{Km}\sigma_2^2 VIF_2}\right) < \frac{(\beta_1^* + \beta_2^*)^2}{\frac{2}{Km}[\sigma_1^2\, VIF_1 + \sigma_2^2\, VIF_2 + 2\sigma_1\sigma_2 VIF_{12}]}.$$

Under the first case, we suppose that $\frac{(\beta_1^*)^2}{\frac{2}{Km}\sigma_1^2 VIF_1} < \frac{(\beta_2^*)^2}{\frac{2}{Km}\sigma_2^2 VIF_2}$, which means $\lambda^{\text{PADJ}} = \frac{(\beta_1^*)^2}{\frac{2}{Km}\sigma_1^2 VIF_1}$. Then, reducing the inequality, we are left with the expression $\beta_1^*\sigma_2 VIF_2 < \beta_2^*\sigma_1 VIF_1$, and from this and the fact that $VIF_{12} < VIF_2$, it follows that $\beta_1^*\sigma_2 VIF_{12} < \beta_2^*\sigma_1 VIF_1$. This inequality and our first supposition both result in the inequality $\lambda^{\text{PADJ}} < \lambda^{\text{COMB}}$ being true, which is shown in Appendix A.2 . Also note that for all p-value adjustment methods, $\alpha^{\text{PADJ}} < \alpha^{\text{COMB}}$, and recall that both of these methods use the $\chi^2$-distribution with 1-DF. A smaller family-wise false-positive rate, $\alpha$, corresponds to smaller statistical power, as does a smaller $\lambda$ value. Thus, since $\lambda^{\text{PADJ}} < \lambda^{\text{COMB}}$, and $\alpha^{\text{PADJ}} < \alpha^{\text{COMB}}$ it follows that $\pi^{\text{PADJ}} < \pi^{\text{COMB}}$, meaning that the p-value adjustment methods will always be less powerful than the combined outcomes approach. For details of the full proof, see Appendix A.2 .

**4.2.2 Method 1: P-value Adjustments vs. Method 3: Single Weighted 1-DF Test** – We examine the equations of the non-centrality parameters for the p-value adjustment method and single weighted 1-DF test. We aim to show:

$$\min\left(\frac{(\beta_1^*)^2}{\frac{2}{Km}\sigma_1^2 VIF_1}, \frac{(\beta_2^*)^2}{\frac{2}{Km}\sigma_2^2 VIF_2}\right) < \left[\frac{\sqrt{\frac{(\beta_1^*)^2}{\frac{2\sigma_1^2}{Km}VIF_1}} + \sqrt{\frac{(\beta_2^*)^2}{\frac{2\sigma_2^2}{Km}VIF_2}}}{\sqrt{2\left(1 + \frac{VIF_{12}}{\sqrt{VIF_1 VIF_2}}\right)}}\right]^2.$$

Under the first case, we suppose that $\frac{(\beta_1^*)^2}{\frac{2}{Km}\sigma_1^2 VIF_1} < \frac{(\beta_2^*)^2}{\frac{2}{Km}\sigma_2^2 VIF_2}$, which means $\lambda^{\text{PADJ}} = \frac{(\beta_1^*)^2}{\frac{2}{Km}\sigma_1^2 VIF_1}$. Then, reducing the inequality, we are left with the expression $\sqrt{2}\sqrt{1 + \frac{VIF_{12}}{\sqrt{VIF_1 VIF_2}}} - 1 < \frac{\beta_2^*\sigma_1\sqrt{VIF_1}}{\beta_1^*\sigma_2\sqrt{VIF_2}}$. From our first supposition, we know that $\beta_1^*\sigma_2\sqrt{VIF_2} < \beta_2^*\sigma_1\sqrt{VIF_1}$, and so it follows that $1 < \frac{\beta_2^*\sigma_1\sqrt{VIF_1}}{\beta_1^*\sigma_2\sqrt{VIF_2}}$. So, to show that the inequality holds, we need to show that $\sqrt{2}\sqrt{1 + \frac{VIF_{12}}{\sqrt{VIF_1 VIF_2}}} - 1 < 1$, which reduces to $VIF_{12}VIF_{12} < VIF_1 VIF_2$. Since $VIF_{12} < VIF_1$ and $VIF_{12} < VIF_2$, it must be the case that $VIF_{12}VIF_{12} < VIF_1 VIF_2$. Thus, we've shown that $\lambda^{\text{PADJ}} < \lambda^{\text{W1DF}}$. Note that for any p-value adjustment method, $\alpha^{\text{PADJ}} < \alpha^{\text{W1DF}}$. Both the p-value adjustment method and the single weighted 1-DF test use the $\chi^2$-distribution with 1-DF. A smaller family-wise false-positive rate, $\alpha$, corresponds to smaller statistical power, as does a smaller value of $\lambda$. Thus, since $\lambda^{\text{PADJ}} < \lambda^{\text{W1DF}}$, and $\alpha^{\text{PADJ}} < \alpha^{\text{W1DF}}$ it follows that $\pi^{\text{PADJ}} < \pi^{\text{W1DF}}$, meaning that the p-value adjustment methods will always be less powerful than the single weighted 1-DF test. For details of the full proof, see Appendix A.3 .

**4.2.3 Method 1: P-value Adjustments vs. Method 4: Disjunctive 2-DF Test** – Examining the equations of the non-centrality parameters for the p-value adjustment method and disjunctive 2-DF test, we aim to show:

$$\min\left(\frac{(\beta_1^*)^2}{\frac{2}{Km}\sigma_1^2 VIF_1}, \frac{(\beta_2^*)^2}{\frac{2}{Km}\sigma_2^2 VIF_2}\right) < \frac{Km[(\beta_1^*)^2\,\sigma_2^2\,VIF_2 \;-\; 2\,\beta_1^*\,\beta_2^*\,\sigma_1\,\sigma_2\,VIF_{12} \;+\; (\beta_2^*)^2\,\sigma_1^2\,VIF_1]}{2\,\sigma_1^2\,\sigma_2^2\,[VIF_1\,VIF_2 \;-\; VIF_{12}^2]}.$$

Under the first case, we suppose that $\frac{(\beta_1^*)^2}{\frac{2}{Km}\sigma_1^2 VIF_1} < \frac{(\beta_2^*)^2}{\frac{2}{Km}\sigma_2^2 VIF_2}$, which means $\lambda^{\mathrm{PADJ}} = \frac{(\beta_1^*)^2}{\frac{2}{Km}\sigma_1^2 VIF_1}$. Then, reducing the inequality, we are left with the expression $0 < (\beta_2\sigma_1 VIF_1 - \beta_1\sigma_2 VIF_{12})^2$, which is always true. This means that the inequality holds, thus proving that $\lambda^{\mathrm{PADJ}} < \lambda^{\mathrm{DIS2DF}}$.

We compare both tests using the $\chi^2$-distribution, and note that the disjunctive test uses 2-DF instead of 1-DF. Due to the differing degrees-of-freedom, $\lambda^{\mathrm{PADJ}} < \lambda^{\mathrm{DIS2DF}}$ and $\alpha^{\mathrm{PADJ}} < \alpha^{\mathrm{DIS2DF}}$ do not necessarily imply that $\pi^{\mathrm{PADJ}} < \pi^{\mathrm{DIS2DF}}$. Similar to our assessment of the single weighted 1-DF test and disjunctive 2-DF test, we can write the power integrals as Marcum Q-Functions, which gives

$$\pi^{\mathrm{PADJ}} = \int_{c^{\mathrm{PADJ}}}^{\infty} \chi^2(x;\ 1, \lambda^{\mathrm{PADJ}})dx = Q_{1/2}\left(\sqrt{\lambda^{\mathrm{PADJ}}}, \sqrt{c^{\mathrm{PADJ}}}\right)$$

$$\pi^{\mathrm{DIS2DF}} = \int_{c^{\mathrm{DIS2DF}}}^{\infty} \chi^2(x;\ 2, \lambda^{\mathrm{DIS2DF}})dx = Q_1\left(\sqrt{\lambda^{\mathrm{DIS2DF}}}, \sqrt{c^{\mathrm{DIS2DF}}}\right).$$

As previously noted, the Marcum Q-Function, $Q_{d/2}(\sqrt{\lambda}, \sqrt{c})$, is strictly increasing in $d/2$ and $\sqrt{\lambda}$ for all $\sqrt{\lambda} \geq 0$ and $\sqrt{c}$, $d/2 > 0$. It is strictly decreasing in $\sqrt{c}$ for all $\sqrt{\lambda}, \sqrt{c} \geq 0$ and $d/2 > 0$. In other words, $Q_{d/2}(\sqrt{\lambda}, \sqrt{c})$ increases as $d/2$ increases, and $Q_{d/2}(\sqrt{\lambda}, \sqrt{c})$ decreases as $\sqrt{c}$ increases. Due to these competing effects, it is not clear how $Q_{1/2}\left(\sqrt{\lambda^{\mathrm{PADJ}}}, \sqrt{c^{\mathrm{PADJ}}}\right)$ compares to $Q_1\left(\sqrt{\lambda^{\mathrm{DIS2DF}}}, \sqrt{c^{\mathrm{DIS2DF}}}\right)$. However, for a specific overall false-positive rate, we can plot the function values (i.e. statistical power) over many possible values for the non-centrality parameters. It was found that for almost every case, $\pi^{\mathrm{DIS2DF}} > \pi^{\mathrm{PADJ}}$, but there are a small number of cases where $\pi^{\mathrm{DIS2DF}} < \pi^{\mathrm{PADJ}}$, namely when the difference between $\lambda^{\mathrm{PADJ}}$ and $\lambda^{\mathrm{DIS2DF}}$ is very small. Though it is possible for $\lambda^{\mathrm{PADJ}} < \lambda^{\mathrm{DIS2DF}}$ due to the differing degrees-of-freedom, it is not globally true that $\pi^{\mathrm{DIS2DF}} > \pi^{\mathrm{PADJ}}$. However, this is not likely to be observed in practice, and across all p-values adjustment methods, $\pi^{\mathrm{DIS2DF}} > \pi^{\mathrm{PADJ}}$ generally. Appendix A.4 explores this in greater detail.

### 4.3 Summary of Theoretical Comparisons

The reason for why the p-value adjustment methods are globally less powerful than the combined outcomes approach and the single weighted 1-DF test can be intuitively understood through the property of the arithmetic mean. It states that for any two real numbers, $a$ and $b$, it is always the case that $\min(a,b) \leq \frac{a+b}{2} \leq \max(a,b)$. In a way, $\lambda^{\mathrm{COMB}}$ and $\lambda^{\mathrm{W1DF}}$ are "averages" of the non-centrality parameters for the first and second outcomes individually, namely $\lambda^{(1)}$ and $\lambda^{(2)}$. The p-value adjustment methods take the "worst" case scenario, resulting in the smallest power possible among the first and second outcome. Thus, it makes sense why they are less powerful than the combined outcomes approach and the single weighted 1-DF test. It's also important to note that the p-value adjustments don't take into account the correlation parameters $\rho_1^{(1,2)}$ or $\rho_2^{(1,2)}$, with the exception of the D/AP method, which uses $\rho_2^{(1,2)}$ to adjust the $\alpha$-level. So, this approach is disregarding important information about the data, possibly contributing to why the statistical power is more conservative. The results in the following numerical evaluation help to confirm these findings; there are no scenarios in which the p-value adjustment methods are more powerful than the combined outcomes approach and the single weighted 1-DF test. The same intuition can apply to why nearly always, the p-value adjustment methods are less powerful than the disjunctive 2-DF test, as this test is also "averaging" statistics between the first and second outcomes. The differing degrees-of-freedom could explain why in theory, the p-value adjustment methods could be more powerful than the disjunctive 2-DF test, though it's not likely to be observed in practice.

To summarize, we've shown that the p-value adjustment methods are globally less powerful than the combined outcomes approach and the single weighted 1-DF test, and that the p-value adjustment methods have a globally smaller non-centrality parameter than the disjunctive 2-DF test, but that it is still theoretically possible for the disjunctive 2-DF test to have a smaller power than the p-value adjustment methods. We've also shown that the combined outcomes approach is equivalent to the single weighted 1-DF test when the outcome specific ICCs and variances between the two outcomes are the same. Lastly, we've shown that the single weighted 1-DF test has the same non-centrality parameter as the disjunctive test when the cluster-corrected standardized effect sizes of the outcomes are equal. In this case, the single weighted 1-DF test will yield more power than the disjunctive 2-DF test. Table 4 shows a summary of these theoretical comparison findings along with the theoretical notation. To more thoroughly examine the performance of the study design methods, we continue on to the numerical evaluation.

Table 4. Summary of theoretical results

| Description | Theoretical Notation |
|---|---|
| **P-Value Adjustment Methods** are always less powerful than the **Combined Outcomes Approach** | $\pi^{\text{PADJ}} < \pi^{\text{COMB}}$ |
| **P-Value Adjustment Methods** are **always less powerful** than the **Single Weighted 1-DF Test**. | $\pi^{\text{PADJ}} < \pi^{\text{W1DF}}$ |
| **P-Value Adjustment Methods** always have a smaller non-centrality parameter than the **Disjunctive 2-DF Test**. However, due to the differing degrees of freedom, there are cases where the P-Value Adjustment Methods can result in higher power than the Disjunctive 2-DF Test (though this is not typically observed in practice). | $\lambda^{\text{PADJ}} < \lambda^{\text{DIS2DF}}$ |
| **Combined Outcomes Approach** is theoretically equivalent to the **Single Weighted 1-DF Test** when the outcome specific ICCs and variances between the two outcomes are the same, resulting in the same statistical power. | If $\rho_0^{(1)} = \rho_0^{(2)}$ and $\sigma_1^2 = \sigma_2^2$, then $\lambda^{\text{COMB}} = \lambda^{\text{W1DF}}$ and $\pi^{\text{COMB}} = \pi^{\text{W1DF}}$ |
| **Single Weighted 1-DF Test** has the same non-centrality parameter as the **Disjunctive 2-DF Test** when the cluster-corrected standardized effect sizes of the first and second outcomes are equal. | If $\frac{\beta_1^*}{\sigma_1\sqrt{VIF_1}} = \frac{\beta_2^*}{\sigma_2\sqrt{VIF_2}}$, then $\lambda^{\text{W1DF}} = \lambda^{\text{DIS2DF}}$ and $\pi^{\text{W1DF}} > \pi^{\text{DIS2DF}}$ for all $\alpha \in (0.01, 0.025, 0.05, 0.1)$ and $\lambda^{\text{W1DF}}, \lambda^{\text{DIS2DF}} \in [0, 30]$ |

## 5. Numerical Evaluation

### 5.1 Overview and Estimands

In the numerical evaluation, we explore the comparison of different testing procedures in terms of power under 1) varying cluster sizes and number of clusters; 2) varying values of the four correlations, namely the endpoint-specific ICC's, inter-subject between-endpoint ICC, and the intra-subject between-endpoint ICC; 3) varying values of the treatment effects on each outcome; and 4) varying values for the total variance of the two outcomes on study design. The goal of the evaluation was to identify which method of the five currently available for hybrid type 2 studies, if any, is globally the most powerful? If none are globally most powerful (which we hypothesize to be the case), under what design assumptions are each of the methods are most powerful? The numerical evaluation was needed because it was not possible to derive fully comprehensive theoretical answers to these questions, particularly for the conjunctive IU test and how it compares to the other methods. The code used to run the numerical evaluation is available on GitHub at https://github.com/melodyaowen/hybrid2numerical. Note that because exact expressions for all quantities of interest are available, there is no need for a simulation study. We discuss this point in greater detail later on.

### 5.2 Methods

The parameters that are varied in the numerical evaluation, along with their values, are displayed in Table 5. Values of each parameter were chosen based on values commonly observed in real CRT data, and to ensure that a wide range of values were considered for each input parameter. We explored treatment effect scenarios where either $\beta_1^* < \beta_2^*$ or $\beta_1^* = \beta_2^*$. Because $\beta_1^* < \beta_2^*$ is symmetric to $\beta_1^* > \beta_2^*$, meaning that the results for $\beta_1^* > \beta_2^*$ would mirror those for $\beta_1^* < \beta_2^*$ if the roles of the two treatment effects were reversed, we did not explore that scenario. We only considered a family-wise false-positive rate of 0.05 and equal treatment allocation.

The number of clusters in each treatment group ($K$) explored in the numerical study was carefully considered, as this is a particularly important parameter when designing CRTs. It was reported by Kahan et al. that the median *total* number of clusters in CRTs is 25 ($K$ = 12 to 13 in each arm for a parallel design).[14] Though hybrid 2 studies are increasing in popularity, the typical number of clusters hybrid 2 studies that are also CRTs is not well known, though a well-known hybrid 2 study by Abbott et al. reported 15 *total* number of clusters in their study, and another study by Clemson et al. reported 28 *total* number of clusters in their study.[15, 16] In these studies, the clusters were clinics and general healthcare practices. In another hybrid 2 study by Galaviz et al, physicians were considered as the clusters, with their patients being the individuals; they reported a *total* of 36 clusters (physicians) in their study.[17] The goal of the numerical study is to evaluate the power that these methods yield over a wide range of scenarios that are commonly found in CRTs, particularly in those in public health and implementation science; to that end, we include a wide range of values for $K$ in order to examine scenarios in which there are a small amount of clusters ($K$ = 4, 6, and 8, i.e. 8, 12, and 16 total), and scenarios where we have the typical total number of clusters in CRTs and hybrid 2 studies or higher ($K$ = 10, 20, and 30, i.e. 20, 40, and 60 total).

The numerical evaluation was conducted using R/RStudio. First, a dataframe of every potential design scenario using all combinations of the inputs displayed in Table 5 was created. Here, a study design "scenario" refers to a unique set of the 10 input parameters that one could use to calculate statistical power. There were a total of 45,000 such input scenarios. For each scenario, the statistical power was calculated using each of the study design methods described in Section 2. The resulting power calculations were assessed in relation to each input parameter, and trends were summarized through figures and tables.

Four separate numerical analyses were conducted, which we refer to as Comparison I, Comparison II, Comparison III, and Comparison IV. The motivation for conducting four separate comparisons was because different probability distributions have been proposed to assess power for different tests. In particular, the disjunctive 2-DF test was derived using the F-distribution, but one can also use the $\chi^2$-distribution for this method. The p-value adjustment methods, combined outcomes approach, and single weighted 1-DF test can also either utilize the F-distribution or $\chi^2$-distribution, and all of these methods use a 2-sided test. The conjunctive IU test differs from the other tests because it is multivariate in nature, where a vector of test statistics is used (one for each outcome). It was originally derived using a multivariate t-distribution, but one can also use a multivariate normal (MVN) distribution. It is not feasible for this method to utilize a $\chi^2$-distribution or F-distribution because it is multivariate in nature, and there is no multivariate F or $\chi^2$-distribution. It is valid to compare the conjunctive IU test under a two-tailed MVN-distribution to the remaining methods under the $\chi^2$-distribution, and to compare the conjunctive IU test under the t-distribution to the remaining methods under the F-distribution. Furthermore, the conjunctive IU test as proposed conducts two 1-sided tests (one for each outcome), but to make this method comparable to the other methods, which are all 2-sided, we consider the conjunctive IU test with two 2-sided tests. On the other hand, recognizing that users may want to understand how the conjunctive IU test compares to the other methods when used as originally proposed (i.e. using two 1-sided tests), we also looked at this case. Thus, this results in four main comparisons – Comparison I: 2-sided method comparison using the F-distribution for Methods 1-4 and the two-sided t-distribution for Method 5; Comparison II: 2-sided method comparison using the $\chi^2$-distribution for Methods 1-4 and the two-sided MVN-distribution for Method 5; Comparison III: "as is" method comparison using the F-distribution for Methods 1-4 and the one-sided t-distribution for Method 5; and Comparison IV: "as is" method comparison using the $\chi^2$-distribution for Methods 1-4 and the one-sided MVN-distribution for Method 5. Table 6 summarizes these comparisons.

Next, we discuss the results of the numerical evaluation under Comparison I in depth. Though we do not display the results for Comparisons II-IV, we outline any key findings or differences in results that were found. We provide the corresponding figures and tables for Comparisons II-IV in the Supplementary Material.

Table 5. Numerical Evaluation Parameters (results in 45,000 unique design scenarios)

| Parameter | Statistical Notation | Description | Considered Values |
|---|---|---|---|
| Number of clusters | K | Number of clusters in treatment group | 4 |
| | | | 6 |
| | | | 8 |
| | | | 10 |
| | | | 20 |
| | | | 30 |
| Cluster size | m | Number of individuals in each cluster | 50 |
| | | | 70 |
| | | | 100 |
| Effects for $Y_1$ and $Y_2$ | $\boldsymbol{\beta}^* = (\beta_1^*, \beta_2^*)$ | Estimated intervention effect vector for the two outcomes ($Y_1$ and $Y_2$) | (0.1, 0.4) |
| | | | (0.2, 0.4) |
| | | | (0.3, 0.4) |
| | | | (0.4, 0.4) |
| Outcome variances | $\boldsymbol{\sigma^2} = (\sigma_1^2, \sigma_2^2)$ | Total variance, $Var(Y_1)$ and $Var(Y_2)$ | (0.5, 1.5) |
| | | | (0.5, 1) |
| | | | (1, 1) |
| | | | (1, 0.5) |
| | | | (1.5, 0.5) |
| Endpoint-specific ICC for $Y_1$ and $Y_2$ | $\boldsymbol{\rho_0} = \left(\rho_0^{(1)}, \rho_0^{(2)}\right)$ | Correlation for $Y_1$ for two different individuals in the same cluster, correlation for $Y_2$ for two different individuals in the same cluster | (0.05, 0.1) |
| | | | (0.07, 0.1) |
| | | | (0.1, 0.1) |
| | | | (0.1, 0.07 |
| | | | (0.1, 0.05) |
| Inter-subject between-endpoint ICC | $\rho_1^{(1,2)}$ | Correlation between $Y_1$ and $Y_2$ for two different individuals in same cluster | 0.005 |
| | | | 0.01 |
| | | | 0.02 |
| | | | 0.05 |
| | | | 0.07 |
| Intra-subject between-endpoint ICC | $\rho_2^{(1,2)}$ | Correlation between $Y_1$ and $Y_2$ for the same individual | 0.1 |
| | | | 0.3 |
| | | | 0.5 |
| | | | 0.7 |
| | | | 0.9 |
| Overall (family-wise) False Positive Rate | $\alpha$ | Probability of one or more Type I error(s) | 0.05 |

Table 6. Summary of Numerical Analysis Comparisons

| | Numerical Analysis Types | | | |
|---|---|---|---|---|
| | **Comparison I** | **Comparison II** | **Comparison III** | **Comparison IV** |
| **Design Method** | "2-sided" method comparison using the F-distribution and t-distribution | "2-sided" method comparison using the $\chi^2$-distribution and MVN-distribution | "As is" method comparison using the F-distribution and t-distribution | "As is" method comparison using the $\chi^2$-distribution and MVN-distribution |
| 1. P-value Adjustment Methods for Multiple Testing | $F$-distribution<br>One 2-sided test | $\chi^2$-distribution<br>One 2-sided test | $F$-distribution<br>One 2-sided test | $\chi^2$-distribution<br>One 2-sided test |
| 2. Combined Outcomes Approach | $F$-distribution<br>One 2-sided test | $\chi^2$-distribution<br>One 2-sided test | $F$-distribution<br>One 2-sided test | $\chi^2$-distribution<br>One 2-sided test |
| 3. Single Weighted 1-DF Combined Test | $F$-distribution<br>One 2-sided test | $\chi^2$-distribution<br>One 2-sided test | $F$-distribution<br>One 2-sided test | $\chi^2$-distribution<br>One 2-sided test |
| 4. Disjunctive 2-DF Test | $F$-distribution<br>One 2-sided test | $\chi^2$-distribution<br>One 2-sided test | $F$-distribution<br>One 2-sided test | $\chi^2$-distribution<br>One 2-sided test |
| 5. Conjunctive Intersection-Union Test | $t$ -distribution<br>Two 2-sided tests | MVN-distribution<br>Two 2-sided tests | $t$ -distribution<br>Two 1-sided tests | MVN-distribution<br>Two 1-sided tests |

*MVN stands for "multivariate normal distribution".

## 5.3 Results

**5.3.1 Distribution of Statistical Power and Overall Method Rankings** – To gain an overall understanding of the statistical power yielded by each method across the 45,000 input scenarios, the distribution of power was examined, averaged over the 10 input parameters varied. Figure 3 displays histograms of statistical power for each method, along with the mean, minimum, and maximum power for each method. The three p-value adjustment methods tended to have lower power compared to the other methods, and the conjunctive test similarly had lower power generally. The combined outcomes approach, single 1-DF test, and disjunctive 2-DF test tended to have higher power among the methods. The ranking of the methods in terms of power was examined in order to better understand overall how the methods compared to one another. For each of the 45,000 input scenarios, each method was ranked 1 through 7; a method with a ranking of 1 means the method had the highest power, while a method with a ranking of 7 means the method had the lowest power. Figure 4 shows a heatmap of these rankings across scenarios – a darker color blue corresponds to a lower ranking (lower power), and a lighter color blue corresponds to a higher ranking (higher power). Included in this figure is a summary table with each of the method's mean ranking. The single weighted 1-DF test had a mean ranking of 1.61, while the disjunctive 2-DF test had a mean ranking of 2.06. The combined outcomes approach had a mean ranking of 2.13, while the conjunctive test had a mean ranking of 4.36. The p-value adjustment methods had the lowest mean rankings, with the Bonferroni method having the lowest mean ranking (6.93). These results give an overview of how the methods measure up against each other in terms of statistical power, but they also demonstrate that no method was globally better than all other methods. Next, we more closely examine trends for how each of the input design parameters affect the statistical power.

Figure 3. Distribution of statistical power for each design method among all 45,000 input scenarios for Comparison I (F and t distributions with 2-sided conjunctive test)

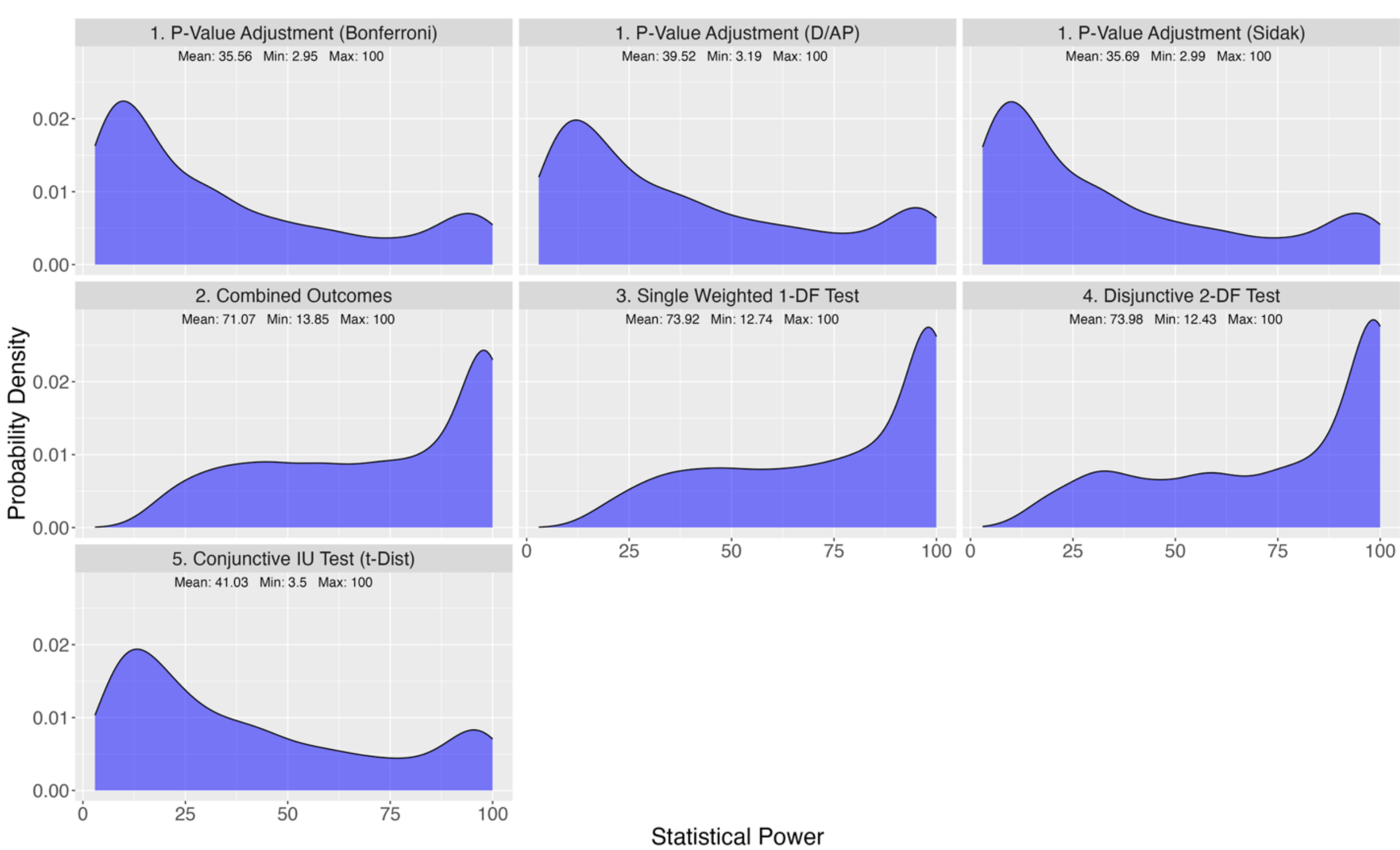

Figure 4. Ranks of all study design methods for each of the 45,000 input scenarios for Comparison I (F and t distributions with 2-sided conjunctive test)

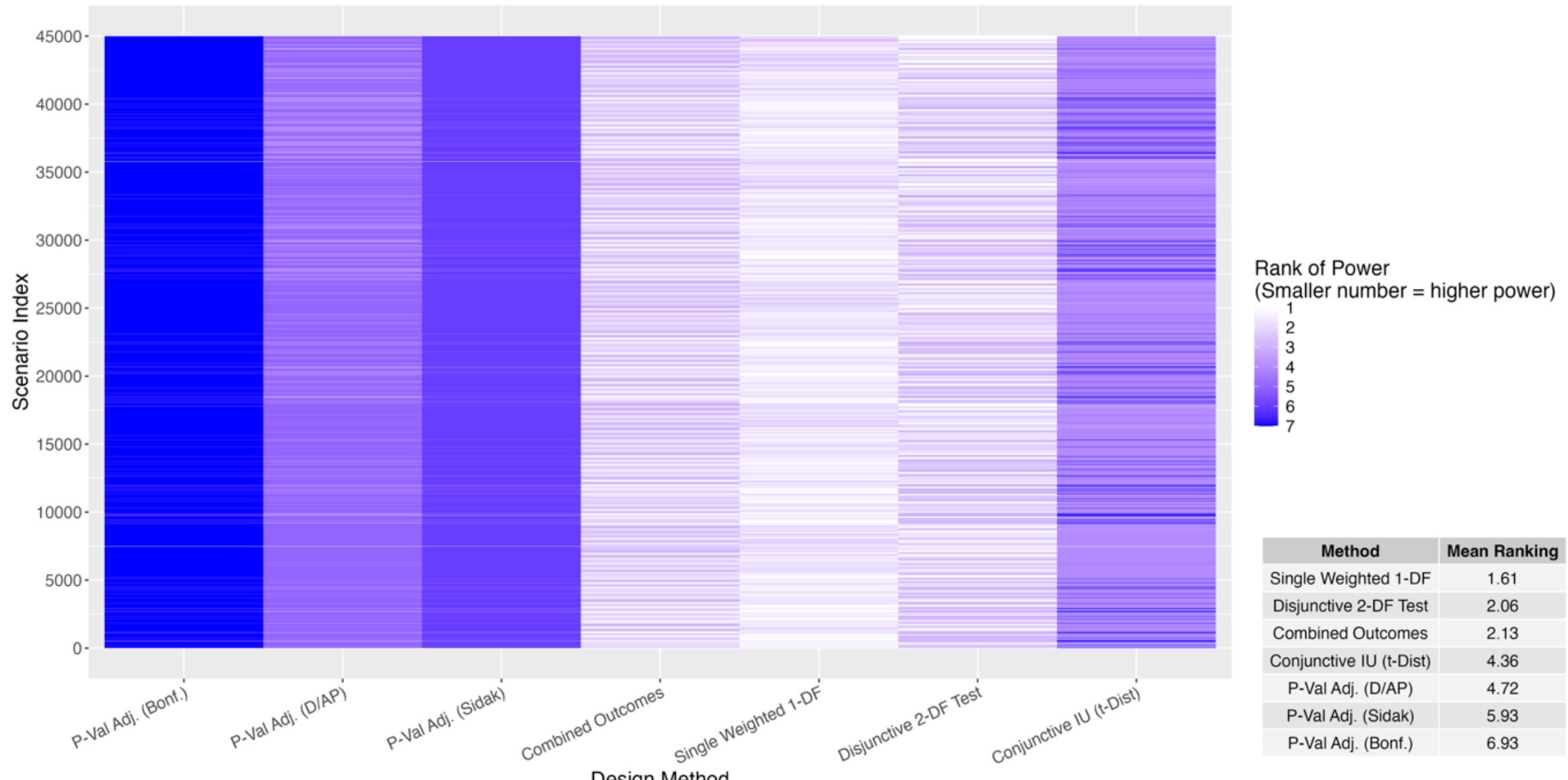

**5.3.2 Power in Relation to Each Input Parameter** – Before evaluating the numerical study results further, we conducted an additional evaluation that examined how each input parameter individually impacts power when allowed to vary over a larger set of values than what was feasible in the numerical evaluation, holding all other input parameters constant. This step was helpful in guiding which aspects of the numerical evaluation to report on. Values for the parameters that did not vary were chosen from the numerical evaluation parameters, and fixed at $K = 8$, $m = 50$, $(\beta_1^*, \beta_2^*) = (0.4, 0.4)$, $(\sigma_1^2, \sigma_2^2) = (1,1)$, $\left(\rho_0^{(1)}, \rho_0^{(2)}\right) = (0.1, 0.1)$, $\rho_1^{(1,2)} = 0.01$, and $\rho_2^{(1,2)} = 0.1$. These values remained the same across the evaluations, with the exception of the input parameter that was allowed to vary. We conducted this additional analysis for each of the four comparisons, but discuss the results for Comparison I only. Results for Comparison II-IV are available in the Supplementary Material. We summarize key findings here, with detailed results and figures given in Appendix B.

As expected, it was found that as the number of clusters increases ($K$), so too will the statistical power. Although a wide range of values for $K$ was examined, increasing the number of clusters did not impact the ranking of the tests in terms of power. However, the power difference between the methods is greater when $K$ is smaller, and it was observed through this additional evaluation, as well as over the wide range of scenarios considered in the numerical study, that when $K$ generally reaches ~20, the differences between the methods become small, at which point only subject matter considerations, rather than power, should dictate the choice of the primary test statistic. The choice of $K$ is important in the design of CRTs, and prior work has addressed the type I error challenges that can arise in CRTs with a small number of clusters. The threshold at which type I error stabilizes depends on the data generating process and the choice of test. For example, Kahan et al. showed that in some scenarios, even single-outcome CRTs may exhibit inflated type I error for as many as 70 total clusters, underscoring that no universal cut-off exists.[14] Our study, however, examines the relative power yielded by each design method. Importantly, the ranking of the tests by power was stable across all values of $K$ we considered; because we evaluated a broad range, we captured both the region where the methods differ most and the region where they converge, ensuring that our conclusions about comparative performance are robust to the number of clusters.

For $m$, the ranking of which method is the most powerful, second most powerful, etc. does not differ as $m$ increases. As the ratio of the treatment effects ($\beta_2^*/\beta_1^*$) increases, so does statistical power, and the method that was the most powerful changed as the ratio increased. As the ratio of the variances ($\sigma_2^2/\sigma_1^2$) increases, the statistical power decreased for all of the methods, and the method that was the most powerful also changed as the ratio increased. This observation motivated the need to look into the treatment effects and variances impacts on power further, which we detail in the next section. As the inter-subject between-endpoint ICC increases ($\rho_1^{(1,2)}$), the power decreases for the single weighted 1-DF test and the disjunctive 2-DF test, and increases for the conjunctive IU test. As the intra-subject between endpoint ICC ($\rho_2^{(1,2)}$) increases, the power very slightly decreased for the single weighted 1-DF test, disjunctive 2-DF test, remained constant for the Bonferroni and Sidak method, and increased slightly for the D/AP method, though these effects are minimal. The outcome specific ICCs, $\rho_0^{(1)}$, and $\rho_0^{(2)}$, had a much more substantial impact on statistical power, changing the ranking of which method is the most powerful as $\rho_0^{(1)}/\rho_0^{(2)}$ increases. So, we further investigated the effects of $\rho_0^{(1)}$ and $\rho_0^{(2)}$ on power.

**5.3.3 When is each method most powerful?** – When examining the statistical power through the numerical analysis of the five study design methods among the 45,000 input scenarios, we narrowed our scope to look the design methods that had at least one scenario where it was the most powerful among all the methods. Among those methods, we examined the conditions of $\beta_1^*$, $\beta_2^*$, $\sigma_1^2$, $\sigma_2^2$, $\rho_0^{(1)}$, and $\rho_0^{(2)}$ that resulted in a method being the most powerful. Examining the results in this way will allow us to give specific recommendations on which methods yield the highest statistical power based on the design parameters that have the biggest impact on power.

Under Comparison I, the combined outcomes approach, single weighted 1-DF test, and disjunctive 2-DF test all had scenarios in which they yielded the highest power, whereas the p-value adjustment methods and conjunctive IU test were never observed to have the highest power. To discern under which conditions these three methods were the most powerful relative to one another, we calculated the frequency each method had the highest power under different scenarios in terms of the standardized effect sizes ($\beta_1^*/\sigma_1$ and $\beta_2^*/\sigma_2$) and ICCs. For example, there are a total of 2,700 of the 45,000 scenarios for which $\beta_2^*/\sigma_2 - \beta_1^*/\sigma_1 \in [0.40, 0.49]$ and $\rho_0^{(1)} > \rho_0^{(2)}$, and in this case, the disjunctive 2-DF test was found to yield the highest statistical power for 92% of these cases. These results are displayed in Table 7; note that because $\beta_1^* < \beta_2^*$ by design of the numerical analysis, there were very few cases where $\beta_2^*/\sigma_2 - \beta_1^*/\sigma_1 < 0$, and so these cases were grouped together.

Based on these results, we see that there are many cases for which the single weighted 1-DF test had the most power out of all the methods more than 80% of the time. In particular, when $\beta_2^*/\sigma_2 - \beta_1^*/\sigma_1 = 0$, this test had higher power than all other methods over 80% of the time.

When $\beta_2^*/\sigma_2 - \beta_1^*/\sigma_1 < 0$, the single weighted 1-DF test does well over 55% of the time, and when $\beta_2^*/\sigma_2 - \beta_1^*/\sigma_1 < 0$ and $\rho_0^{(1)} = \rho_0^{(2)}$, it is most powerful in over 86% of the time. From this, we conclude that the single weighted 1-DF test tends to have higher power when $\beta_2^*/\sigma_2 - \beta_1^*/\sigma_1 \leq 0$, especially when the outcome specific ICCs are the same. In contrast, the disjunctive 2-DF test tended to have the highest power most frequently when the difference between the standardized treatment effects was greater. For example, when $\beta_2^*/\sigma_2 - \beta_1^*/\sigma_1 \in [0.30, 0.39]$ and $\rho_0^{(1)} > \rho_0^{(2)}$, this method had the highest power over 80% of the time. When $\beta_2^*/\sigma_2 - \beta_1^*/\sigma_1 \in [0.40, 0.49]$, an even greater difference between the treatment effects, this method had the highest power up to 92%. The results based on the unstandardized effect sizes were also examined, and this table is available in the Supplementary Material (S.5). These findings were similar to the unstandardized effect sizes for the single weighted 1-DF test; this test tends to have the highest power when the effect sizes are the same. We also see that the disjunctive 2-DF test does well when the effect sizes are different, in particular when $\beta_1^* < \beta_2^*$ and $\sigma_1^2 > \sigma_2^2$. Table 8 summarizes the results for both the standardized and unstandardized effect sizes, showing when these methods are most powerful 50-80% of the time, and >80% of the time.

**5.3.4 Conjunctive Test vs. Bonferroni P-Value Adjustment** – Though the conjunctive test was never the most powerful among all the methods considered, it was important to better understand the situations in which this test is more powerful than arguably the most popular method, the Bonferroni p-value adjustment method. This is because this test is the only test that explores a conjunctive hypothesis, which many researchers may be interested in utilizing in the context of a hybrid 2 study. To this end, we narrowed our scope and looked solely at the numerical evaluation results for the conjunctive test and the Bonferroni adjustment method alone. Table 9 shows summary statistics for the power difference between the two methods, namely $\pi^{CONJ} - \pi_{Bonferroni}^{PADJ}$, across the levels of the 10 input parameters. Generally, as $K$, $m$, $\rho_1^{(1,2)}$ and $\rho_2^{(1,2)}$ increase in value, the conjunctive test becomes increasingly more powerful than the Bonferroni. When $\rho_0^{(1)} \neq \rho_0^{(2)}$, the conjunctive test is more powerful than the Bonferroni to a higher degree than when $\rho_0^{(1)} = \rho_0^{(2)}$, and across the other design parameters, over all values of $\beta_2^*/\sigma_2 - \beta_1^*/\sigma_1$, the conjunctive test performs better than the Bonferroni. In general, $\pi^{CONJ} > \pi_{Bonferroni}^{PADJ}$ in 70% to 100% of the scenarios considered, except when $\beta_2^*/\sigma_2 = \beta_1^*/\sigma_1$ and $\rho_0^{(1)} = \rho_0^{(2)}$, where $\pi^{CONJ} < \pi_{Bonferroni}^{PADJ}$ (see online Supplementary Material, S.1).

**5.3.5 Conjunctive Test vs. Single Weighted 1-DF Test** – Similarly, we explored the degree to which the single weighted 1-DF test was more powerful than the conjunctive test. Although the single weighted 1-DF test had the highest average ranking of power among all the methods, it may be worth the loss of power when pursuing the conjunctive test due to its strength in hypothesis setup. Table 10 displays the percentage of scenarios where $\pi^{\text{W1DF}} - \pi^{\text{CONJ}}$ was between 0-5%, 5-10%, 10-20%, or >20, stratified by all of the different values of the 10 input parameters. Though for the majority of all scenarios, $\pi^{\text{W1DF}} - \pi^{\text{CONJ}} > 20\%$, there were some noteworthy findings. For a very small number of clusters in each treatment group ($K = 4$), $\pi^{\text{W1DF}} - \pi^{\text{CONJ}} < 20\%$ about 42% of the time. Similarly, when $\rho_1^{(1,2)}$ was at its highest value of 0.07, $\pi^{\text{W1DF}} - \pi^{\text{CONJ}} < 20\%$ about 51% of the time. Lastly, when $\beta_2^*/\sigma_2 = \beta_1^*/\sigma_1$, $\pi^{\text{W1DF}} - \pi^{\text{CONJ}} < 20\%$ about 51% of the time. This suggests that under these scenarios, the conjunctive test could be a viable option as opposed to the most powerful method, the single weighted 1-DF test. When the conjunctive test uses a 1-sided tail as in Comparisons III and IV, the power difference between these two methods is even less – these results are shown in the online Supplementary Material (S.3 and S.4).

**5.3.6 Comparisons II, III, and IV** – Overall, the results did not change much between the different comparisons. The cases for when the combined outcomes approach, the single weighted 1-DF test, and the disjunctive 2-DF test are best remained largely the same, and are exactly equivalent the majority of the time. This is consistent for both the unstandardized and standardized results. We conclude that the choice of distribution ($\chi^2$-distribution vs. F-distribution, and MVN-distribution vs. t-distribution), does not largely impact which methods are most powerful. Similarly, whether or not the conjunctive IU test is 2-sided or 1-sided does not change the fact that it is more powerful than the Bonferroni p-value adjustment method. These results are available in the Supplementary Material (S.2-S.4).

**5.3.7 Summary of Results** – Overall, the combined outcomes approach, single weighted 1-DF test, and disjunctive 2-DF test had the highest statistical power across the 45,000 design input scenarios. No method is globally most powerful, but we showed that the p-value adjustment methods are always less powerful than the combined outcomes approach and single weighted 1-DF test. When the outcome specific ICCs and variances between the two outcomes are the same, the combined outcomes approach is equivalent to the single weighted 1-DF test. Based on the average rankings of power for Comparison I, the most powerful methods in order are as follows: the single weighted 1-DF test, disjunctive 2-DF test, combined outcomes approach, conjunctive IU test, D/AP p-value adjustment, Sidak p-value adjustment, and Bonferroni p-value adjustment.

Treatment effects, outcome variances, and outcome-specific ICCs tend to have the largest impact on statistical power. The choice of reference distribution for hypothesis tests did not largely change which design methods were more or less powerful, though we note that the choice of reference distribution could potentially have an impact on the type I error rate, particularly in scenarios with a small number of clusters; future work could examine this. Table 8 summarizes the numerical study findings from Comparison I based on both the standardized effect sizes and unstandardized effect sizes, specifying under which cases a method has the highest power 50-80% of the time, and which cases a method has the highest power > 80% of the time. In general, we concluded that when the treatment effect sizes are the same, or when the standardized treatment effects are the same or close, the single weighted 1-DF test tends to have the highest power. When the treatment effects are different, the disjunctive 2-DF test tends to perform better, and as the difference between the standardized treatment effects increases, the disjunctive 2-DF test will yield higher power than the remaining methods. We revealed that the conjunctive test is more powerful than the Bonferroni method in the vast majority of scenarios. We also examined the degree to which the conjunctive test is less powerful than the most powerful method (single weighted 1-DF test). Table 11 summarizes both the results of the mathematical comparisons and numerical study by showing the relationship between each pairing of study design methods.

Table 7. Numerical evaluation results – proportion of scenarios where study design methods are most powerful based on standardized effect sizes, summarized for **Comparison I** (F and t distributions with 2-sided conjunctive test)

| $\frac{\beta_2^*}{\sigma_2} - \frac{\beta_1^*}{\sigma_1}$ | $\left(\rho_0^{(1)}, \rho_0^{(2)}\right)$ | Proportion of scenarios where the method was most powerful | | | | # Total Scenarios |
|---|---|---|---|---|---|---|
| | | **Combined Outcomes** | **Combined Outcomes = Single 1-DF** | **Single 1-DF** | **Disjunctive 2-DF** | |
| $< 0$ | $\rho_0^{(1)} < \rho_0^{(2)}$ | 0% | 10% | **78%** | 39% | 3600 |
| | $\rho_0^{(1)} = \rho_0^{(2)}$ | 0% | 8% | **94%** | 12% | 1800 |
| | $\rho_0^{(1)} > \rho_0^{(2)}$ | 21% | 21% | **67%** | 16% | 3600 |
| $0$ | $\rho_0^{(1)} < \rho_0^{(2)}$ | 0% | 14% | **89%** | 15% | 900 |
| | $\rho_0^{(1)} = \rho_0^{(2)}$ | 0% | **100%** | 0% | 0% | 450 |
| | $\rho_0^{(1)} > \rho_0^{(2)}$ | 0% | 14% | **89%** | 15% | 900 |
| $[0.05, 0.19]$ | $\rho_0^{(1)} < \rho_0^{(2)}$ | 38% | 10% | **54%** | 8% | 4500 |
| | $\rho_0^{(1)} = \rho_0^{(2)}$ | 45% | 23% | 31% | 7% | 2250 |
| | $\rho_0^{(1)} > \rho_0^{(2)}$ | 43% | 6% | 31% | 33% | 4500 |
| $[0.20, 0.29]$ | $\rho_0^{(1)} < \rho_0^{(2)}$ | 38% | 8% | 41% | 26% | 3600 |
| | $\rho_0^{(1)} = \rho_0^{(2)}$ | 16% | 23% | 40% | 30% | 1800 |
| | $\rho_0^{(1)} > \rho_0^{(2)}$ | 6% | 5% | 40% | **65%** | 3600 |
| $[0.30, 0.39]$ | $\rho_0^{(1)} < \rho_0^{(2)}$ | 14% | 3% | 38% | **54%** | 2700 |
| | $\rho_0^{(1)} = \rho_0^{(2)}$ | 0% | 5% | 29% | **69%** | 1350 |
| | $\rho_0^{(1)} > \rho_0^{(2)}$ | 0% | 0% | 22% | **87%** | 2700 |
| $[0.40, 0.49]$ | $\rho_0^{(1)} < \rho_0^{(2)}$ | 0% | 0% | 13% | **87%** | 2700 |
| | $\rho_0^{(1)} = \rho_0^{(2)}$ | 0% | 0% | 8% | **92%** | 1350 |
| | $\rho_0^{(1)} > \rho_0^{(2)}$ | 0% | 0% | 8% | **97%** | 2700 |

* Percentages are rounded to the nearest integer

Table 8. Summary of trends for highest statistical power from numerical evaluation based on standardized and unstandardized effect sizes for **Comparison I** (F and t distributions with 2-sided conjunctive test)

| **Design Method** | **Results based on standardized effect sizes** | | **Results based on unstandardized effect sizes** | |
|---|---|---|---|---|
| | Most powerful<br>50-80% of the time | Most powerful<br>>80% of the time | Most powerful<br>50-80% of the time | Most powerful<br>>80% of the time |
| Combined Outcomes Approach | -- | $\frac{\beta_2}{\sigma_2} - \frac{\beta_1}{\sigma_1} = 0, \rho_0^{(1)} = \rho_0^{(2)}$ | $\beta_1 < \beta_2, \sigma_1^2 < \sigma_2^2, \rho_0^{(1)} = \rho_0^{(2)}$<br>$\beta_1 < \beta_2, \sigma_1^2 < \sigma_2^2, \rho_0^{(1)} > \rho_0^{(2)}$<br>$\beta_1 < \beta_2, \sigma_1^2 = \sigma_2^2, \rho_0^{(1)} < \rho_0^{(2)}$<br>$\beta_1 < \beta_2, \sigma_1^2 = \sigma_2^2, \rho_0^{(1)} = \rho_0^{(2)}$ | $\beta_1 = \beta_2, \sigma_1^2 = \sigma_2^2, \rho_0^{(1)} = \rho_0^{(2)}$ |
| Single Weighted 1-DF Test | $\frac{\beta_2}{\sigma_2} - \frac{\beta_1}{\sigma_1} < 0, \rho_0^{(1)} < \rho_0^{(2)}$<br>$\frac{\beta_2}{\sigma_2} - \frac{\beta_1}{\sigma_1} < 0, \rho_0^{(1)} > \rho_0^{(2)}$<br>$\frac{\beta_2}{\sigma_2} - \frac{\beta_1}{\sigma_1} \in [0.05, 0.19], \rho_0^{(1)} < \rho_0^{(2)}$ | $\frac{\beta_2}{\sigma_2} - \frac{\beta_1}{\sigma_1} < 0, \rho_0^{(1)} = \rho_0^{(2)}$<br>$\frac{\beta_2}{\sigma_2} - \frac{\beta_1}{\sigma_1} = 0, \rho_0^{(1)} < \rho_0^{(2)}$<br>$\frac{\beta_2}{\sigma_2} - \frac{\beta_1}{\sigma_1} = 0, \rho_0^{(1)} = \rho_0^{(2)}$<br>$\frac{\beta_2}{\sigma_2} - \frac{\beta_1}{\sigma_1} = 0, \rho_0^{(1)} > \rho_0^{(2)}$ | $\beta_1 < \beta_2, \sigma_1^2 < \sigma_2^2, \rho_0^{(1)} < \rho_0^{(2)}$<br>$\beta_1 < \beta_2, \sigma_1^2 = \sigma_2^2, \rho_0^{(1)} = \rho_0^{(2)}$<br>$\beta_1 = \beta_2, \sigma_1^2 < \sigma_2^2, \rho_0^{(1)} < \rho_0^{(2)}$<br>$\beta_1 = \beta_2, \sigma_1^2 > \sigma_2^2, \rho_0^{(1)} > \rho_0^{(2)}$ | $\beta_1 = \beta_2, \sigma_1^2 < \sigma_2^2, \rho_0^{(1)} = \rho_0^{(2)}$<br>$\beta_1 = \beta_2, \sigma_1^2 < \sigma_2^2, \rho_0^{(1)} > \rho_0^{(2)}$<br>$\beta_1 = \beta_2, \sigma_1^2 = \sigma_2^2, \rho_0^{(1)} < \rho_0^{(2)}$<br>$\beta_1 = \beta_2, \sigma_1^2 = \sigma_2^2, \rho_0^{(1)} = \rho_0^{(2)}$<br>$\beta_1 = \beta_2, \sigma_1^2 = \sigma_2^2, \rho_0^{(1)} > \rho_0^{(2)}$<br>$\beta_1 = \beta_2, \sigma_1^2 > \sigma_2^2, \rho_0^{(1)} < \rho_0^{(2)}$<br>$\beta_1 = \beta_2, \sigma_1^2 > \sigma_2^2, \rho_0^{(1)} = \rho_0^{(2)}$ |
| Disjunctive 2-DF Test | $\frac{\beta_2}{\sigma_2} - \frac{\beta_1}{\sigma_1} \in [0.20, 0.29], \rho_0^{(1)} > \rho_0^{(2)}$<br>$\frac{\beta_2}{\sigma_2} - \frac{\beta_1}{\sigma_1} \in [0.30, 0.39], \rho_0^{(1)} = \rho_0^{(2)}$<br>$\frac{\beta_2}{\sigma_2} - \frac{\beta_1}{\sigma_1} \in [0.30, 0.39], \rho_0^{(1)} < \rho_0^{(2)}$ | $\frac{\beta_2}{\sigma_2} - \frac{\beta_1}{\sigma_1} \in [0.30, 0.39], \rho_0^{(1)} > \rho_0^{(2)}$<br>$\frac{\beta_2}{\sigma_2} - \frac{\beta_1}{\sigma_1} \in [0.40, 0.49], \rho_0^{(1)} < \rho_0^{(2)}$<br>$\frac{\beta_2}{\sigma_2} - \frac{\beta_1}{\sigma_1} \in [0.40, 0.49], \rho_0^{(1)} = \rho_0^{(2)}$<br>$\frac{\beta_2}{\sigma_2} - \frac{\beta_1}{\sigma_1} \in [0.40, 0.49], \rho_0^{(1)} > \rho_0^{(2)}$ | $\beta_1 < \beta_2, \sigma_1^2 = \sigma_2^2, \rho_0^{(1)} > \rho_0^{(2)}$<br>$\beta_1 < \beta_2, \sigma_1^2 > \sigma_2^2, \rho_0^{(1)} < \rho_0^{(2)}$<br>$\beta_1 < \beta_2, \sigma_1^2 > \sigma_2^2, \rho_0^{(1)} = \rho_0^{(2)}$<br>$\beta_1 = \beta_2, \sigma_1^2 < \sigma_2^2, \rho_0^{(1)} < \rho_0^{(2)}$<br>$\beta_1 = \beta_2, \sigma_1^2 > \sigma_2^2, \rho_0^{(1)} > \rho_0^{(2)}$ | $\beta_1 < \beta_2, \sigma_1^2 > \sigma_2^2, \rho_0^{(1)} > \rho_0^{(2)}$ |

*To save on space, the "*"'s are dropped from the $\beta$ terms in this table

Table 9. Summary statistics for $\pi^{CONJ} - \pi^{PADJ}_{Bonferroni}$ stratified by input parameter values for **Comparison I** (F and t distributions with 2-sided conjunctive test)

| Input Parameter | Value | Mean Difference | Minimum | Maximum | # Total Scenarios |
|---|---|---|---|---|---|
| $K$ | 4 | 3.3 | -3.2 | 13.6 | 7500 |
| | 6 | 4.7 | -7.0 | 14.2 | 7500 |
| | 8 | 5.9 | -8.9 | 13.4 | 7500 |
| | 10 | 6.7 | -9.5 | 12.8 | 7500 |
| | 20 | 6.7 | -4.3 | 11.9 | 7500 |
| | 30 | 5.4 | -0.6 | 11.6 | 7500 |
| $m$ | 50 | 5.6 | -8.6 | 14.2 | 15000 |
| | 70 | 5.4 | -9.5 | 13.6 | 15000 |
| | 100 | 5.5 | -9.0 | 13.9 | 15000 |
| $\frac{\beta_2^*}{\sigma_2} - \frac{\beta_1^*}{\sigma_1}$ | $< 0$ | 5.3 | -6.4 | 14.2 | 9000 |
| | 0 | 3.2 | -7.8 | 13.4 | 2250 |
| | $[0.05, 0.19]$ | 4.2 | -9.5 | 13.6 | 11250 |
| | $[0.20, 0.29]$ | 6.9 | -0.6 | 14.2 | 9000 |
| | $[0.30, 0.39]$ | 6.9 | 0.4 | 12.8 | 6750 |
| | $[0.40, 0.49]$ | 5.3 | 1.6 | 11.9 | 6750 |
| $\rho_0^{(1)}, \rho_0^{(2)}$ | $\rho_0^{(1)} < \rho_0^{(2)}$ | 5.4 | -9.5 | 14.2 | 18000 |
| | $\rho_0^{(1)} = \rho_0^{(2)}$ | 4.7 | -7.8 | 12.6 | 9000 |
| | $\rho_0^{(1)} > \rho_0^{(2)}$ | 5.9 | -6.4 | 14.2 | 18000 |
| $\rho_1^{(1,2)}$ | 0.005 | 4.4 | -9.5 | 13.1 | 9000 |
| | 0.01 | 4.6 | -8.8 | 13.1 | 9000 |
| | 0.02 | 5.0 | -7.4 | 13.2 | 9000 |
| | 0.05 | 6.3 | -2.9 | 13.4 | 9000 |
| | 0.07 | 7.1 | 0 | 14.2 | 9000 |
| $\rho_2^{(1,2)}$ | 0.1 | 5.2 | -9.5 | 13.4 | 9000 |
| | 0.3 | 5.3 | -8.9 | 13.7 | 9000 |
| | 0.5 | 5.5 | -8.4 | 13.9 | 9000 |
| | 0.7 | 5.6 | -7.8 | 14.1 | 9000 |
| | 0.9 | 5.7 | -7.3 | 14.2 | 9000 |

Table 10. Numerical evaluation results – proportion of scenarios for ranges of the power difference between the single weighted 1-DF test and conjunctive IU test, summarized for **Comparison I** (F and t distributions with 2-sided conjunctive test)

| Parameter | Value | $\pi^{W1DF} - \pi^{CONJ}$ 0% to 5% | $\pi^{W1DF} - \pi^{CONJ}$ 5% to 10% | $\pi^{W1DF} - \pi^{CONJ}$ 10% to 20% | $\pi^{W1DF} - \pi^{CONJ}$ > 20% | # Total Scenarios |
|---|---|---|---|---|---|---|
| $K$ | 4 | 1% | 7% | 34% | 58% | 7500 |
| | 6 | 0% | 1% | 12% | 87% | 7500 |
| | 8 | 0% | 1% | 7% | 92% | 7500 |
| | 10 | 0% | 1% | 9% | 90% | 7500 |
| | 20 | 19% | 11% | 16% | 55% | 7500 |
| | 30 | 42% | 10% | 8% | 40% | 7500 |
| $m$ | 50 | 9% | 6% | 15% | 70% | 15000 |
| | 70 | 11% | 4% | 14% | 71% | 15000 |
| | 100 | 12% | 5% | 14% | 70% | 15000 |
| $\frac{\beta_2^*}{\sigma_2} - \frac{\beta_1^*}{\sigma_1}$ | $< 0$ | 25% | 5% | 16% | 54% | 9000 |
| | 0 | 26% | 9% | 16% | 49% | 2250 |
| | $[0.05, 0.19]$ | 10% | 10% | 20% | 60% | 11250 |
| | $[0.20, 0.29]$ | 7% | 5% | 14% | 74% | 9000 |
| | $[0.30, 0.39]$ | 1% | 2% | 12% | 86% | 6750 |
| | $[0.40, 0.49]$ | 0% | 0% | 4% | 96% | 6750 |
| $\rho_0^{(1)}, \rho_0^{(2)}$ | $\rho_0^{(1)} < \rho_0^{(2)}$ | 13% | 5% | 17% | 65% | 18000 |
| | $\rho_0^{(1)} = \rho_0^{(2)}$ | 7% | 5% | 15% | 73% | 9000 |
| | $\rho_0^{(1)} > \rho_0^{(2)}$ | 10% | 5% | 11% | 74% | 18000 |
| $\rho_1^{(1,2)}$ | 0.005 | 10% | 4% | 7% | 80% | 9000 |
| | 0.01 | 10% | 4% | 7% | 80% | 9000 |
| | 0.02 | 10% | 4% | 9% | 78% | 9000 |
| | 0.05 | 10% | 5% | 21% | 64% | 9000 |
| | 0.07 | 12% | 10% | 28% | 49% | 9000 |
| $\rho_2^{(1,2)}$ | 0.1 | 10% | 5% | 13% | 72% | 9000 |
| | 0.3 | 10% | 5% | 14% | 71% | 9000 |
| | 0.5 | 10% | 5% | 14% | 70% | 9000 |
| | 0.7 | 11% | 5% | 15% | 69% | 9000 |
| | 0.9 | 11% | 6% | 16% | 68% | 9000 |

Table 11. Matrix comparison of all statistical design methods

| | PADJ Sidak | PADJ D/AP | COMB | W1DF | DIS2DF | CONJ |
|---|---|---|---|---|---|---|
| **PADJ Bonferroni** | $\pi_{\text{Bonf}}^{\text{PADJ}} < \pi_{\text{Sidak}}^{\text{PADJ}}$ globally | $\pi_{\text{Bonf}}^{\text{PADJ}} < \pi_{\text{D/AP}}^{\text{PADJ}}$ globally | $\pi_{\text{Bonf}}^{\text{PADJ}} < \pi^{\text{COMB}}$ globally | $\pi_{\text{Bonf}}^{\text{PADJ}} < \pi^{\text{W1DF}}$ globally | $\pi_{\text{Bonf}}^{\text{PADJ}} < \pi^{\text{DIS2DF}}$ generally | $\pi_{\text{Bonf}}^{\text{PADJ}} < \pi^{\text{CONJ}}$ generally |
| **PADJ Sidak** | | $\pi_{\text{Sidak}}^{\text{PADJ}} < \pi_{\text{D/AP}}^{\text{PADJ}}$ globally | $\pi_{\text{Sidak}}^{\text{PADJ}} < \pi^{\text{COMB}}$ globally | $\pi_{\text{Sidak}}^{\text{PADJ}} < \pi^{\text{W1DF}}$ globally | $\pi_{\text{Sidak}}^{\text{PADJ}} < \pi^{\text{DIS2DF}}$ generally | Context dependent |
| **PADJ D/AP** | | | $\pi_{\text{D/AP}}^{\text{PADJ}} < \pi^{\text{COMB}}$ globally | $\pi_{\text{D/AP}}^{\text{PADJ}} < \pi^{\text{W1DF}}$ globally | $\pi_{\text{D/AP}}^{\text{PADJ}} < \pi^{\text{DIS2DF}}$ generally | Context dependent |
| **COMB** | | | | If $\rho_0^{(1)} = \rho_0^{(2)}$ and $\sigma_1^2 = \sigma_2^2$, then $\pi^{\text{COMB}} = \pi^{\text{W1DF}}$ globally | Context dependent | $\pi^{\text{CONJ}} < \pi^{\text{COMB}}$ generally |
| **W1DF** | | | | | If $\frac{\beta_1^*}{\sigma_1\sqrt{VIF_1}} = \frac{\beta_2^*}{\sigma_2\sqrt{VIF_2}}$, then $\pi^{\text{DIS2DF}} < \pi^{\text{W1DF}}$ globally | $\pi^{\text{CONJ}} < \pi^{\text{W1DF}}$ generally |
| **DIS2DF** | | | | | | $\pi^{\text{CONJ}} < \pi^{\text{DIS2DF}}$ generally |

* “Globally” – proven theoretically; holds true for all parameter values. “Generally” – dominant in most numerical scenarios; exceptions exist. “Context-dependent” – dominance switches across the parameter space; no single method is uniformly superior.

## 6. Discussion

In this manuscript, we examined the performance of a number of valid design methods for hybrid type 2 cluster randomized trials with continuous co-primary outcomes. These included the p-value adjustment methods, combined outcomes approach, single weighted 1-DF test, disjunctive 2-DF test, and the conjunctive test. A theoretical comparison of the power equations was conducted. It was proven that the p-value adjustment methods are globally less powerful than the combined outcomes approach and the single weighted 1-DF test. It was also shown that the non-centrality parameter for the p-value adjustment methods is always smaller than the non-centrality parameter for the disjunctive 2-DF test, but due to the differing degrees of freedom, there are theoretical cases where the p-value adjustment methods can result in higher power than the disjunctive 2-DF test. We also showed that when the cluster-corrected standardized effect sizes of the first and second outcomes are equal, then the single weighted 1-DF test will be more powerful than the disjunctive test. Lastly, when the outcome specific ICCs and variances between the two outcomes are the same, then combined outcomes approach is theoretically equivalent to the single weighted 1-DF test.[3]

In the numerical evaluation, we explored comparisons that could not be obtained theoretically. We conducted four comparisons that differed in the choice of distribution and whether or not the conjunctive test utilized a 1-tail or 2-tail hypothesis. Results for all comparisons are given in the Supplementary Material, and we discussed the results from Comparison I in depth, which looked at a 2-tailed conjunctive IU test with the t-distribution, and all other methods using an F-distribution. It was found that the treatment effects, outcome variances, and outcome specific ICCs had the largest effect on power and which methods were more powerful than other methods. The combined outcomes approach, single weighted 1-DF test, and disjunctive 2-DF test had the most power, while the p-value adjustment methods and conjunctive IU test never had situations where they yielded the highest power. No method was found to be globally more powerful than another method. In general, the disjunctive 2-DF test did well when the treatment effects were not equal, while the single weighted 1-DF test did well when the treatment effects were equal. We also quantified the extent the extent to which the conjunctive test was more powerful than the popular Bonferroni p-value adjustment method, and also examined it in comparison to the most powerful method – the single weighted 1-DF test.

The choice of conducting a numerical evaluation was made because exact expressions are available for the quantities under consideration. These results can be interpretable as those when asymptotic inference is valid. It is beyond the scope of this paper to investigate relative finite sample performance of these methods, although such a study could be of further interest. Though our numerical evaluation covered 45,000 scenarios that included a wide range of input values for each parameter, and the additional analysis further allowed each individual parameter to vary over an even wider range of values, there are important limitations. We did not examine different values of the overall family-wise false positive rates, and limited our scope to $\alpha = 0.05$. We also did not consider differing treatment allocation ratios, and looked only at equal allocation (though our R package does accommodate unequal treatment allocation ratios). Future work could expand on these results, and look at other input scenarios. Furthermore, we note that these results are restricted to CRTs with two continuous co-primary outcomes. Though these methods have not formally been explored for the case of two binary co-primary outcomes, it is possible to use these design methods to approximate study design specifications for binary outcomes. In fact, in Owen et al., a common approximation is used to obtain the variance for the difference between two binomial proportions from independent data. This is then used instead of the variance for the difference between two means and accounts for clustering.[18] Further work is needed to formally extend and examine these methods for binary data. It is also needed to generalize these methods and their comparisons to stepped wedge hybrid type 2 designs.[19] Despite these limitations, our theoretical comparisons and numerical studies shed light on important relationships between the study design methods for hybrid type 2 studies, allowing strong conclusions to be made.

Table 8 shows in which scenarios a certain method has the highest power, to serve as a guide for researchers to better understand which methods to look into when designing their own hybrid 2 CRT. Though certain methods yielded higher power, we do not dismiss any of the methods as options altogether, for there may be situations where any of the methods could be most fitting depending on one's study and research goals. So, we also encourage the reader to make use of the `crt2power` R package that we introduced[4], or the user-friendly `crt2powerApplication` ShinyApp. This new R package is currently available on CRAN, and allows the user to enact any of the methods to calculate $K$, $m$, or power, for their own studies; likewise, the ShinyApp that uses this package is also currently available using the link provided earlier. This software package and accompanying application allow the user to make an informed decision about which method fits their research goals and needs. The theoretical comparisons, numerical evaluation, and software, have made it possible to better understand the performance of these methods, contributing to the knowledge base of hybrid type 2 studies, and CRTs with continuous co-primary endpoints more generally.

## Conflict of Interest

The authors declare no potential conflict of interests.

# Appendix

## A. Theoretical Comparisons of Study Design Methods

### A.1 Comparison of Method 3: Single Weighted 1-DF Combined Test and Method 4: Disjunctive 2-DF Test

To compare these two methods, we set the equations of the non-centrality parameters equal and see if the expressions reduce. The steps are shown below.

$$\lambda^{\mathrm{W1DF}} = \left[\frac{\sqrt{\frac{(\beta_1^*)^2}{\frac{2\sigma_1^2}{Km}VIF_1}} + \sqrt{\frac{(\beta_2^*)^2}{\frac{2\sigma_2^2}{Km}VIF_2}}}{\sqrt{2\left(1 + \frac{VIF_{12}}{\sqrt{VIF_1 VIF_2}}\right)}}\right]^2$$

$$\lambda^{\mathrm{DIS2DF}} = \left[\frac{Km[(\beta_1^*)^2\,\sigma_2^2\,VIF_2 \;-\; 2\,\beta_1^*\,\beta_2^*\,\sigma_1\,\sigma_2\,VIF_{12} \;+\; (\beta_2^*)^2\,\sigma_1^2\,VIF_1]}{2\,\sigma_1^2\,\sigma_2^2\,[VIF_1\,VIF_2 \;-\; VIF_{12}^2]}\right]$$

$$VIF_1 = 1 + (m-1)\rho_0^{(1)};\; VIF_2 = 1 + (m-1)\rho_0^{(2)};\; VIF_{12} = \rho_2^{(1,2)} + (m-1)\rho_1^{(1,2)}$$

$$\lambda^{\mathrm{W1DF}} \stackrel{?}{=} \lambda^{\mathrm{DIS2DF}}$$

$$\left[\frac{\sqrt{\frac{(\beta_1^*)^2}{\frac{2\sigma_1^2}{Km}VIF_1}} + \sqrt{\frac{(\beta_2^*)^2}{\frac{2\sigma_2^2}{Km}VIF_2}}}{\sqrt{2\left(1 + \frac{VIF_{12}}{\sqrt{VIF_1 VIF_2}}\right)}}\right]^2 \stackrel{?}{=} \left[\frac{Km[(\beta_1^*)^2\,\sigma_2^2\,VIF_2 \;-\; 2\,\beta_1^*\,\beta_2^*\,\sigma_1\,\sigma_2\,VIF_{12} \;+\; (\beta_2^*)^2\,\sigma_1^2\,VIF_1]}{2\,\sigma_1^2\,\sigma_2^2\,[VIF_1\,VIF_2 \;-\; VIF_{12}^2]}\right]$$

$$\frac{\frac{Km(\beta_1^*)^2}{2\sigma_1^2 VIF_1} + \frac{Km(\beta_2^*)^2}{2\sigma_2^2 VIF_2} + \frac{Km\beta_1^*\beta_2^*}{\sigma_1\sigma_2\sqrt{VIF_1 VIF_2}}}{1 + \frac{VIF_{12}}{\sqrt{VIF_1 VIF_2}}} \stackrel{?}{=} \frac{Km[(\beta_1^*)^2\,\sigma_2^2\,VIF_2 \;-\; 2\,\beta_1^*\,\beta_2^*\,\sigma_1\,\sigma_2\,VIF_{12} \;+\; (\beta_2^*)^2\,\sigma_1^2\,VIF_1]}{\sigma_1^2\sigma_2^2\,VIF_1 VIF_2 - \sigma_1^2\sigma_2^2\,VIF_{12}^2}$$

$$\frac{\dfrac{(\beta_1^*)^2}{2\sigma_1^2VIF_1}+\dfrac{(\beta_2^*)^2}{2\sigma_2^2VIF_2}+\dfrac{\beta_1^*\beta_2^*}{\sigma_1\sigma_2\sqrt{VIF_1VIF_2}}}{1+\dfrac{VIF_{12}}{\sqrt{VIF_1VIF_2}}}\stackrel{?}{=}\frac{(\beta_1^*)^2\sigma_2^2VIF_2-2\beta_1^*\beta_2^*\sigma_1\sigma_2VIF_{12}+(\beta_2^*)^2\sigma_1^2VIF_1}{\sigma_1^2\sigma_2^2\ VIF_1VIF_2-\sigma_1^2\sigma_2^2\ VIF_{12}^2}$$

$$\frac{(\beta_1^*)^2(2\sigma_2^2VIF_2)\left(\sigma_1\sigma_2\sqrt{VIF_1VIF_2}\right)+(\beta_2^*)^2(2\sigma_1^2VIF_1)\left(\sigma_1\sigma_2\sqrt{VIF_1VIF_2}\right)+\beta_1^*\beta_2^*(2\sigma_1^2VIF_1)(2\sigma_2^2VIF_2)}{2\sigma_1^2VIF_1(2\sigma_2^2VIF_2)\left(\sigma_1\sigma_2\sqrt{VIF_1VIF_2}\right)+\dfrac{VIF_{12}2\sigma_1^2VIF_1(2\sigma_2^2VIF_2)\left(\sigma_1\sigma_2\sqrt{VIF_1VIF_2}\right)}{\sqrt{VIF_1VIF_2}}}$$
$$\stackrel{?}{=}\frac{(\beta_1^*)^2\sigma_2^2VIF_2-2\beta_1^*\beta_2^*\sigma_1\sigma_2VIF_{12}+(\beta_2^*)^2\sigma_1^2VIF_1}{\sigma_1^2\sigma_2^2\ VIF_1VIF_2-\sigma_1^2\sigma_2^2\ VIF_{12}^2}$$

$$\frac{\sigma_2^2(\beta_1^*)^2VIF_2\sqrt{VIF_1VIF_2}+\sigma_1^2(\beta_2^*)^2VIF_1\sqrt{VIF_1VIF_2}+2\sigma_1\sigma_2\beta_1^*\beta_2^*VIF_1VIF_2}{2\sigma_1^2\sigma_2^2VIF_1VIF_2\sqrt{VIF_1VIF_2}+2\sigma_1^2\sigma_2^2VIF_1VIF_2VIF_{12}}$$
$$\stackrel{?}{=}\frac{(\beta_1^*)^2\sigma_2^2VIF_2-2\beta_1^*\beta_2^*\sigma_1\sigma_2VIF_{12}+(\beta_2^*)^2\sigma_1^2VIF_1}{\sigma_1^2\sigma_2^2\ VIF_1VIF_2-\sigma_1^2\sigma_2^2\ VIF_{12}^2}$$

$$\frac{\sigma_2^2(\beta_1^*)^2VIF_2+\sigma_1^2(\beta_2^*)^2VIF_1+2\sigma_1\sigma_2\beta_1^*\beta_2^*\sqrt{VIF_1VIF_2}}{2\sigma_1^2\sigma_2^2VIF_1VIF_2+2\sigma_1^2\sigma_2^2VIF_{12}\sqrt{VIF_1VIF_2}}$$
$$\stackrel{?}{=}\frac{\sigma_2^2(\beta_1^*)^2VIF_2+\sigma_1^2(\beta_2^*)^2VIF_1-2\sigma_1\sigma_2\beta_1^*\beta_2^*VIF_{12}}{\sigma_1^2\sigma_2^2\ VIF_1VIF_2-\sigma_1^2\sigma_2^2\ VIF_{12}^2}$$

$$\frac{\left[\sigma_2^2(\beta_1^*)^2VIF_2+\sigma_1^2(\beta_2^*)^2VIF_1+2\sigma_1\sigma_2\beta_1^*\beta_2^*\sqrt{VIF_1VIF_2}\ \right](\sigma_1^2\sigma_2^2\ VIF_1VIF_2-\sigma_1^2\sigma_2^2\ VIF_{12}^2)}{\left(2\sigma_1^2\sigma_2^2VIF_1VIF_2+2\sigma_1^2\sigma_2^2VIF_{12}\sqrt{VIF_1VIF_2}\right)(\sigma_1^2\sigma_2^2\ VIF_1VIF_2-\sigma_1^2\sigma_2^2\ VIF_{12}^2)}\stackrel{?}{=}$$
$$\frac{[\sigma_2^2(\beta_1^*)^2VIF_2+\sigma_1^2(\beta_2^*)^2VIF_1-2\sigma_1\sigma_2\beta_1^*\beta_2^*VIF_{12}]\left(2\sigma_1^2\sigma_2^2VIF_1VIF_2+2\sigma_1^2\sigma_2^2VIF_{12}\sqrt{VIF_1VIF_2}\right)}{\left(2\sigma_1^2\sigma_2^2VIF_1VIF_2+2\sigma_1^2\sigma_2^2VIF_{12}\sqrt{VIF_1VIF_2}\right)(\sigma_1^2\sigma_2^2\ VIF_1VIF_2-\sigma_1^2\sigma_2^2\ VIF_{12}^2)}$$

$$\left[\sigma_2^2(\beta_1^*)^2VIF_2+\sigma_1^2(\beta_2^*)^2VIF_1+2\sigma_1\sigma_2\beta_1^*\beta_2^*\sqrt{VIF_1VIF_2}\ \right](\sigma_1^2\sigma_2^2\ VIF_1VIF_2-\sigma_1^2\sigma_2^2\ VIF_{12}^2)$$
$$\stackrel{?}{=}[\sigma_2^2(\beta_1^*)^2VIF_2+\sigma_1^2(\beta_2^*)^2VIF_1-2\sigma_1\sigma_2\beta_1^*\beta_2^*VIF_{12}]\left(2\sigma_1^2\sigma_2^2VIF_1VIF_2\right.$$
$$\left.+\,2\sigma_1^2\sigma_2^2VIF_{12}\sqrt{VIF_1VIF_2}\right)$$

$$\sigma_2^2(\beta_1^*)^2VIF_2\sigma_1^2\sigma_2^2\ VIF_1VIF_2+\sigma_1^2(\beta_2^*)^2VIF_1\sigma_1^2\sigma_2^2\ VIF_1VIF_2$$
$$+\,2\sigma_1\sigma_2\beta_1^*\beta_2^*\sqrt{VIF_1VIF_2}\ \sigma_1^2\sigma_2^2\ VIF_1VIF_2-\sigma_1^2\sigma_2^2\ VIF_{12}^2\sigma_2^2(\beta_1^*)^2VIF_2$$
$$-\,\sigma_1^2\sigma_2^2\ VIF_{12}^2\sigma_1^2(\beta_2^*)^2VIF_1-\sigma_1^2\sigma_2^2\ VIF_{12}^2 2\sigma_1\sigma_2\beta_1^*\beta_2^*\sqrt{VIF_1VIF_2}$$
$$\stackrel{?}{=}\sigma_2^2(\beta_1^*)^2VIF_2 2\sigma_1^2\sigma_2^2VIF_1VIF_2+\sigma_1^2(\beta_2^*)^2VIF_1 2\sigma_1^2\sigma_2^2VIF_1VIF_2$$
$$-\,2\sigma_1\sigma_2\beta_1^*\beta_2^*VIF_{12}2\sigma_1^2\sigma_2^2VIF_1VIF_2+\sigma_2^2(\beta_1^*)^2VIF_2 2\sigma_1^2\sigma_2^2VIF_{12}\sqrt{VIF_1VIF_2}$$
$$+\,\sigma_1^2(\beta_2^*)^2VIF_1 2\sigma_1^2\sigma_2^2VIF_{12}\sqrt{VIF_1VIF_2}$$
$$-\,2\sigma_1\sigma_2\beta_1^*\beta_2^*VIF_{12}2\sigma_1^2\sigma_2^2VIF_{12}\sqrt{VIF_1VIF_2}$$

$$
\begin{aligned}
&\sigma_2^2(\beta_1^*)^2 VIF_1 VIF_2^2 + \sigma_1^2(\beta_2^*)^2 VIF_1^2 VIF_2 + 2\sigma_1\sigma_2\beta_1^*\beta_2^* VIF_1 VIF_2\sqrt{VIF_1 VIF_2} \\
&\qquad - \sigma_2^2(\beta_1^*)^2 VIF_2 VIF_{12}^2 - \sigma_1^2(\beta_2^*)^2 VIF_1 VIF_{12}^2 - 2\sigma_1\sigma_2\beta_1^*\beta_2^* VIF_{12}^2\sqrt{VIF_1 VIF_2} \\
&\stackrel{?}{=} 2\sigma_2^2(\beta_1^*)^2 VIF_1 VIF_2^2 + 2\sigma_1^2(\beta_2^*)^2 VIF_1^2 VIF_2 - 4\sigma_1\sigma_2\beta_1^*\beta_2^* VIF_1 VIF_2 VIF_{12} \\
&\qquad + 2\sigma_2^2(\beta_1^*)^2 VIF_2 VIF_{12}\sqrt{VIF_1 VIF_2} + 2\sigma_1^2(\beta_2^*)^2 VIF_1 VIF_{12}\sqrt{VIF_1 VIF_2} \\
&\qquad - 4\sigma_1\sigma_2\beta_1^*\beta_2^* VIF_{12}^2\sqrt{VIF_1 VIF_2}
\end{aligned}
$$

$$
\begin{aligned}
&2\sigma_1\sigma_2\beta_1^*\beta_2^* VIF_1 VIF_2\sqrt{VIF_1 VIF_2} - 2\sigma_1\sigma_2\beta_1^*\beta_2^* VIF_{12}^2\sqrt{VIF_1 VIF_2} + 4\sigma_1\sigma_2\beta_1^*\beta_2^* VIF_{12}^2\sqrt{VIF_1 VIF_2} \\
&\qquad + 4\sigma_1\sigma_2\beta_1^*\beta_2^* VIF_1 VIF_2 VIF_{12} \\
&\stackrel{?}{=} \sigma_2^2(\beta_1^*)^2 VIF_2 VIF_{12}^2 + \sigma_2^2(\beta_1^*)^2 VIF_1 VIF_2^2 + 2\sigma_2^2(\beta_1^*)^2 VIF_2 VIF_{12}\sqrt{VIF_1 VIF_2} \\
&\qquad + \sigma_1^2(\beta_2^*)^2 VIF_1 VIF_{12}^2 + \sigma_1^2(\beta_2^*)^2 VIF_1^2 VIF_2 + 2\sigma_1^2(\beta_2^*)^2 VIF_1 VIF_{12}\sqrt{VIF_1 VIF_2}
\end{aligned}
$$

$$
\begin{aligned}
&2\sigma_1\sigma_2\beta_1^*\beta_2^*\sqrt{VIF_1 VIF_2}\left(VIF_1 VIF_2 + VIF_{12}^2 + 2VIF_{12}\sqrt{VIF_1 VIF_2}\right) \\
&\stackrel{?}{=} \sigma_2^2(\beta_1^*)^2 VIF_2\left(VIF_1 VIF_2 + VIF_{12}^2 + 2VIF_{12}\sqrt{VIF_1 VIF_2}\right) \\
&\qquad + \sigma_1^2(\beta_2^*)^2 VIF_1\left(VIF_1 VIF_2 + VIF_{12}^2 + 2VIF_{12}\sqrt{VIF_1 VIF_2}\right)
\end{aligned}
$$

$$
\begin{aligned}
&2\sigma_1\sigma_2\beta_1^*\beta_2^*\sqrt{VIF_1 VIF_2}\left(VIF_{12} + \sqrt{VIF_1 VIF_2}\right)^2 \\
&\stackrel{?}{=} \sigma_2^2(\beta_1^*)^2 VIF_2\left(VIF_{12} + \sqrt{VIF_1 VIF_2}\right)^2 + \sigma_1^2(\beta_2^*)^2 VIF_1\left(VIF_{12} + \sqrt{VIF_1 VIF_2}\right)^2
\end{aligned}
$$

$$2\sigma_1\sigma_2\beta_1^*\beta_2^*\sqrt{VIF_1 VIF_2} \stackrel{?}{=} \sigma_2^2(\beta_1^*)^2 VIF_2 + \sigma_1^2(\beta_2^*)^2 VIF_1$$

$$0 \stackrel{?}{=} \sigma_2^2(\beta_1^*)^2 VIF_2 + \sigma_1^2(\beta_2^*)^2 VIF_1 - 2\sigma_1\sigma_2\beta_1^*\beta_2^*\sqrt{VIF_1 VIF_2}$$

$$0 \leq \left(\sigma_2\beta_1^*\sqrt{VIF_2} - \sigma_1\beta_2^*\sqrt{VIF_1}\right)^2$$

$$\Rightarrow \lambda^{\mathrm{W1DF}} \leq \lambda^{\mathrm{DIS2DF}}$$

We attempted to compare the exact theoretical quantities between the power for the single weighted 1-DF test and the power for the disjunctive 2-DF test, namely

$$\pi^{\mathrm{W1DF}} = \int_{c^{\mathrm{W1DF}}}^{\infty} \chi^2(x;\ 1, \lambda^{\mathrm{W1DF}})dx = Q_{\frac{1}{2}}\left(\sqrt{\lambda^{\mathrm{W1DF}}}, \sqrt{c^{\mathrm{W1DF}}}\right)$$

$$\pi^{\mathrm{DIS2DF}} = \int_{c^{\mathrm{DIS2DF}}}^{\infty} \chi^2(x;\ 2, \lambda^{(\mathrm{DIS2DF})})dx = Q_1\left(\sqrt{\lambda^{\mathrm{DIS2DF}}}, \sqrt{c^{\mathrm{DIS2DF}}}\right)$$

We show that the power for the single weighted 1-DF test can be reduced nicely to a closed-form expression, but the power for the disjunctive 2-DF test cannot. This motivates and justifies our use for using numerical analyses to explore the relationship between these methods.

$$
\begin{aligned}
&\pi^{\mathrm{W1DF}} = \int_{c^{\mathrm{W1DF}}}^{\infty} \chi^2(x;\ 1, \lambda^{\mathrm{W1DF}})dx \\
&= F(\infty; 1, \lambda^{\mathrm{W1DF}}) - F(c^{\mathrm{W1DF}}; 1, \lambda^{\mathrm{W1DF}}) \\
&= 1 - F(c^{\mathrm{W1DF}}; 1, \lambda^{\mathrm{W1DF}}) \\
&= 1 - \left[1 - Q_{\frac{1}{2}}\left(\sqrt{\lambda^{\mathrm{W1DF}}}, \sqrt{c^{\mathrm{W1DF}}}\right)\right]
\end{aligned}
$$

$$= Q_{\frac{1}{2}}\left(\sqrt{\lambda^{\mathrm{W1DF}}}, \sqrt{c^{\mathrm{W1DF}}}\right) \text{ (Marcum Q} - \text{Function)}$$

$$= \frac{1}{\left(\sqrt{\lambda^{\mathrm{W1DF}}}\right)^{\frac{1}{2}-1}} \int_{\sqrt{c^{\mathrm{W1DF}}}}^{\infty} x^{\frac{1}{2}} \exp\left(-\frac{x^2 + \left(\sqrt{\lambda^{\mathrm{W1DF}}}\right)^2}{2}\right) I_{\frac{1}{2}-1}\left(x\sqrt{\lambda^{\mathrm{W1DF}}}\right) dx$$

$$= \left(\lambda^{\mathrm{W1DF}}\right)^{\frac{1}{4}} \int_{\sqrt{c^{\mathrm{W1DF}}}}^{\infty} x^{\frac{1}{2}} \exp\left(-\frac{x^2 + \lambda^{\mathrm{W1DF}}}{2}\right) I_{-\frac{1}{2}}\left(x\sqrt{\lambda^{\mathrm{W1DF}}}\right) dx$$

(Invoke property of modified Bessel function)

$$= \left(\lambda^{\mathrm{W1DF}}\right)^{\frac{1}{4}} \int_{\sqrt{c^{\mathrm{W1DF}}}}^{\infty} x^{\frac{1}{2}} \exp\left(-\frac{x^2 + \lambda^{\mathrm{W1DF}}}{2}\right) \sqrt{\frac{2}{\pi x \sqrt{\lambda^{\mathrm{W1DF}}}}} \cosh\left(x\sqrt{\lambda^{\mathrm{W1DF}}}\right) dx$$

$$= \left(\lambda^{\mathrm{W1DF}}\right)^{\frac{1}{4}} \frac{\sqrt{2}}{\sqrt{\pi}} \frac{1}{\left(\lambda^{\mathrm{W1DF}}\right)^{\frac{1}{4}}} \int_{\sqrt{c^{\mathrm{W1DF}}}}^{\infty} \exp\left(-\frac{x^2 + \lambda^{\mathrm{W1DF}}}{2}\right) \cosh\left(x\sqrt{\lambda^{\mathrm{W1DF}}}\right) dx$$

$$= \frac{\sqrt{2}}{\sqrt{\pi}} \int_{\sqrt{c^{\mathrm{W1DF}}}}^{\infty} \exp\left(-\frac{x^2 + \lambda^{\mathrm{W1DF}}}{2}\right) \cosh\left(x\sqrt{\lambda^{\mathrm{W1DF}}}\right) dx$$

$$= \frac{\sqrt{2}}{\sqrt{\pi}} \int_{\sqrt{c^{\mathrm{W1DF}}}}^{\infty} \exp\left(-\frac{x^2 + \lambda^{\mathrm{W1DF}}}{2}\right) \frac{\exp\left(x\sqrt{\lambda^{\mathrm{W1DF}}}\right) + \exp\left(-x\sqrt{\lambda^{\mathrm{W1DF}}}\right)}{2} dx$$

$$= \frac{\sqrt{2}}{2\sqrt{\pi}} \int_{\sqrt{c^{\mathrm{W1DF}}}}^{\infty} \left[\exp\left(-\frac{x^2 + \lambda^{\mathrm{W1DF}}}{2}\right) \exp\left(x\sqrt{\lambda^{\mathrm{W1DF}}}\right) + \exp\left(-\frac{x^2 + \lambda^{\mathrm{W1DF}}}{2}\right) \exp\left(-x\sqrt{\lambda^{\mathrm{W1DF}}}\right)\right] dx$$

$$= \frac{\sqrt{2}}{2\sqrt{\pi}} \int_{\sqrt{c^{\mathrm{W1DF}}}}^{\infty} \exp\left(-\frac{x^2 + \lambda^{\mathrm{W1DF}}}{2}\right) \exp\left(x\sqrt{\lambda^{\mathrm{W1DF}}}\right) dx$$

$$+ \frac{\sqrt{2}}{2\sqrt{\pi}} \int_{\sqrt{c^{\mathrm{W1DF}}}}^{\infty} \exp\left(-\frac{x^2 + \lambda^{\mathrm{W1DF}}}{2}\right) \exp\left(-x\sqrt{\lambda^{\mathrm{W1DF}}}\right) dx$$

$$= \frac{\sqrt{2}}{2\sqrt{\pi}} \int_{\sqrt{c^{\mathrm{W1DF}}}}^{\infty} \exp\left(x\sqrt{\lambda^{\mathrm{W1DF}}} - \frac{x^2 + \lambda^{\mathrm{W1DF}}}{2}\right) dx + \frac{\sqrt{2}}{2\sqrt{\pi}} \int_{\sqrt{c^{\mathrm{W1DF}}}}^{\infty} \exp\left(-x\sqrt{\lambda^{\mathrm{W1DF}}} - \frac{x^2 + \lambda^{\mathrm{W1DF}}}{2}\right) dx$$

$$= \frac{\sqrt{2}}{2\sqrt{\pi}} \int_{\sqrt{c^{\mathrm{W1DF}}}}^{\infty} \exp\left(\frac{2x\sqrt{\lambda^{\mathrm{W1DF}}}}{2} - \frac{x^2 + \lambda^{\mathrm{W1DF}}}{2}\right) dx + \frac{\sqrt{2}}{2\sqrt{\pi}} \int_{\sqrt{c^{\mathrm{W1DF}}}}^{\infty} \exp\left(-\frac{2x\sqrt{\lambda^{\mathrm{W1DF}}}}{2} - \frac{x^2 + \lambda^{\mathrm{W1DF}}}{2}\right) dx$$

$$= \frac{\sqrt{2}}{2\sqrt{\pi}} \int_{\sqrt{c^{\mathrm{W1DF}}}}^{\infty} \exp\left(\frac{2x\sqrt{\lambda^{\mathrm{W1DF}}} - x^2 - \lambda^{\mathrm{W1DF}}}{2}\right) dx + \frac{\sqrt{2}}{2\sqrt{\pi}} \int_{\sqrt{c^{\mathrm{W1DF}}}}^{\infty} \exp\left(\frac{-2x\sqrt{\lambda^{\mathrm{W1DF}}} - x^2 - \lambda^{\mathrm{W1DF}}}{2}\right) dx$$

$$= \frac{\sqrt{2}}{2\sqrt{\pi}} \exp\left(-\frac{\lambda^{\mathrm{W1DF}}}{2}\right) \int_{\sqrt{c^{\mathrm{W1DF}}}}^{\infty} \exp\left(\frac{2x\sqrt{\lambda^{\mathrm{W1DF}}} - x^2}{2}\right) dx$$

$$+ \frac{\sqrt{2}}{2\sqrt{\pi}} \exp\left(-\frac{\lambda^{S1\mathrm{DF}}}{2}\right) \int_{\sqrt{c^{\mathrm{W1DF}}}}^{\infty} \exp\left(\frac{-2x\sqrt{\lambda^{\mathrm{W1DF}}} - x^2}{2}\right) dx$$

$$= \frac{\sqrt{2}}{2\sqrt{\pi}} \exp\left(-\frac{\lambda^{\mathrm{W1DF}}}{2}\right) \int_{\sqrt{c^{\mathrm{W1DF}}}}^{\infty} \exp\left(\frac{2x\sqrt{\lambda^{\mathrm{W1DF}}} - x^2}{2}\right) dx$$

$$+ \frac{\sqrt{2}}{2\sqrt{\pi}} \exp\left(-\frac{\lambda^{\mathrm{W1DF}}}{2}\right) \int_{\sqrt{c^{\mathrm{W1DF}}}}^{\infty} \exp\left(\frac{-2x\sqrt{\lambda^{\mathrm{W1DF}}} - x^2}{2}\right) dx$$

$$= \frac{\sqrt{2}}{2\sqrt{\pi}} \exp\left(-\frac{\lambda^{\mathrm{W1DF}}}{2}\right) \left[\frac{\sqrt{\pi}}{\sqrt{2}} \exp\left(\frac{\lambda^{\mathrm{W1DF}}}{2}\right) \operatorname{erfc}\left(\frac{\sqrt{c^{\mathrm{W1DF}}} - \sqrt{\lambda^{\mathrm{W1DF}}}}{\sqrt{2}}\right)\right]$$

$$+ \frac{\sqrt{2}}{2\sqrt{\pi}} \exp\left(-\frac{\lambda^{\mathrm{W1DF}}}{2}\right) \left[\frac{\sqrt{\pi}}{\sqrt{2}} \exp\left(\frac{\lambda^{\mathrm{W1DF}}}{2}\right) \operatorname{erfc}\left(\frac{\sqrt{c^{\mathrm{W1DF}}} + \sqrt{\lambda^{\mathrm{W1DF}}}}{\sqrt{2}}\right)\right]$$

$$= \frac{1}{2} \operatorname{erfc}\left(\frac{\sqrt{c^{\mathrm{W1DF}}} - \sqrt{\lambda^{\mathrm{W1DF}}}}{\sqrt{2}}\right) + \frac{1}{2} \operatorname{erfc}\left(\frac{\sqrt{c^{\mathrm{W1DF}}} + \sqrt{\lambda^{\mathrm{W1DF}}}}{\sqrt{2}}\right)$$

$$
\begin{aligned}
\pi^{\text{DIS2DF}} &= \int_{c^{\text{DIS2DF}}}^{\infty} \chi^2(y;\ 2, \lambda^{\text{DIS2DF}}) dy \\
&= F(\infty; 2, \lambda^{\text{DIS2DF}}) - F(c^{\text{DIS2DF}}; 2, \lambda^{\text{DIS2DF}}) \\
&= 1 - F(c^{\text{DIS2DF}}; 2, \lambda^{\text{DIS2DF}}) \\
&= 1 - \left[1 - Q_1\left(\sqrt{\lambda^{\text{DIS2DF}}}, \sqrt{c^{\text{DIS2DF}}}\right)\right] \\
&= Q_1\left(\sqrt{\lambda^{\text{DIS2DF}}}, \sqrt{c^{\text{DIS2DF}}}\right) \\
&= \frac{1}{\left(\sqrt{\lambda^{\text{DIS2DF}}}\right)^{1-1}} \int_{\sqrt{c^{\text{DIS2DF}}}}^{\infty} y^1 \exp\left(-\frac{y^2 + \left(\sqrt{\lambda^{\text{DIS2DF}}}\right)^2}{2}\right) I_{1-1}\left(y\sqrt{\lambda^{\text{DIS2DF}}}\right) dy \\
&= \int_{\sqrt{c^{\text{DIS2DF}}}}^{\infty} y \exp\left(-\frac{y^2 + \lambda^{\text{DIS2DF}}}{2}\right) I_0\left(y\sqrt{\lambda^{\text{DIS2DF}}}\right) dy \\
&= \int_{\sqrt{c^{\text{DIS2DF}}}}^{\infty} y \exp\left(-\frac{y^2 + \lambda^{\text{DIS2DF}}}{2}\right) \left[\frac{1}{\pi}\int_0^{\pi} \exp\left(y\sqrt{\lambda^{\text{DIS2DF}}}\cos(t)\right) dt\right] dy
\end{aligned}
$$

Figure 5 shows the statistical power, $\pi^{\text{W1DF}}$ and $\pi^{\text{DIS2DF}}$, when $\lambda^{\text{W1DF}} = \lambda^{\text{DIS2DF}} = \lambda \in (0,30)$. For this assessment, we set $c^{\text{W1DF}} = \chi^2_{1-\alpha}(1)$ and $c^{\text{DIS2DF}} = \chi^2_{1-\alpha}(2)$ for common family-wise false positive rates, namely $\alpha = 0.01, 0.025, 0.05,$ and 0.1. We observe that for this case when the non-centrality parameters are the same, the single weighted 1-DF test approach always yields more power than the disjunctive 2-DF test approach, regardless of the choice of $\alpha$. Thus, although the relationship between these two methods is not simple to discern when directly comparing their power equations, using R we've shown that the single weighted 1-DF test is always more powerful in practice when $\lambda^{\text{W1DF}} = \lambda^{\text{DIS2DF}}$ for $\lambda \in (0,30)$ and $\alpha = 0.01, 0.025, 0.05,$ and 0.1.

Figure 5. Theoretical comparison of power for the Single Weighted 1-DF Test and the Disjunctive 2-DF Test, $\pi^{W1DF}$ $vs.\ \pi^{DIS2DF}$, for varying overall false-positive rate values ($\alpha$) when $\lambda^{W1DF} = \lambda^{DIS2DF}$ using the $\chi^2$-distribution

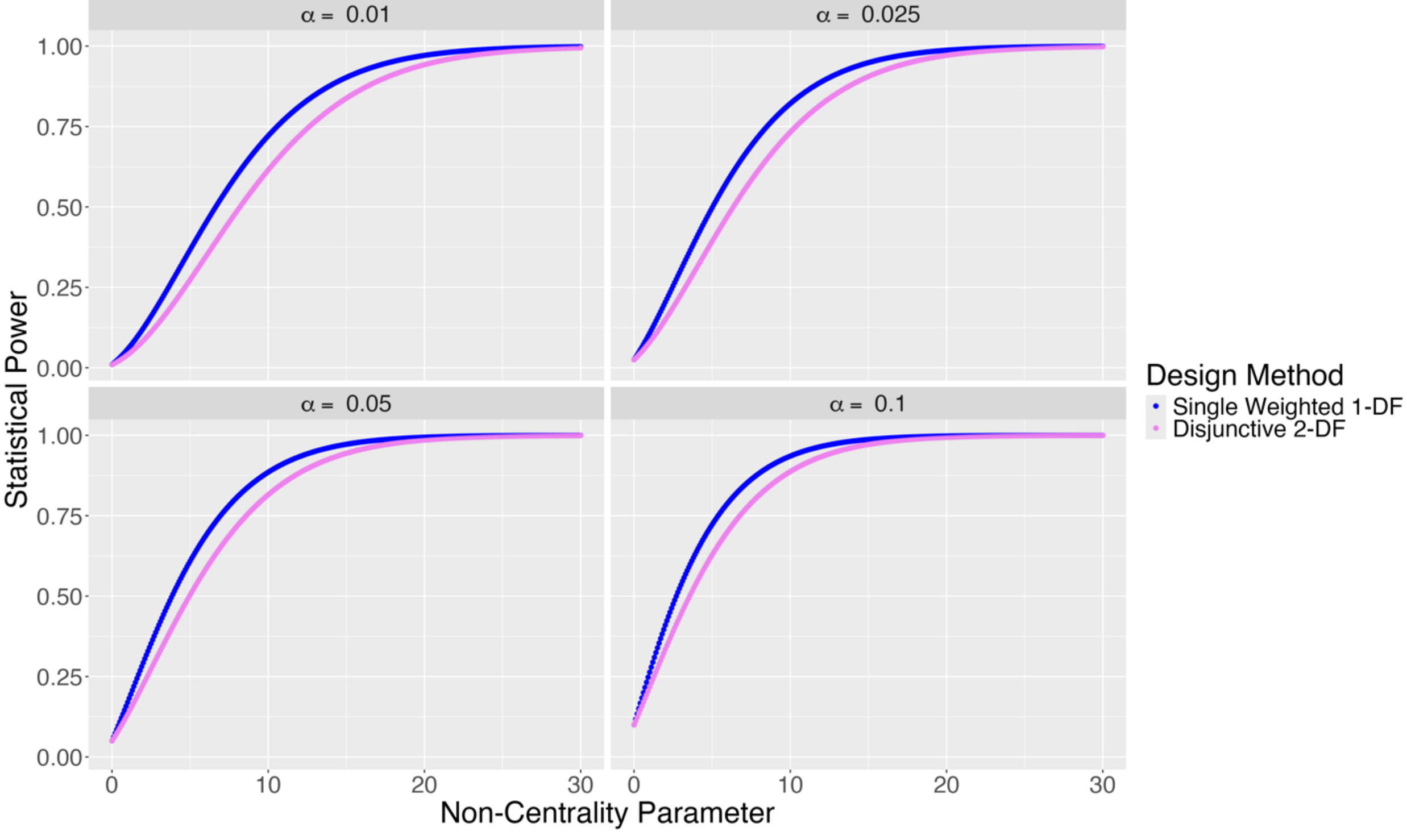

### A.2 Comparison of Method 1: P-value Adjustments and Method 2: Combined Outcomes

$$\lambda^{\text{PADJ}} = \min\left( \frac{(\beta_1^*)^2}{2\,\frac{\sigma_1^2}{Km}\left[1 + (m-1)\rho_0^{(1)}\right]}, \frac{(\beta_2^*)^2}{2\,\frac{\sigma_2^2}{Km}\left[1 + (m-1)\rho_0^{(2)}\right]} \right)$$

$$\lambda^{\text{COMB}} = \frac{(\beta_1^* + \beta_2^*)^2}{2\,\frac{\sigma_1^2 + \sigma_2^2 + 2\rho_2^{(1,2)}\sigma_1\sigma_2}{Km}\left[1 + (m-1)\frac{\rho_0^{(1)}\sigma_1^2 + \rho_0^{(2)}\sigma_2^2 + 2\rho_1^{(1,2)}\sigma_1\sigma_2}{\sigma_1^2 + \sigma_2^2 + 2\rho_2^{(1,2)}\sigma_1\sigma_2}\right]}$$

Can also be written in terms of variance inflation factor terms:

$$\lambda^{\text{PADJ}} = \min\left( \frac{(\beta_1^*)^2}{\frac{2}{Km}\sigma_1^2 VIF_1}, \frac{(\beta_2^*)^2}{\frac{2}{Km}\sigma_2^2 VIF_2} \right)$$

$$\lambda^{\text{COMB}} = \frac{(\beta_1^* + \beta_2^*)^2}{\frac{2}{Km}\left[\sigma_1^2\, VIF_1 + \sigma_2^2\, VIF_2 + 2\sigma_1\sigma_2 VIF_{12}\right]}$$

Recall that $VIF_1 = 1 + (m-1)\rho_0^{(1)}$; $VIF_2 = 1 + (m-1)\rho_0^{(2)}$; $VIF_{12} = \rho_2^{(1,2)} + (m-1)\rho_1^{(1,2)}$. We also have that $\rho_1^{(1,2)} < \rho_0^{(1)}$; $\rho_1^{(1,2)} < \rho_0^{(2)}$; $\rho_2^{(1,2)} < 1$. This implies $VIF_{12} < VIF_1$ and $VIF_{12} < VIF_2$.

Note that for the p-value adjustment method, $\pi^{\text{PADJ}} = \min\left(\pi^{(1)}, \pi^{(2)}\right) = \min\left(1 - \chi^2\left[\lambda^{(1)}, 1\right], 1 - \chi^2\left[\lambda^{(2)}, 1\right]\right)$, which can also be written as $\pi^{\text{PADJ}} = 1 - \chi^2\left[\min\left(\lambda^{(1)}, \lambda^{(2)}\right), 1\right]$. For the combined outcomes approach, $\pi^{\text{COMB}} = 1 - \chi^2[\lambda^{\text{COMB}}, 1]$. These two methods for calculating power use the same distribution and degrees-of-freedom.

Also note that for any adjustment method, it is always the case that $\alpha^{\text{PADJ}} < \alpha^{\text{COMB}}$. A smaller $\alpha$ value corresponds to smaller statistical power (keeping all other terms constant), and similarly, a smaller non-centrality parameter corresponds to smaller statistical power (keeping all other terms constant). So, to show that $\pi^{\text{PADJ}} < \pi^{\text{COMB}}$, it will suffice to show that $\lambda^{\text{PADJ}} < \lambda^{\text{COMB}}$.

For these proofs, we also assume that the treatment effects are non-negative. If this is not the case, they can be transformed in order to meet this assumption.

<u>Case 1</u>

Suppose $\frac{(\beta_1^*)^2}{\frac{2}{Km}\sigma_1^2 VIF_1} < \frac{(\beta_2^*)^2}{\frac{2}{Km}\sigma_2^2 VIF_2}$. This implies that $\lambda^{\text{PADJ}} = \frac{(\beta_1^*)^2}{\frac{2}{Km}\sigma_1^2 VIF_1}$. We want to show that $\frac{(\beta_1^*)^2}{\frac{2}{Km}\sigma_1^2 VIF_1} < \frac{(\beta_1^* + \beta_2^*)^2}{\frac{2}{Km}\left[\sigma_1^2\, VIF_1 + \sigma_2^2\, VIF_2 + 2\sigma_1\sigma_2 VIF_{12}\right]}$. Use the fact that:

$$\frac{(\beta_1^*)^2}{\frac{2}{Km}\sigma_1^2 VIF_1} < \frac{(\beta_2^*)^2}{\frac{2}{Km}\sigma_2^2 VIF_2} \Rightarrow \frac{(\beta_1^*)^2}{\sigma_1^2 VIF_1} < \frac{(\beta_2^*)^2}{\sigma_2^2 VIF_2}$$

$$\Rightarrow (\beta_1^*)^2\sigma_2^2 VIF_2 < (\beta_2^*)^2\sigma_1^2 VIF_1$$
$$\Rightarrow \sqrt{(\beta_1^*)^2\sigma_2^2 VIF_2} < \sqrt{(\beta_2^*)^2\sigma_1^2 VIF_1}$$
$$\Rightarrow \beta_1^*\sigma_2\sqrt{VIF_2} < \beta_2^*\sigma_1\sqrt{VIF_1}$$

Since we already know that $VIF_{12} < VIF_2$ and $VIF_{12} < VIF_1$, then clearly $\sqrt{VIF_{12}} < \sqrt{VIF_2}$ and $\sqrt{VIF_{12}} < \sqrt{VIF_1}$. From this fact, it follows that $\beta_1^*\sigma_2\sqrt{VIF_{12}} < \beta_2^*\sigma_1\sqrt{VIF_1}$. Looking at this inequality now, we have:

$$\beta_1^*\sigma_2\sqrt{VIF_{12}} < \beta_2^*\sigma_1\sqrt{VIF_1}$$
$$\Rightarrow \beta_1^*\sigma_2\sqrt{VIF_{12}}\sqrt{VIF_{12}} < \beta_2^*\sigma_1\sqrt{VIF_1}\sqrt{VIF_{12}}$$
$$\Rightarrow \beta_1^*\sigma_2 VIF_{12} < \beta_2^*\sigma_1\sqrt{VIF_1}\sqrt{VIF_{12}} < \beta_2^*\sigma_1\sqrt{VIF_1}\sqrt{VIF_1}$$
$$\Rightarrow \beta_1^*\sigma_2 VIF_{12} < \beta_2^*\sigma_1\sqrt{VIF_1}\sqrt{VIF_{12}} < \beta_2^*\sigma_1 VIF_1$$
$$\Rightarrow \beta_1^*\sigma_2 VIF_{12} < \beta_2^*\sigma_1 VIF_1$$

and thus we've shown that the inequality $\beta_1^*\sigma_2 VIF_{12} < \beta_2^*\sigma_1 VIF_1$ holds. Now, we want to show the inequality:

$$\frac{(\beta_1^*)^2}{\frac{2}{Km}\sigma_1^2 VIF_1} \overset{?}{<} \frac{(\beta_1^* + \beta_2^*)^2}{\frac{2}{Km}[\sigma_1^2\, VIF_1 + \sigma_2^2\, VIF_2 + 2\sigma_1\sigma_2 VIF_{12}]}$$
$$\frac{(\beta_1^*)^2}{\sigma_1^2 VIF_1} \overset{?}{<} \frac{(\beta_1^*)^2 + (\beta_2^*)^2 + 2\beta_1^*\beta_2^*}{\sigma_1^2\, VIF_1 + \sigma_2^2\, VIF_2 + 2\sigma_1\sigma_2 VIF_{12}}$$
$$(\beta_1^*)^2[\sigma_1^2\, VIF_1 + \sigma_2^2\, VIF_2 + 2\sigma_1\sigma_2 VIF_{12}] \overset{?}{<} [(\beta_1^*)^2 + (\beta_2^*)^2 + 2\beta_1^*\beta_2^*]\sigma_1^2 VIF_1$$
$$(\beta_1^*)^2\sigma_1^2\, VIF_1 + (\beta_1^*)^2\sigma_2^2\, VIF_2 + 2(\beta_1^*)^2\sigma_1\sigma_2 VIF_{12} \overset{?}{<} (\beta_1^*)^2\sigma_1^2 VIF_1 + (\beta_2^*)^2\sigma_1^2 VIF_1 + 2\beta_1^*\beta_2^*\sigma_1^2 VIF_1$$
$$(\beta_1^*)^2\sigma_2^2\, VIF_2 + 2(\beta_1^*)^2\sigma_1\sigma_2 VIF_{12} \overset{?}{<} (\beta_2^*)^2\sigma_1^2 VIF_1 + 2\beta_1^*\beta_2^*\sigma_1^2 VIF_1$$

We already know that $(\beta_1^*)^2\sigma_2^2\, VIF_2 < (\beta_2^*)^2\sigma_1^2 VIF_1$, so we just need to show that $2(\beta_1^*)^2\sigma_1\sigma_2 VIF_{12} < 2\beta_1^*\beta_2^*\sigma_1^2 VIF_1$.

$$2(\beta_1^*)^2\sigma_1\sigma_2 VIF_{12} \overset{?}{<} 2\beta_1^*\beta_2^*\sigma_1^2 VIF_1$$
$$\beta_1^*\sigma_2 VIF_{12} \overset{?}{<} \beta_2^*\sigma_1 VIF_1$$

We've just shown that this holds from our first assumption that $\frac{(\beta_1^*)^2}{\frac{2}{Km}\sigma_1^2 VIF_1} < \frac{(\beta_2^*)^2}{\frac{2}{Km}\sigma_2^2 VIF_2}$. So, because $(\beta_1^*)^2\sigma_2^2\, VIF_2 < (\beta_2^*)^2\sigma_1^2 VIF_1$ and $\beta_1^*\sigma_2 VIF_{12} < \beta_2^*\sigma_1 VIF_1$, it follows that

$$\lambda^{\text{PADJ}} = \frac{(\beta_1^*)^2}{\frac{2}{Km}\sigma_1^2 VIF_1} < \frac{(\beta_1^* + \beta_2^*)^2}{\frac{2}{Km}[\sigma_1^2\, VIF_1 + \sigma_2^2\, VIF_2 + 2\sigma_1\sigma_2 VIF_{12}]} = \lambda^{\text{COMB}}$$

Then in terms of power, since $\lambda^{\text{PADJ}} < \lambda^{\text{COMB}}$, and $\alpha^{\text{PADJ}} < \alpha^{\text{COMB}}$ it follows that $\pi^{\text{PADJ}} < \pi^{\text{COMB}}$, meaning that the p-value adjustment methods will always be less powerful than the combined outcomes approach.

Case 2

For this case, we suppose instead that $\frac{(\beta_2^*)^2}{\frac{2}{Km}\sigma_2^2 VIF_2} < \frac{(\beta_1^*)^2}{\frac{2}{Km}\sigma_1^2 VIF_1}$.

This implies that $\lambda^{\mathrm{PADJ}} = \frac{(\beta_2^*)^2}{\frac{2}{Km}\sigma_2^2 VIF_2}$. The proof follows from the same logic as Case 1.

## A.3 Comparison of Method 1: P-value Adjustments and Method 3: Single Weighted 1-DF Test

$$\lambda^{\text{PADJ}} = \min\left(\frac{(\beta_1^*)^2}{2\,\frac{\sigma_1^2}{Km}\left[1+(m-1)\rho_0^{(1)}\right]}, \frac{(\beta_2^*)^2}{2\,\frac{\sigma_2^2}{Km}\left[1+(m-1)\rho_0^{(2)}\right]}\right)$$

$$\lambda^{\text{W1DF}} = \left[\frac{\sqrt{\frac{(\beta_1^*)^2}{\frac{2\sigma_1^2}{Km}\left[1+(m-1)\rho_0^{(1)}\right]}} + \sqrt{\frac{(\beta_2^*)^2}{\frac{2\sigma_2^2}{Km}\left[1+(m-1)\rho_0^{(2)}\right]}}}{\sqrt{2\left(1+\frac{\left(\rho_2^{(1,2)}+(m-1)\rho_1^{(1,2)}\right)}{\sqrt{\left(1+(m-1)\rho_0^{(1)}\right)\left(1+(m-1)\rho_0^{(2)}\right)}}\right)}}\right]^2$$

Can also be written in terms of variance inflation factor terms:

$$\lambda^{\text{PADJ}} = \min\left(\frac{(\beta_1^*)^2}{\frac{2}{Km}\sigma_1^2 VIF_1}, \frac{(\beta_2^*)^2}{\frac{2}{Km}\sigma_2^2 VIF_2}\right);\ \lambda^{\text{W1DF}} = \left[\frac{\sqrt{\frac{(\beta_1^*)^2}{\frac{2\sigma_1^2}{Km}VIF_1}} + \sqrt{\frac{(\beta_2^*)^2}{\frac{2\sigma_2^2}{Km}VIF_2}}}{\sqrt{2\left(1+\frac{VIF_{12}}{\sqrt{VIF_1 VIF_2}}\right)}}\right]^2$$

Recall that $VIF_1 = 1 + (m-1)\rho_0^{(1)}; VIF_2 = 1 + (m-1)\rho_0^{(2)}; VIF_{12} = \rho_2^{(1,2)} + (m-1)\rho_1^{(1,2)}$. We also have that $\rho_1^{(1,2)} < \rho_0^{(1)}; \rho_1^{(1,2)} < \rho_0^{(2)}; \rho_2^{(1,2)} < 1$. This implies $VIF_{12} < VIF_1$ and $VIF_{12} < VIF_2$.

Note that for the p-value adjustment method, $\pi^{\text{PADJ}} = \min\left(\pi^{(1)}, \pi^{(2)}\right) = \min\left(1 - \chi^2\left[\lambda^{(1)}, 1\right], 1 - \chi^2\left[\lambda^{(2)}, 1\right]\right)$, which can also be written as $\pi^{\text{PADJ}} = 1 - \chi^2\left[\min\left(\lambda^{(1)}, \lambda^{(2)}\right), 1\right]$. For the single weighted 1-DF test, $\pi^{\text{W1DF}} = 1 - \chi^2[\lambda^{\text{W1DF}}, 1]$. These two methods for calculating power use the same distribution and degrees-of-freedom.

Also note that for any adjustment method, it is always the case that $\alpha^{\text{PADJ}} < \alpha^{\text{W1DF}}$. A smaller $\alpha$ value corresponds to smaller statistical power (keeping all other terms constant), and similarly, a smaller non-centrality parameter corresponds to smaller statistical power (keeping all other terms constant). So, to show that $\pi^{\text{PADJ}} < \pi^{\text{W1DF}}$, it will suffice to show that $\lambda^{\text{PADJ}} < \lambda^{\text{W1DF}}$.

For these proofs, we also assume that the treatment effects are non-negative. If this is not the case, they can be transformed in order to meet this assumption.

Case 1

Suppose $\frac{(\beta_1^*)^2}{\frac{2}{Km}\sigma_1^2 VIF_1} < \frac{(\beta_2^*)^2}{\frac{2}{Km}\sigma_2^2 VIF_2}$. This implies that $\lambda^{\text{PADJ}} = \frac{(\beta_1^*)^2}{\frac{2}{Km}\sigma_1^2 VIF_1}$. We want to show that $\frac{(\beta_1^*)^2}{\frac{2}{Km}\sigma_1^2 VIF_1} < \left[\frac{\sqrt{\frac{(\beta_1^*)^2}{\frac{2\sigma_1^2}{Km}VIF_1}} + \sqrt{\frac{(\beta_2^*)^2}{\frac{2\sigma_2^2}{Km}VIF_2}}}{\sqrt{2\left(1+\frac{VIF_{12}}{\sqrt{VIF_1 VIF_2}}\right)}}\right]^2$. Use the fact that:

$$\frac{(\beta_1^*)^2}{\frac{2}{Km}\sigma_1^2 VIF_1} < \frac{(\beta_2^*)^2}{\frac{2}{Km}\sigma_2^2 VIF_2} \Rightarrow \frac{(\beta_1^*)^2}{\sigma_1^2 VIF_1} < \frac{(\beta_2^*)^2}{\sigma_2^2 VIF_2}$$
$$\Rightarrow (\beta_1^*)^2\sigma_2^2 VIF_2 < (\beta_2^*)^2\sigma_1^2 VIF_1$$
$$\Rightarrow \sqrt{(\beta_1^*)^2\sigma_2^2 VIF_2} < \sqrt{(\beta_2^*)^2\sigma_1^2 VIF_1}$$
$$\Rightarrow \beta_1^*\sigma_2\sqrt{VIF_2} < \beta_2^*\sigma_1\sqrt{VIF_1}$$

Since we already know that $VIF_{12} < VIF_2$ and $VIF_{12} < VIF_1$, then clearly $\sqrt{VIF_{12}} < \sqrt{VIF_2}$ and $\sqrt{VIF_{12}} < \sqrt{VIF_1}$. From this fact, it follows that $\beta_1^*\sigma_2\sqrt{VIF_{12}} < \beta_2^*\sigma_1\sqrt{VIF_1}$. Looking at this inequality now, we have:

$$\beta_1^*\sigma_2\sqrt{VIF_{12}} < \beta_2^*\sigma_1\sqrt{VIF_1}$$
$$\Rightarrow \beta_1^*\sigma_2\sqrt{VIF_{12}}\sqrt{VIF_{12}} < \beta_2^*\sigma_1\sqrt{VIF_1}\sqrt{VIF_{12}}$$
$$\Rightarrow \beta_1^*\sigma_2 VIF_{12} < \beta_2^*\sigma_1\sqrt{VIF_1}\sqrt{VIF_{12}} < \beta_2^*\sigma_1\sqrt{VIF_1}\sqrt{VIF_1}$$
$$\Rightarrow \beta_1^*\sigma_2 VIF_{12} < \beta_2^*\sigma_1\sqrt{VIF_1}\sqrt{VIF_{12}} < \beta_2^*\sigma_1 VIF_1$$
$$\Rightarrow \beta_1^*\sigma_2 VIF_{12} < \beta_2^*\sigma_1 VIF_1$$

and thus we have shown that the inequality $\beta_1^*\sigma_2 VIF_{12} < \beta_2^*\sigma_1 VIF_1$ holds. Now, we want to show the inequality:

$$\frac{(\beta_1^*)^2}{\frac{2}{Km}\sigma_1^2 VIF_1} \overset{?}{<} \left[\frac{\sqrt{\frac{(\beta_1^*)^2}{\frac{2\sigma_1^2}{Km}VIF_1}} + \sqrt{\frac{(\beta_2^*)^2}{\frac{2\sigma_2^2}{Km}VIF_2}}}{\sqrt{2\left(1+\frac{VIF_{12}}{\sqrt{VIF_1 VIF_2}}\right)}}\right]^2$$

$$\frac{\beta_1^*}{\sqrt{\frac{2}{Km}\sigma_1^2 VIF_1}} \overset{?}{<} \frac{\sqrt{\frac{(\beta_1^*)^2}{\frac{2\sigma_1^2}{Km}VIF_1}} + \sqrt{\frac{(\beta_2^*)^2}{\frac{2\sigma_2^2}{Km}VIF_2}}}{\sqrt{2\left(1+\frac{VIF_{12}}{\sqrt{VIF_1 VIF_2}}\right)}}$$

$$\frac{\beta_1^*}{\sqrt{\frac{2}{Km}}\sqrt{\sigma_1^2 VIF_1}} \overset{?}{<} \frac{\frac{\beta_1^*}{\sqrt{\frac{2}{Km}}\sqrt{\sigma_1^2 VIF_1}} + \frac{\beta_2^*}{\sqrt{\frac{2}{Km}}\sqrt{\sigma_2^2 VIF_2}}}{\sqrt{2\left(1 + \frac{VIF_{12}}{\sqrt{VIF_1 VIF_2}}\right)}}$$

$$\frac{\beta_1^*}{\sqrt{\sigma_1^2 VIF_1}} \overset{?}{<} \frac{\frac{\beta_1^*}{\sqrt{\sigma_1^2 VIF_1}} + \frac{\beta_2^*}{\sqrt{\sigma_2^2 VIF_2}}}{\sqrt{2}\sqrt{1 + \frac{VIF_{12}}{\sqrt{VIF_1 VIF_2}}}}$$

$$\frac{\beta_1^*}{\sqrt{\sigma_1^2 VIF_1}}\sqrt{2}\sqrt{1 + \frac{VIF_{12}}{\sqrt{VIF_1 VIF_2}}} \overset{?}{<} \frac{\beta_1^*}{\sqrt{\sigma_1^2 VIF_1}} + \frac{\beta_2^*}{\sqrt{\sigma_2^2 VIF_2}}$$

$$\beta_1^*\sqrt{2}\sqrt{1 + \frac{VIF_{12}}{\sqrt{VIF_1 VIF_2}}} \overset{?}{<} \frac{\beta_1^*\sqrt{\sigma_1^2 VIF_1}}{\sqrt{\sigma_1^2 VIF_1}} + \frac{\beta_2^*\sqrt{\sigma_1^2 VIF_1}}{\sqrt{\sigma_2^2 VIF_2}}$$

$$\beta_1^*\sqrt{2}\sqrt{1 + \frac{VIF_{12}}{\sqrt{VIF_1 VIF_2}}} \overset{?}{<} \beta_1^* + \beta_2^*\frac{\sqrt{\sigma_1^2 VIF_1}}{\sqrt{\sigma_2^2 VIF_2}}$$

$$\beta_1^*\sqrt{2}\sqrt{1 + \frac{VIF_{12}}{\sqrt{VIF_1 VIF_2}}} - \beta_1^* \overset{?}{<} \beta_2^*\frac{\sqrt{\sigma_1^2 VIF_1}}{\sqrt{\sigma_2^2 VIF_2}}$$

$$\beta_1^*\sigma_2\sqrt{VIF_2}\sqrt{2}\sqrt{1 + \frac{VIF_{12}}{\sqrt{VIF_1 VIF_2}}} - \beta_1^*\sigma_2\sqrt{VIF_2} \overset{?}{<} \beta_2^*\sigma_1\sqrt{VIF_1}$$

$$\beta_1^*\sigma_2\sqrt{VIF_2}\left(\sqrt{2}\sqrt{1 + \frac{VIF_{12}}{\sqrt{VIF_1 VIF_2}}} - 1\right) \overset{?}{<} \beta_2^*\sigma_1\sqrt{VIF_1}$$

$$\sqrt{2}\sqrt{1 + \frac{VIF_{12}}{\sqrt{VIF_1 VIF_2}}} - 1 \overset{?}{<} \frac{\beta_2^*\sigma_1\sqrt{VIF_1}}{\beta_1^*\sigma_2\sqrt{VIF_2}}$$

Since we know that $\beta_1^*\sigma_2\sqrt{VIF_2} < \beta_2^*\sigma_1\sqrt{VIF_1}$, we know that $1 < \frac{\beta_2^*\sigma_1\sqrt{VIF_1}}{\beta_1^*\sigma_2\sqrt{VIF_2}}$. So, to prove the inequality above holds, we just need to show that $\sqrt{2}\sqrt{1 + \frac{VIF_{12}}{\sqrt{VIF_1 VIF_2}}} - 1 < 1$.

$$\sqrt{2}\sqrt{1 + \frac{VIF_{12}}{\sqrt{VIF_1 VIF_2}}} - 1 \overset{?}{<} 1$$

$$\sqrt{1 + \frac{VIF_{12}}{\sqrt{VIF_1 VIF_2}}} \overset{?}{<} \frac{2}{\sqrt{2}}$$

$$1 + \frac{VIF_{12}}{\sqrt{VIF_1 VIF_2}} \overset{?}{<} \frac{4}{2}$$

$$\frac{VIF_{12}}{\sqrt{VIF_1 VIF_2}} \overset{?}{<} 1$$

$$VIF_{12} \overset{?}{<} \sqrt{VIF_1 VIF_2}$$

$$VIF_{12} VIF_{12} \overset{?}{<} VIF_1 VIF_2$$

Now, since $VIF_{12} < VIF_1$ and $VIF_{12} < VIF_2$, it must be the case that $VIF_{12}VIF_{12} < VIF_1VIF_2$. Thus, the inequality holds, and we've shown that

$$\lambda^{\text{PADJ}} = \frac{(\beta_1^*)^2}{\frac{2}{Km}\sigma_1^2 VIF_1} < \left[ \frac{\sqrt{\frac{(\beta_1^*)^2}{\frac{2\sigma_1^2}{Km} VIF_1}} + \sqrt{\frac{(\beta_2^*)^2}{\frac{2\sigma_2^2}{Km} VIF_2}}}{\sqrt{2\left(1 + \frac{VIF_{12}}{\sqrt{VIF_1 VIF_2}}\right)}} \right]^2 = \lambda^{\text{W1DF}}$$

Then in terms of power, since $\lambda^{\text{PADj}} < \lambda^{\text{W1DF}}$, and $\alpha^{\text{PADJ}} < \alpha^{\text{W1DF}}$ it follows that $\pi^{\text{PADJ}} < \pi^{\text{W1DF}}$, meaning that the p-value adjustment methods will always be less powerful than the single weighted 1-DF test.

<u>Case 2</u>

For this case, we suppose instead that $\frac{(\beta_2^*)^2}{\frac{2}{Km}\sigma_2^2 VIF_2} < \frac{(\beta_1^*)^2}{\frac{2}{Km}\sigma_1^2 VIF_1}$. This implies that $\lambda^{\text{PADJ}} = \frac{(\beta_2^*)^2}{\frac{2}{Km}\sigma_2^2 VIF_2}$. The proof follows from the same logic as Case 1.

### A.4 Comparison of Method 1: P-value Adjustments and Method 4: Disjunctive 2-DF Test

$$\lambda^{\text{PADJ}} = \min\left(\frac{(\beta_1^*)^2}{2\,\frac{\sigma_1^2}{Km}\left[1 + (m-1)\rho_0^{(1)}\right]}, \frac{(\beta_2^*)^2}{2\,\frac{\sigma_2^2}{Km}\left[1 + (m-1)\rho_0^{(2)}\right]}\right)$$

$$\lambda^{\text{DIS2DF}} = \left[\frac{Km\left[(\beta_1^*)^2\,\sigma_2^2\,\left(1 + (m-1)\rho_0^{(2)}\right) - 2\,\beta_1^*\,\beta_2^*\,\sigma_1\,\sigma_2\,\left(\rho_2^{(1,2)} + (m-1)\rho_1^{(1,2)}\right) + (\beta_2^*)^2\,\sigma_1^2\,\left(1 + (m-1)\rho_0^{(1)}\right)\right]}{2\,\sigma_1^2\,\sigma_2^2\,\left[\left(1 + (m-1)\rho_0^{(1)}\right)\left(1 + (m-1)\rho_0^{(2)}\right) - \left(\rho_2^{(1,2)} + (m-1)\rho_1^{(1,2)}\right)^2\right]}\right]$$

Can also be written in terms of variance inflation factor terms:

$$\lambda^{\text{PADJ}} = \min\left(\frac{(\beta_1^*)^2}{\frac{2}{Km}\sigma_1^2 VIF_1}, \frac{(\beta_2^*)^2}{\frac{2}{Km}\sigma_2^2 VIF_2}\right)$$

$$\lambda^{\text{DIS2DF}} = \frac{Km[(\beta_1^*)^2\,\sigma_2^2\,VIF_2 - 2\,\beta_1^*\,\beta_2^*\,\sigma_1\,\sigma_2\,VIF_{12} + (\beta_2^*)^2\,\sigma_1^2\,VIF_1]}{2\,\sigma_1^2\,\sigma_2^2\,[VIF_1\,VIF_2 - VIF_{12}^2]}$$

Recall that $VIF_1 = 1 + (m-1)\rho_0^{(1)}; VIF_2 = 1 + (m-1)\rho_0^{(2)}; VIF_{12} = \rho_2^{(1,2)} + (m-1)\rho_1^{(1,2)}$. We also have that $\rho_1^{(1,2)} < \rho_0^{(1)}$; $\rho_1^{(1,2)} < \rho_0^{(2)}$; $\rho_2^{(1,2)} < 1$. This implies $VIF_{12} < VIF_1$ and $VIF_{12} < VIF_2$.

Note that for the p-value adjustment method, $\pi^{\text{PADJ}} = \min\left(\pi^{(1)}, \pi^{(2)}\right) = \min\left(1 - \chi^2\left[\lambda^{(1)}, 1\right], 1 - \chi^2\left[\lambda^{(2)}, 1\right]\right)$, which can also be written as $\pi^{\text{PADJ}} = 1 - \chi^2\left[\min\left(\lambda^{(1)}, \lambda^{(2)}\right), 1\right]$. For the disjunctive 2-DF test, $\pi^{\text{DIS2DF}} = 1 - \int_0^{5.99} \chi^2(x;\ 2, \lambda^{\text{DIS2DF}})dx$. These two methods for calculating power use the same distribution but with differing degrees of freedom.

We want to understand the relationship between $\lambda^{\text{PADJ}}$ and $\lambda^{\text{DIS2DF}}$. If it is the case that $\lambda^{\text{PADJ}} < \lambda^{\text{DIS2DF}}$, we will further examine whether or not this implies that $\pi^{\text{PADJ}} < \pi^{\text{DIS2DF}}$.

For these proofs, we also assume that the treatment effects are non-negative. If this is not the case, they can be transformed in order to meet this assumption.

Case 1

Suppose $\frac{(\beta_1^*)^2}{\frac{2}{Km}\sigma_1^2 VIF_1} < \frac{(\beta_2^*)^2}{\frac{2}{Km}\sigma_2^2 VIF_2}$. This implies that $\lambda^{\text{PADJ}} = \frac{(\beta_1^*)^2}{\frac{2}{Km}\sigma_1^2 VIF_1}$. We want to show that $\frac{(\beta_1^*)^2}{\frac{2}{Km}\sigma_1^2 VIF_1} < \frac{Km[(\beta_1^*)^2\,\sigma_2^2\,VIF_2 - 2\,\beta_1^*\,\beta_2^*\,\sigma_1\,\sigma_2\,VIF_{12} + (\beta_2^*)^2\,\sigma_1^2\,VIF_1]}{2\,\sigma_1^2\,\sigma_2^2\,[VIF_1\,VIF_2 - VIF_{12}^2]}$. Use the fact that:

$$\frac{(\beta_1^*)^2}{\frac{2}{Km}\sigma_1^2 VIF_1} < \frac{(\beta_2^*)^2}{\frac{2}{Km}\sigma_2^2 VIF_2} \Rightarrow \frac{(\beta_1^*)^2}{\sigma_1^2 VIF_1} < \frac{(\beta_2^*)^2}{\sigma_2^2 VIF_2}$$

$$\Rightarrow (\beta_1^*)^2\sigma_2^2 VIF_2 < (\beta_2^*)^2\sigma_1^2 VIF_1$$

$$\Rightarrow \sqrt{(\beta_1^*)^2\sigma_2^2 VIF_2} < \sqrt{(\beta_2^*)^2\sigma_1^2 VIF_1}$$

$$\Longrightarrow \beta_1^* \sigma_2 \sqrt{VIF_2} < \ \beta_2^* \sigma_1 \sqrt{VIF_1}$$

Since we already know that $VIF_{12} < VIF_2$ and $VIF_{12} < VIF_1$, then clearly $\sqrt{VIF_{12}} < \sqrt{VIF_2}$ and $\sqrt{VIF_{12}} < \sqrt{VIF_1}$. From this fact, it follows that $\beta_1^* \sigma_2 \sqrt{VIF_{12}} < \beta_2^* \sigma_1 \sqrt{VIF_1}$. Looking at this inequality now, we have:

$$\beta_1^* \sigma_2 \sqrt{VIF_{12}} < \beta_2^* \sigma_1 \sqrt{VIF_1}$$
$$\Longrightarrow \beta_1^* \sigma_2 \sqrt{VIF_{12}} \sqrt{VIF_{12}} < \beta_2^* \sigma_1 \sqrt{VIF_1} \sqrt{VIF_{12}}$$
$$\Longrightarrow \beta_1^* \sigma_2 VIF_{12} < \beta_2^* \sigma_1 \sqrt{VIF_1} \sqrt{VIF_{12}} < \beta_2^* \sigma_1 \sqrt{VIF_1} \sqrt{VIF_1}$$
$$\Longrightarrow \beta_1^* \sigma_2 VIF_{12} < \beta_2^* \sigma_1 \sqrt{VIF_1} \sqrt{VIF_{12}} < \beta_2^* \sigma_1 VIF_1$$
$$\Longrightarrow \beta_1^* \sigma_2 VIF_{12} < \beta_2^* \sigma_1 VIF_1$$

and thus we've shown that the inequality $\beta_1^* \sigma_2 VIF_{12} < \beta_2^* \sigma_1 VIF_1$ holds. Now, we want to show the inequality:

$$\frac{(\beta_1^*)^2}{\frac{2}{Km} \sigma_1^2 VIF_1} < \frac{Km[(\beta_1^*)^2 \ \sigma_2^2 \ VIF_2 \ - \ 2 \ \beta_1^* \ \beta_2^* \ \sigma_1 \ \sigma_2 \ VIF_{12} \ + \ (\beta_2^*)^2 \ \sigma_1^2 \ VIF_1]}{2 \ \sigma_1^2 \ \sigma_2^2 \ [VIF_1 \ VIF_2 \ - \ VIF_{12}^2]}$$
$$\frac{(\beta_1^*)^2}{VIF_1} < \frac{(\beta_1^*)^2 \ \sigma_2^2 \ VIF_2 \ - \ 2 \ \beta_1^* \ \beta_2^* \ \sigma_1 \ \sigma_2 \ VIF_{12} \ + \ (\beta_2^*)^2 \ \sigma_1^2 \ VIF_1}{\sigma_2^2 \ [VIF_1 \ VIF_2 \ - \ VIF_{12}^2]}$$

$$\begin{aligned}&(\beta_1^*)^2 \sigma_2^2 \ [VIF_1 \ VIF_2 \ - \ VIF_{12}^2] \\ &\qquad < (\beta_1^*)^2 \ \sigma_2^2 \ VIF_1 VIF_2 \ - \ 2 \ \beta_1^* \ \beta_2^* \ \sigma_1 \ \sigma_2 VIF_1 \ VIF_{12} \ + \ (\beta_2^*)^2 \ \sigma_1^2 VIF_1 \ VIF_1\end{aligned}$$

$$\begin{aligned}&(\beta_1^*)^2 \sigma_2^2 \ VIF_1 \ VIF_2 - \ (\beta_1^*)^2 \sigma_2^2 VIF_{12}^2 \\ &\qquad < (\beta_1^*)^2 \ \sigma_2^2 \ VIF_1 VIF_2 \ - \ 2 \ \beta_1^* \ \beta_2^* \ \sigma_1 \ \sigma_2 VIF_1 \ VIF_{12} \ + \ (\beta_2^*)^2 \ \sigma_1^2 VIF_1 \ VIF_1\end{aligned}$$

$$-\ (\beta_1^*)^2 \sigma_2^2 VIF_{12}^2 < -\ 2 \ \beta_1^* \ \beta_2^* \ \sigma_1 \ \sigma_2 VIF_1 \ VIF_{12} \ + \ (\beta_2^*)^2 \ \sigma_1^2 VIF_1 \ VIF_1$$

$$0 < (\beta_2^*)^2 \ \sigma_1^2 VIF_1^2 + (\beta_1^*)^2 \sigma_2^2 VIF_{12}^2 - 2 \ \beta_1^* \ \beta_2^* \ \sigma_1 \ \sigma_2 VIF_1 \ VIF_{12}$$

$$0 < (\beta_2^* \sigma_1 VIF_1 - \beta_1^* \sigma_2 VIF_{12})^2$$

We've already shown that $\beta_1^* \sigma_2 VIF_{12} < \ \beta_2^* \sigma_1 VIF_1$, so $\beta_2^* \sigma_1 VIF_1 - \beta_1^* \sigma_2 VIF_{12} > 0$, and clearly $(\beta_2^* \sigma_1 VIF_1 - \beta_1^* \sigma_2 VIF_{12})^2 > 0$. Thus, the inequality holds, and we've shown that

$$\begin{aligned}\lambda^{\text{PADJ}} = \frac{(\beta_1^*)^2}{\frac{2}{Km} \sigma_1^2 VIF_1} &< \frac{Km[(\beta_1^*)^2 \ \sigma_2^2 \ VIF_2 \ - \ 2 \ \beta_1^* \ \beta_2^* \ \sigma_1 \ \sigma_2 \ VIF_{12} \ + \ (\beta_2^*)^2 \ \sigma_1^2 \ VIF_1]}{2 \ \sigma_1^2 \ \sigma_2^2 \ [VIF_1 \ VIF_2 \ - \ VIF_{12}^2]} \\ &= \lambda^{\text{DIS2DF}}\end{aligned}$$

Now, since the degrees-of-freedom differ between these tests, we want to examine if $\pi^{\text{PADJ}} = 1 - \chi^2[\lambda^{\text{PADJ}}, 1] < \pi^{\text{DIS2DF}} = 1 - \chi^2[\lambda^{\text{DIS2DF}}, 2]$ while accounting for the change in degrees-of-freedom. Their integrations can be written as

$$\pi^{\text{PADJ}} = \int_{c^{\text{PADJ}}}^{\infty} \chi^2(x; \ 1, \lambda^{\text{PADJ}}) dx = Q_{1/2}\left(\sqrt{\lambda^{\text{PADJ}}}, \sqrt{c^{\text{PADJ}}}\right)$$

$$\pi^{\mathrm{DIS2DF}} = \int_{c^{\mathrm{DIS2DF}}}^{\infty} \chi^2(x;\ 2, \lambda^{\mathrm{DIS2DF}})dx = Q_1\left(\sqrt{\lambda^{\mathrm{DIS2DF}}}, \sqrt{c^{\mathrm{DIS2DF}}}\right)$$

Even though we've shown that $\lambda^{\mathrm{PADJ}} < \lambda^{\mathrm{DIS2DF}}$, this does not necessarily imply that $\pi^{\mathrm{PADJ}} < \pi^{\mathrm{DIS2DF}}$. Similar to the comparison of the single weighted 1-DF test and the disjunctive 2-DF test, the differing degrees-of-freedom result in differing orders of their Marcum Q-Function forms. The Marcum Q-Function, $Q_{d/2}(\sqrt{\lambda}, \sqrt{c})$, is strictly increasing in $d/2$ and $\sqrt{\lambda}$ for all $\sqrt{\lambda} \geq 0$ and $\sqrt{c}$, $d/2 > 0$. It is strictly decreasing in $\sqrt{c}$ for all $\sqrt{\lambda}$, $\sqrt{c} \geq 0$ and $d/2 > 0$. In other words, $Q_{d/2}(\sqrt{\lambda}, \sqrt{c})$ increases as $d/2$ increases, and $Q_{d/2}(\sqrt{\lambda}, \sqrt{c})$ decreases as $\sqrt{c}$ increases. Due to these competing effects, it is not clear how $Q_{1/2}\left(\sqrt{\lambda^{\mathrm{PADJ}}}, \sqrt{c^{\mathrm{PADJ}}}\right)$ compares to $Q_1\left(\sqrt{\lambda^{\mathrm{DIS2DF}}}, \sqrt{c^{\mathrm{DIS2DF}}}\right)$ even though $\lambda^{\mathrm{PADJ}} < \lambda^{\mathrm{DIS2DF}}$. However, for a specific overall false-positive rate, we can plot the function values (i.e. statistical power) over many possible values for the non-centrality parameters. See main text for final results and analysis.

The derivation for Case 2: $\frac{(\beta_2^*)^2}{\frac{2}{Km}\sigma_2^2 VIF_2} < \frac{(\beta_1^*)^2}{\frac{2}{Km}\sigma_1^2 VIF_1}$, follows with the same logic as Case 1.

To examine the relationship between the p-value adjustment methods and the disjunctive 2-DF test further, we set the overall false-positive rate to be $\alpha = 0.05$. Then, letting $\lambda^{\mathrm{PADJ}} \in (0,30)$ and $\lambda^{\mathrm{DIS2DF}} \in (0,30)$, we found each combination of $\lambda^{\mathrm{PADJ}}$ and $\lambda^{\mathrm{DIS2DF}}$ with the constraint that $\lambda^{\mathrm{PADJ}} < \lambda^{\mathrm{DIS2DF}}$. Then, we plotted $\pi^{\mathrm{DIS2DF}} - \pi^{\mathrm{PADJ}}$ against $\lambda^{\mathrm{DIS2DF}} - \lambda^{\mathrm{PADJ}}$ for the Bonferroni and Sidak Correction, and the D/AP correction for $\rho_2^{(1,2)} \in [0.1, 0.3, 0.5, 0.7]$. For almost every case, $\pi^{\mathrm{DIS2DF}} > \pi^{\mathrm{PADJ}}$, but there are a small number of cases where $\pi^{\mathrm{DIS2DF}} < \pi^{\mathrm{PADJ}}$, namely when the difference between $\lambda^{\mathrm{PADJ}}$ and $\lambda^{\mathrm{DIS2DF}}$ is very small. For the D/AP method, as $\rho_2^{(1,2)}$ decreases, $\lambda^{\mathrm{DIS2DF}} - \lambda^{\mathrm{PADJ}}$ needs to be even smaller for $\pi^{\mathrm{DIS2DF}} < \pi^{\mathrm{PADJ}}$ to occur. To summarize, we've shown that $\lambda^{\mathrm{PADJ}} < \lambda^{\mathrm{DIS2DF}}$, but due to the differing degrees-of-freedom, it is not globally true that $\pi^{\mathrm{DIS2DF}} > \pi^{\mathrm{PADJ}}$. Figure 6 and Figure 7 show that for most values of $\lambda^{\mathrm{PADJ}}$ and $\lambda^{\mathrm{DIS2DF}}$, it is generally observed that $\pi^{\mathrm{DIS2DF}} > \pi^{\mathrm{PADJ}}$ for each p-value adjustment method (with precision of 4 decimal places), but this relationship is affected by the value of $\rho_2^{(1,2)}$ for the D/AP method. We've shown that it is theoretically possible for $\pi^{\mathrm{DIS2DF}} = \pi^{\mathrm{PADJ}}$ or $\pi^{\mathrm{DIS2DF}} < \pi^{\mathrm{PADJ}}$, but these cases are not likely to be observed in practice.

Figure 6. Theoretical comparison of statistical power between the P-value Adjustment Methods (Bonferroni and Sidak) and the Disjunctive 2-DF Test for varying non-centrality parameters using $\chi^2$-distribution for $\lambda^{PADJ} < \lambda^{DIS2DF}$

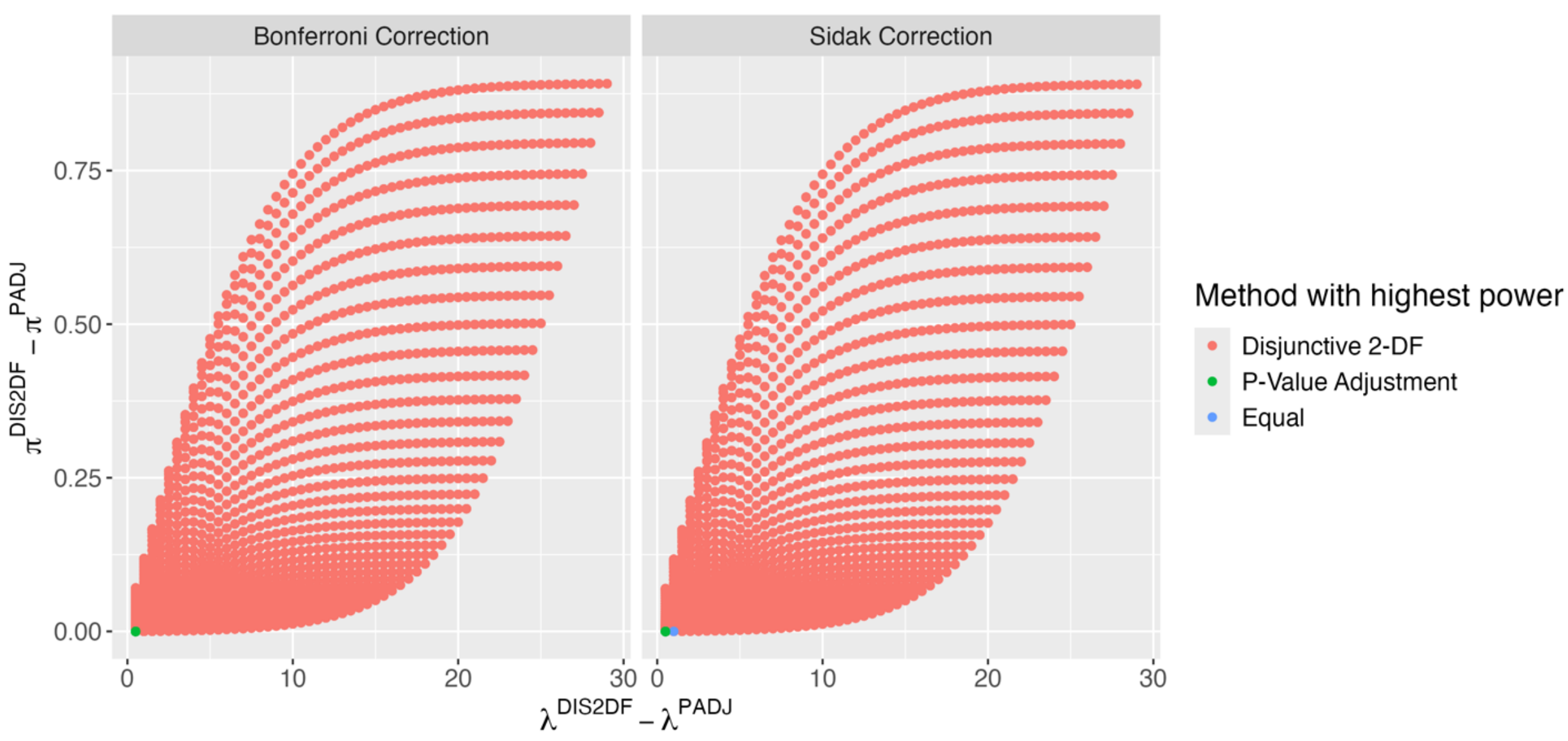

Figure 7. Theoretical comparison of statistical power between the P-value Adjustment Methods (D/AP Correction) and the Disjunctive 2-DF Test for varying non-centrality parameters using $\chi^2$-distribution for $\lambda^{PADJ} < \lambda^{DIS2DF}$

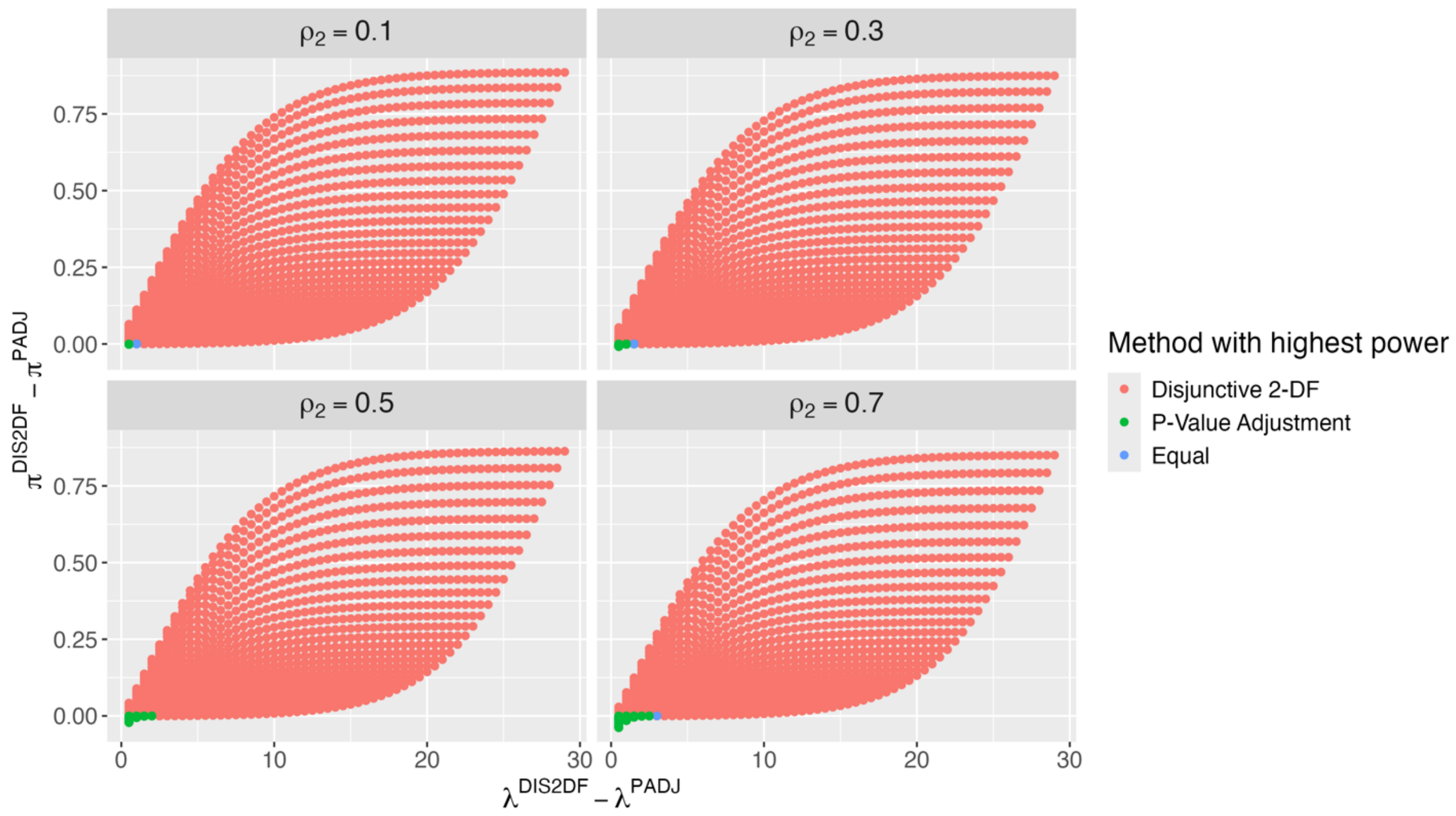

### A.5 Method 4: Disjunctive 2-DF Test as a 1-DF Test

For the disjunctive 2-DF test, it is shown in Yang et al. that under equal treatment allocation, the noncentrality parameter is written as $2K(\boldsymbol{L}\beta^*)^T \left(\boldsymbol{L}\,\Omega_{\beta^*}\boldsymbol{L}^T\right)^{-1}\boldsymbol{L}\beta^*$, where $\Omega_{\beta^*}$ is the variance-covariance matrix for the treatment effect estimate $\beta^*$.[5] For two co-primary outcomes, the contrast matrix is a $2 \times 2$ matrix, $\boldsymbol{L} = \begin{bmatrix} 1 & 0 \\ 0 & 1 \end{bmatrix}$, and, under equal treatment allocation,

$$\omega_1^2 = \frac{4\,\sigma_1^2\left[1 + (m-1)\rho_0^{(1)}\right]}{m} = \frac{4\,\sigma_1^2\,VIF_1}{m};\quad \omega_2^2 = \frac{4\,\sigma_2^2\left[1 + (m-1)\rho_0^{(2)}\right]}{m} = \frac{4\,\sigma_2^2 VIF_2}{m};$$

$$\omega_{1,2} = \frac{4\,\sigma_1\sigma_2\,[\rho_2^{(1,2)} + (m-1)\rho_1^{(1,2)}]}{m} = \frac{4\,\sigma_1\sigma_2 VIF_{12}}{m}$$

where $\omega_1^2$ and $\omega_2^2$ are the elements on the main diagonal of the $2 \times 2$ variance-covariance matrix, and $\omega_{qq'}$ is the element of the off-diagonal. The non-centrality parameter is thus

$$\lambda^{\mathrm{DISJ}} = \frac{Km[(\beta_1^*)^2\sigma_2^2 VIF_2 + (\beta_2^*)^2\sigma_1^2\,VIF_1 - 2\beta_1^*\beta_2^*\sigma_1\sigma_2 VIF_{12}]}{2\sigma_1^2\sigma_2^2[VIF_1 VIF_2 - VIF_{12}^2]}.$$

Recall that the variance inflation factor terms are as follows:

$$VIF_1 = 1 + (m-1)\rho_0^{(1)};\ VIF_2 = 1 + (m-1)\rho_0^{(2)};\ \ VIF_{12} = \rho_2^{(1,2)} + (m-1)\rho_1^{(1,2)}.$$

Recall also that the non-centrality parameter of the combined outcomes approach can be written as

$$\lambda^{\mathrm{COMB}} = \frac{(\beta_1^* + \beta_2^*)^2}{\frac{2}{Km}[\sigma_1^2\,VIF_1 + \sigma_2^2\,VIF_2 + 2\sigma_1\sigma_2 VIF_{12}]}.$$

If instead we collapsed the contrast matrix $\boldsymbol{L}$ into a single row, i.e. $\boldsymbol{L} = [1\ 1]$, this test would no longer be a 2-DF test, and the non-centrality parameter reduces to $\lambda^{\mathrm{COMB}}$. This derivation is shown below.

$$\begin{aligned}
\lambda^{\mathrm{DISJ,DF=1}} &= 2K(\boldsymbol{L}\beta^*)^T \left(\boldsymbol{L}\,\Omega_{\beta^*}\boldsymbol{L}^T\right)^{-1}\boldsymbol{L}\beta^* \\
&= 2K\left(\begin{bmatrix} 1 & 1 \end{bmatrix}\begin{bmatrix} \beta_1^* \\ \beta_2^* \end{bmatrix}\right)^T \left(\begin{bmatrix} 1 & 1 \end{bmatrix}\begin{bmatrix} \omega_1^2 & \omega_{1,2} \\ \omega_{1,2} & \omega_2^2 \end{bmatrix}\begin{bmatrix} 1 \\ 1 \end{bmatrix}\right)^{-1} \begin{bmatrix} 1 & 1 \end{bmatrix}\begin{bmatrix} \beta_1^* \\ \beta_2^* \end{bmatrix} \\
&= 2K(\beta_1^* + \beta_2^*)\left(\begin{bmatrix} \omega_1^2 + \omega_{1,2} & \omega_2^2 + \omega_{1,2} \end{bmatrix}\begin{bmatrix} 1 \\ 1 \end{bmatrix}\right)^{-1}(\beta_1^* + \beta_2^*) \\
&= 2K\frac{(\beta_1^* + \beta_2^*)^2}{\omega_1^2 + \omega_2^2 + 2\omega_{1,2}} \\
&= 2K\frac{(\beta_1^* + \beta_2^*)^2}{\frac{4\,\sigma_1^2\,VIF_1}{m} + \frac{4\,\sigma_2^2 VIF_2}{m} + \frac{(2)4\,\sigma_1\sigma_2 VIF_{12}}{m}}
\end{aligned}$$

$$= \frac{(\beta_1^* + \beta_2^*)^2}{\frac{2}{Km}[\sigma_1^2\, VIF_1 + \sigma_2^2 VIF_2 + 2\sigma_1\sigma_2 VIF_{12}]}$$
$$= \lambda^{\text{COMB}}$$

Thus, this shows that when the disjunctive 2-DF test is reduced to a disjunctive 1-DF test, it is algebraically equivalent to the combined outcomes approach. Because the combined outcomes approach is theoretically equivalent to the single weighted 1-DF test when the outcome specific ICCs and variances between the two outcomes are the same (shown in Owen et al.[3]), this means that this relationship holds between the disjunctive 2-DF test and the single weighted 1-DF test when the disjunctive 2-DF test is instead a 1-DF test with $\boldsymbol{L} = [1\ 1]$. Importantly, collapsing the contrast matrix to a single row does not change the underlying hypothesis from disjunctive to conjunctive. In the 1-DF setting, the null hypothesis is rejected when the combined linear contrast differs from zero, meaning that a nonzero effect on either outcome is sufficient to reject. A conjunctive test, in contrast, would require evidence that both treatment effects are simultaneously nonzero, which is not what the single-row contrast evaluates.

## B. Statistical Power in Relation to Each Input Parameter

An evaluation was conducted that examined ho each input parameter individually impacts power when allowed to vary over a larger set of values than what was feasible in the numerical evaluation, holding all other input parameters constant. Values for the parameters that did not vary were chosen from the numerical evaluation parameters, and fixed at $K = 8$, $m = 50$, $(\beta_1^*, \beta_2^*) = (0.4, 0.4)$, $(\sigma_1^2, \sigma_2^2) = (1,1)$, $\left(\rho_0^{(1)}, \rho_0^{(2)}\right) = (0.1, 0.1)$, $\rho_1^{(1,2)} = 0.01$, and $\rho_2^{(1,2)} = 0.1$. These values remained the same for Figure 8 and Figure 9 with the exception of the input parameter that was allowed to vary. We conducted this additional analysis for each of the four comparisons, but discuss the results for Comparison I only. Results for Comparison II-IV are available in the Supplementary Material.

Figure 8 shows the statistical power based on varying values for $\beta_2^*/\beta_1^*$, $\sigma_2^2/\sigma_1^2$, $K$, and $m$. Naturally, as the number of clusters increases ($K$), so too will the statistical power, and drastically so for each design method. Similarly, as the number of individuals in each cluster increases ($m$), so too will the statistical power, though this increase is less extreme for each of the design methods. Because increasing $K$ and $m$ did not impact which method was the most powerful, second most powerful, etc., it was clear that these input parameters were less important to report on in the numerical evaluation.

Examining the ratio of the treatment effects, namely $\beta_2^*/\beta_1^*$, we see that as this ratio increases, so does the power, and the method that was most powerful changed as the ratio increased. Similarly, as the ratio of the variances, namely $\sigma_2^2/\sigma_1^2$, increased, the statistical power decreased for all of the methods, and the method that was the most powerful also changed as the ratio increased. This observation motivated the need to look into the treatment effects and variances impacts on power further, which we detail in the main manuscript. We also note that curves for the p-value adjustment methods are not smooth in regard to these ratios – this is because the p-value adjustment methods take the minimum power over the two outcomes separately, and when $\beta_2^* > \beta_1^*$, the final power will always be $\pi^{\text{PADJ}} = \min\left(\pi^{(1)}, \pi^{(2)}\right) = \pi^{(1)}$. For example, for the Bonferroni adjustment, when $\beta_1^* = 0.4$ and $\beta_2^* = 0.3$, $\pi^{\text{PADJ}} = \min(44\%, 25\%) = 25\%$; when $\beta_1^* = 0.4$ and $\beta_2^* = 0.4$, $\pi^{\text{PADJ}} = \min(44\%, 44\%) = 44\%$; when $\beta_1^* = 0.4$ and $\beta_2^* = 0.5$, $\pi^{\text{PADJ}} = \min(44\%, 64\%) = 44\%$. No matter how much $\pi^{(2)}$ grows, the power cannot exceed $\pi^{(1)}$ since $\pi^{(1)} < \pi^{(2)}$ in this situation where $\beta_2^*$ is allowed to vary but $\beta_1^*$ is constant; the same logic applies for the variances and the outcome specific ICCs.

We also examined the impact of the correlations in the same way, and Figure 9 shows the statistical power based on varying values for $\rho_1^{(1,2)}$, $\rho_2^{(1,2)}$, and $\rho_0^{(2)}/\rho_0^{(1)}$. As the inter-subject between-endpoint ICC increases ($\rho_1^{(1,2)}$), the power decreases for the single weighted 1-DF test and disjunctive 2-DF test, and increases for conjunctive IU test. As the intra-subject between endpoint ICC increases ($\rho_2^{(1,2)}$), the power very slightly decreased for the single weighted 1-DF test, disjunctive 2-DF test, remained constant for the Bonferroni and Sidak method, and increased slightly for the D/AP p-value adjustment method. It also increases for the conjunctive test – a phenomenon that is discussed in Yang et al.[5]. We speculate that as correlation between the endpoints increase, large values on one outcome will tend to coincide with large values on the other, making it easier to satisfy the conjunctive requirement and thus boosting power. However, these effects are slight, and it is clear that this variable does not have a large impact on the statistical power. Contrastingly, the outcome specific ICCs, $\rho_0^{(1)}$, and $\rho_0^{(2)}$, have a much more substantial impact on statistical power, changing the ranking of which method is the most powerful as $\rho_0^{(1)}/\rho_0^{(2)}$ increases. So, we further investigated the effects of $\rho_0^{(1)}$ and $\rho_0^{(2)}$ on power.

Figure 8. Statistical power based on varying values for $\beta_1^*$, $\beta_2^*$, $\sigma_1^2$, $\sigma_2^2$, $K$, and $m$ for Comparison I (F and t distributions with 2-sided conjunctive test)

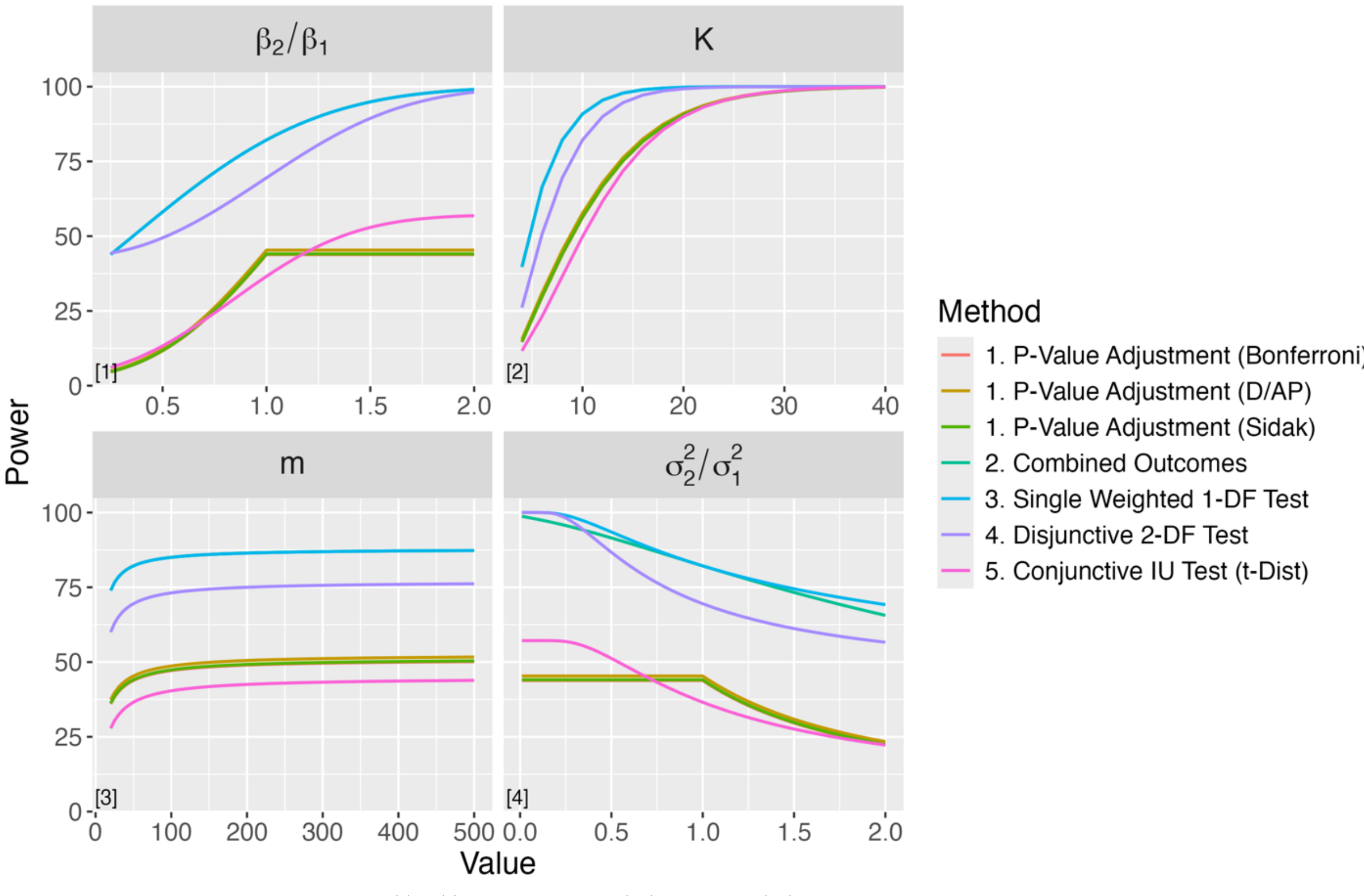

[1] $K = 8$, $m = 50$, $(\sigma_1^2, \sigma_2^2) = (1,1)$, $\left(\rho_0^{(1)}, \rho_0^{(2)}\right) = (0.1, 0.1)$, $\rho_1^{(1,2)} = 0.01$, $\rho_2^{(1,2)} = 0.1$, $\beta_1^* = 0.4$ and $\beta_2^* \in [0.1, 0.8]$
[2] $m = 50$, $(\beta_1^*, \beta_2^*) = (0.4, 0.4)$, $(\sigma_1^2, \sigma_2^2) = (1,1)$, $\left(\rho_0^{(1)}, \rho_0^{(2)}\right) = (0.1, 0.1)$, $\rho_1^{(1,2)} = 0.01$, $\rho_2^{(1,2)} = 0.1$, and $K \in [4, 24]$
[3] $K = 8$, $(\beta_1^*, \beta_2^*) = (0.4, 0.4)$, $(\sigma_1^2, \sigma_2^2) = (1,1)$, $\left(\rho_0^{(1)}, \rho_0^{(2)}\right) = (0.1, 0.1)$, $\rho_1^{(1,2)} = 0.01$, $\rho_2^{(1,2)} = 0.1$, and $m \in [20, 500]$
[4] $K = 8$, $m = 50$, $(\beta_1^*, \beta_2^*) = (0.4, 0.4)$, $\left(\rho_0^{(1)}, \rho_0^{(2)}\right) = (0.1, 0.1)$, $\rho_1^{(1,2)} = 0.01$, $\rho_2^{(1,2)} = 0.1$, $\sigma_1^2 = 1$, and $\sigma_2^2 \in [0.01, 2]$

Figure 9. Statistical power based on varying values for $\rho_1^{(1,2)}$, $\rho_2^{(1,2)}$, $\rho_0^{(1)}$, and $\rho_0^{(2)}$ for Comparison I (F and t distributions with 2-sided conjunctive test)

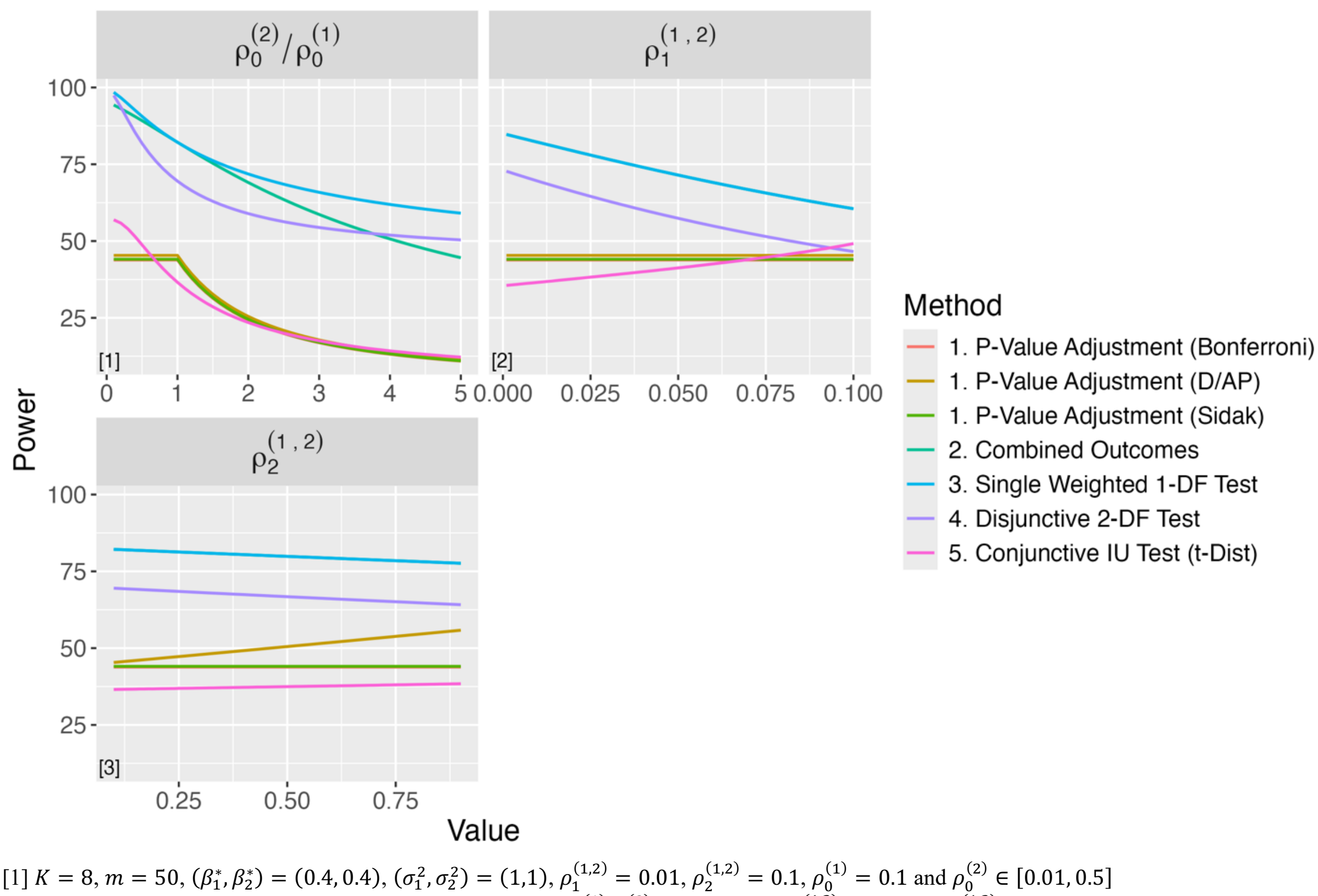

[1] $K = 8$, $m = 50$, $(\beta_1^*, \beta_2^*) = (0.4, 0.4)$, $(\sigma_1^2, \sigma_2^2) = (1,1)$, $\rho_1^{(1,2)} = 0.01$, $\rho_2^{(1,2)} = 0.1$, $\rho_0^{(1)} = 0.1$ and $\rho_0^{(2)} \in [0.01, 0.5]$
[2] $K = 8$, $m = 50$, $(\beta_1^*, \beta_2^*) = (0.4, 0.4)$, $(\sigma_1^2, \sigma_2^2) = (1,1)$, $\left(\rho_0^{(1)}, \rho_0^{(2)}\right) = (0.1, 0.1)$, $\rho_2^{(1,2)} = 0.1$, and $\rho_1^{(1,2)} \in [0.001, 0.1]$
[3] $K = 8$, $m = 50$, $(\beta_1^*, \beta_2^*) = (0.4, 0.4)$, $(\sigma_1^2, \sigma_2^2) = (1,1)$, $\left(\rho_0^{(1)}, \rho_0^{(2)}\right) = (0.1, 0.1)$, $\rho_1^{(1,2)} = 0.01$, and $\rho_2^{(1,2)} \in [0.1, 0.9]$